\documentclass[a4paper]{article}
\usepackage[a4paper, left=2.5cm, right=2.5cm, top=2.5cm, bottom=2.5cm]{geometry}
\usepackage{graphicx}
\usepackage{subcaption}    
\usepackage{multicol} 
\usepackage{multirow}
\usepackage{comment}
\usepackage{cite}
\usepackage{amsmath}  
\usepackage{booktabs}    
\usepackage{amssymb}    
\usepackage{float}      
\usepackage{array}
\usepackage{soul,xcolor,tcolorbox}
\usepackage{lineno}

\usepackage{hyperref}
\hypersetup{
	colorlinks=true,     
	linkcolor=blue,        
	urlcolor=blue,         
	filecolor=magenta,     
	citecolor=blue         
}

\date{}

\usepackage[font=small]{caption}  
\newcommand{\supercite}[1]{\!\mbox{\textsuperscript{\cite{#1}}}}
 
\def\le{\leqslant}

\soulregister{\cite}{7} 
\soulregister{\supercite}{7} 
\soulregister{\ref}{7}

\newif\ifhighlightFirst
\highlightFirstfalse  

\newtcolorbox{highlightedGreen}{colback=green!30, colframe=green!80!black, sharp corners}
\newtcolorbox{highlightedYellow}{colback=yellow!30, colframe=yellow!80!black, sharp corners}

\begin{document} 
	
	\title{\large \textbf{A Novel Hierarchy of Quantum Kernel Networks on Smoothed Particle Hydrodynamics}}
	
	\author{
		\small{Yudong Li$^{1,2,}$\footnotemark[1] ,\quad Wenkui Shi$^{1}$,\quad Chunfa Wang$^{3}$,\quad Zhihao Qian$^{1,2}$,\quad Zhiqiang Feng$^{4,5}$,\quad Moubin Liu$^{1,2,}$\footnotemark[1]}
		\\[2mm]
		\footnotesize{1. School of Mechanics and Engineering Science, Peking University, Beijing 100871, PR China}
		\\
		\footnotesize{2. Nanchang Institute of Innovation, Peking University, Nanchang 330038, PR China}
		\\
		\footnotesize{3. School of Mechanics and Aerospace Engineering, Dalian University of Technology, Dalian 116024, PR China}
		\\
		\footnotesize{4. School of Mechanics and Aerospace Engineering, Southwest Jiaotong University, Chengdu 611756, PR China}
		\\
		\footnotesize{5. Laboratory of Mechanics and Energy, Univ-Evry, Paris-Saclay University, Evry 91190, France}
	}
	
	\renewcommand{\thefootnote}{\fnsymbol{footnote}}
	\footnotetext[1]{\hspace{0.5mm}Corresponding authors. \textit{E-mail address:} yudongli@pku.edu.cn (Y.D. Li); mbliu@pku.edu.cn (M.B. Liu).}
	\renewcommand{\thefootnote}{\arabic{footnote}}
	
	\maketitle	
	
	\thispagestyle{plain}
	
	\small
	\begin{abstract}
		\noindent \textbf{Abstract~~~}This study proposed the hierarchy of quantum kernel networks by combing multi quantum networks with smoothed particle hydrodynamics (SPH). The Lagrangian quantum network model was further developed based on an improved quantum multilayer perceptron (QMLP). A sequential hybrid quantum‑classical framework was constructed to ensure robust particle gradient‑based optimization and mitigate barren plateaus for computational particle dynamics. This approach combines smoothing kernels with quantum learning, establishing a novel quantum intelligent particle paradigm. The framework was validated through some benchmarks on multifarious quantum neural networks, static multi‑level vortex reconstructions and transient scalar advective transports. Numerical results show that while elementary quantum circuits struggle with generalization in unstructured domains, the hybrid crossed‑QMLP matches the fitting accuracy of classical SPH in quantum optimized space. Despite current limitations in computational efficiency and hardware implementation, this work paves the way for a new investigation on quantum-particle approach by mapping unstructured Lagrangian particle topologies into integrated quantum networks.
		\\[2mm]
		\textbf{Key words~~~}Quantum-intelligence fusion; Quantum computing; Lagrangian quantum kernel networks; Computational particle dynamics; Meshfree/particle methods; Smoothed particle hydrodynamics  
	\end{abstract}
	
	\section{Introduction}
	\normalsize \hspace{10pt}
	With the advent of novel revolutionary technologies such as quantum computing and artificial intelligence, computational paradigms are undergoing a fundamental transformation, opening unprecedented opportunities for solving complex scientific and engineering problems\supercite{WOS:000435674600001,WOS:001329781900001,WOS:000750754200006,WOS:001693338100001,WOS:001745771900001,YANG2025118348}. Emerging research explores the integration of artificial intelligence for data-driven modeling and adaptive resolution, while visionary investigations probe its foundational synergy with quantum computing principles, aiming to transcend the classical von Neumann bottleneck and unlock unprecedented capabilities in simulating ultra-complex, multi-scale systems.
	
	As a meshfree numerical method, smoothed particle hydrodynamics (SPH) has matured into a robust, versatile, and highly adaptive computational framework on computational physics, continuously expanding its frontiers\supercite{monaghan1992smoothed,2010Smoothed,2019Smoothed}. By discarding computational grids in favor of discrete, interacting particles, SPH inherently captures the Lagrangian evolution of physical fields. It possesses unique advantages in resolving highly complex topological changes, enabling widespread applications in marine engineering for wave-structure interaction, additive manufacturing for powder-scale process and biomechanics for modeling soft tissues\supercite{WOS:000463794900001,WOS:001418113800001,WOS:001629494500001,WOS:001299461100001}. Despite continuous improvements such as regularized and variational formulations that resolve consistency and tensile instability issues, SPH still encounters a critical computational bottleneck. Evaluating particle interactions entails dynamic neighbor searching\supercite{lxxb2021-439}, spatial kernel differentiation\supercite{10.1063/5.0197088,LI2023112213}, and large-scale equation solving\supercite{LI2025119962}. The complexity of these operations scales nonlinearly with the number of particles, imposing a stringent ceiling on ultra-large-scale industrial simulations running on classical silicon chip.To transcend the limitations of classical von Neumann architectures, computational fluid dynamics is increasingly turning to the disruptive potential of quantum computing\supercite{AUYEUNG2024108909,WOS:000552379800001}.
	
	Computational fluid dynamics (CFD) typically involving to solve complex systems of nonlinear partial differential equations that are high-dimensional and multi-physical computing. The superposition and entanglement properties of quantum computing offer potential exponential speedup for core tasks such as matrix inversion and large-scale optimization\supercite{RIEFFEL2024598}. Consequently, with advancements in quantum information technology, research into quantum computational fluid dynamics (QCFD) has gradually emerged in recent years\supercite{thomson2024unravelling,CHEN2024117428}. Current research primarily unfolds along two major algorithmic pathways to address the contradiction between the linear characteristic of qubits and the nonlinearity of fluid dynamics as follows.
	
	The first is Hamiltonian simulation and quantum state mapping, which is the dominant theoretical path. This approach employs mathematical techniques such as the generalized Madelung transformation to map fluid equations (e.g., the Navier-Stokes equations) onto the Schrödinger flow equation of a quantum system\supercite{WOS:001757011700001,WOS:001510363200001}, thereby enabling the direct simulation of fluid evolution on a quantum computer. Building on this, the team led by Yang et al.\supercite{2024Simulating} achieved an end-to-end digital simulation of unsteady flow on a superconducting quantum processor. 
	
	The second path involves the design of specialized quantum algorithms, with research in nonlinear CFD broadly categorized into three types\supercite{tennie2025quantum}: hybrid quantum-classical algorithms, mean-field solvers, and linearization-based solvers. Special details are as follows.
	
	\textbf{Hybrid quantum-classical algorithms:} In nonlinear dynamics research, these algorithms typically transform the solution of differential equations into a minimization problem. The loss function is evaluated on a quantum processor (e.g., using variational quantum circuits or quantum neural networks), while parameter updates are performed by a classical optimizer. Its implementations are diverse\supercite{tennie2025quantum,WOS:001315373400001}, including function approximation based on tensor networks and quantum feature maps, data-driven reservoir computing and time-stepping algorithms combined with specific quantum oracles. Specific applications have been extended to fluid mechanics, such as enhancing equation-solving capabilities by modifying the HHL algorithm\supercite{2021Solving}, proposing a quantum lattice Boltzmann method (QLBM)\supercite{TODOROVA2020109347,WOS:001650371600001}, developing variational quantum linear solvers and quantum physics-informed neural networks (QPINN)\supercite{WOS:001362623400015}. Furthermore, researches into quantum spectral method, finite difference method and their Hamiltonian simulation, equation solving based on the quantum linear system algorithm (QLSA)\supercite{WOS:001284392600001,WOS:001575455600001} and variational quantum computing for various boundary conditions\supercite{OVER2025106508} are also progressing.
	
	\textbf{Mean-field quantum solving:} This method achieves approximated nonlinear evolution of a single copy through symmetric interactions among identical copies of quantum state. Dynamic variables are encoded using amplitude states. The original formulation of this algorithm in reference\supercite{WOS:000270672100002} employed a matrix inversion scheme, realized via QLSA for a series of forward Euler steps. The output is a large historical state composed of multi-copy states at different discrete time steps.
	
	\textbf{Linearization-based quantum solving:} The core idea is to transform the nonlinear system into a linear one for processing\supercite{WOS:000552379800001}. One technique uses the Carleman linearization, converting finite-dimensional nonlinear differential equations into (truncated) infinite-dimensional linear systems, which can then be solved using QLSA. Another method is based on the Fokker-Planck equation, which describes the evolution of distributional function. After spatial semi-discretization, the discrete function values can be efficiently stored in a quantum register via amplitude encoding, while time integration is achieved through repeated Hamiltonian simulation and measurement. Overall, the strength of nonlinear terms is crucial for performance. For both linearization-based and mean-field solving, the computational cost increases with the strength of nonlinearity, a result consistent with established analytical bounds. Furthermore, the derivation of explicit numerical solutions remains an open topic for future research in quantum computing.
	
	In this study, given that pure quantum circuits still struggle with deep nonlinearities and inherent noise of the noisy intermediate-scale quantum (NISQ) era, hybrid classical-quantum models offer the most pragmatic and immediate approach to resolve complex SPH kernel evaluations. It is organized as follows: \textit{Section 2} introduces the fundamental concepts of quantum computing, including quantum bits, variational quantum circuits, and the architectures of general QNNs, improved QMLPs, and QCNNs, followed by a discussion of hybrid quantum‑classical optimization paradigms. \textit{Section 3} presents the proposed hierarchy of quantum kernel networks on smoothed particle hydrodynamics (SPH), briefly revisiting the SPH method, its kernel approximations and the discretization of computational theme. Subsequently, describes the Lagrangian quantum network models detailing the single quantum circuit baseline, forward hierarchies and hybrid crossed architectures. \textit{Section 4} reports numerical results, comprising performance comparisons among quantum network processors, static quantum kernel tests on complex vortex fields and transient scalar advection benchmarks. Finally, we concludes the paper with a summary of key findings, current limitations and future research directions.
	
	\section{Quantum Networks}
	\subsection{Quantum Basics}
	\normalsize \hspace{10pt}
	To clarify the fundamental distinctions between classical von Neumann computers and universal quantum computers, a comparative analysis of their basic information units is essential. In classical computing, a bit is definitively in a state of 0 or 1, enabling deterministic instruction processing and data storage. Conversely, in quantum computing, the fundamental unit is the quantum bit (qubit). A single qubit generally exists in a superposition state, denoted as $\left | \psi \right \rangle$. This indeterminate state persists until a measurement is performed, causing it to collapse to a definite classical outcome. The property of superposition, particularly combined with entanglement across multiple qubits, embodies the potential for exponential parallelism and superior data encoding efficiency.
	
	\begin{figure}[H]
		\centering
		\includegraphics[scale=0.3]{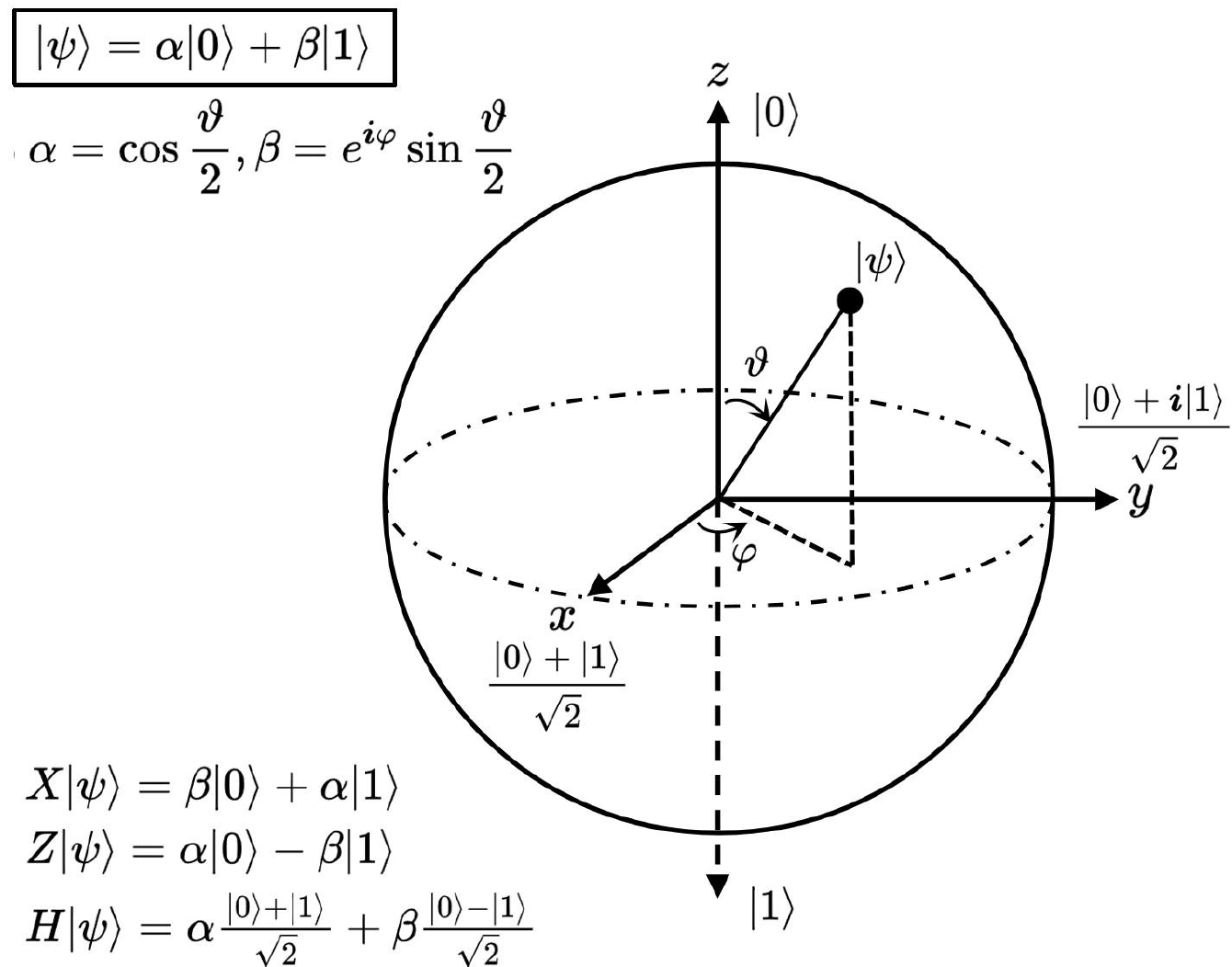}
		\caption{\small Bloch sphere representation of the quantum state $\left | \psi \right \rangle$ with fundamental gate operations.}
		\label{fig_Quantum:bloch}
	\end{figure}
	
	\begin{figure}[H]
		\centering
		\includegraphics[scale=0.65]{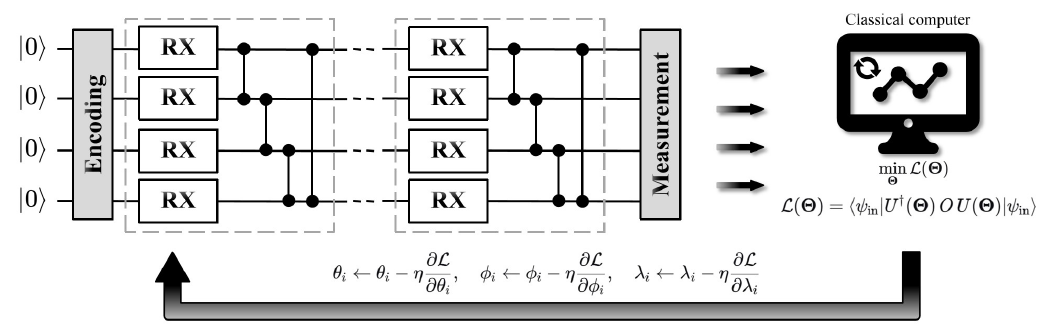}
		\caption{\small Generic architecture of quantum neural networks.}
		\label{fig_Quantum:QNNmodel}
	\end{figure}
	
	\textbf{Quantum neural networks (QNNs):} Fig.(\ref{fig_Quantum:QNNmodel}) illustrates a generic QNN architecture constructed from trainable variational quantum circuits (VQC)\supercite{Cerezo2021Variationalquantum,WOS:001584542600001}. Currently, these training tasks are implemented as a hybrid quantum-classical computational process. Given a classical training set ${\cal T} = \{ ({{\boldsymbol{x}}_i},{{\boldsymbol{y}}_i})\}$ comprising \(n\)-dimensional feature vectors \(\boldsymbol{x}_i\) and their corresponding multi-labels \(\boldsymbol{y}_i\), a quantum encoding layer ${\rm E}({{\boldsymbol{x}}_i})$ is applied to the predefined base state for generating a quantum input feature map (QIFP), denoted as \(|\phi_0 \rangle\). This QIFP is subsequently processed by a parameterized VQC ansatz via these unitary transformations.
	\begin{equation}\label{eq_Quantum:bloch}
		|\phi \rangle  = \Phi ({\boldsymbol{\alpha }}){\rm E}|{\phi _0}({\boldsymbol{x}})\rangle 
	\end{equation}
	where $|\phi \rangle$ denotes the final quantum state on specified algorithm, $\Phi ({\boldsymbol{\alpha }})$ represents a set of unitary transformations with free parameters ${\boldsymbol{\alpha }} = \{ \theta ,\phi ,\lambda \}$ subject to adaptive optimization. As show in Fig.(\ref{fig_Quantum:QNNmodel}), the VQC ansatz, composed of a structured sequence of quantum gates, fundamentally defines the Hilbert space of quantum networks. Similar to the layered structure of classical neural networks, the QNNs can be constructed by stacking multiple VQC ansatz layers, i.e., $\Phi ({\boldsymbol{\alpha }}) = {U_L}({{\boldsymbol{\alpha }}_L}) \cdots {U_2}({{\boldsymbol{\alpha }}_2}){U_1}({{\boldsymbol{\alpha }}_1})$. The final predictive output is obtained by performing measurements on the resulting quantum state, followed by necessary classical post-processing. A predefined loss function ${\mathcal{L}}\left( {\boldsymbol{\Theta }} \right)$ quantifies the discrepancy between the QNNs' prediction and the true target \(\boldsymbol{y}_i\). It notes that training the QNNs involves a hybrid quantum-classical optimization loop that iteratively searches for the optimal parameter set ${{\boldsymbol{\alpha }}^ * }$.
	
	\textbf{VQC ansatz design:} The design of VQC ansatz is typically empirical, guided by prior knowledge of the target problem, as the performance of different circuit architectures across diverse datasets is generally difficult to predict the prior results. As depicted in Fig.(\ref{fig_Quantum:QNNmodel}), a widely adopted and effective ansatz consists of parameterized single-qubit rotation gates (e.g., ${\rm{RX,RY,RZ}}$, or their combinations), followed by fixed two-qubit CNOT gates arranged to enforce nearest-neighbor coupling among all qubits. Its neighboring physical mechanism is similar to smoothed particle hydrodynamics (SPH)\supercite{2010Smoothed,WOS:001501337900001}. This specific design has demonstrated superior predictive accuracy in comparative studies and it is frequently employed as a foundational ansatz for various training tasks\supercite{Cerezo2021Variationalquantum,2020Variational,PhysRevA.104.052409}. Its underlying principle is that single-qubit rotations provide a tunable parameter space, while the entangling two-qubit gates introduce essential quantum correlations, which are critical for the model's expressive power.
	
	\textbf{Errors in NISQ devices:} The practical implementation of QNNs is constrained by inherent noise in current noisy intermediate-scale quantum (NISQ) processors. Its noise originates from multiple sources, including imperfect control pulses, inevitable crosstalk among with qubits, and environmental decoherence\supercite{WOS:001672321600001}. A primary limitation of NISQ devices is the accumulation of decoherence errors over time, which severely restricts the feasible depth (number of sequential gates) and width (number of qubits) of executable quantum circuits, thereby rendering complex, deep quantum circuits impractical. Furthermore, quantum gate operations on themselves are not perfectly accurate and introduce operational errors such as bit-flip and phase-flip errors into current computation\supercite{WOS:001133244600002}. To mitigate these effects and sustain operational fidelity, quantum hardware necessitates frequent characterization and calibration. This noisy operational environment fundamentally limits the scale and complexity of quantum algorithms that can be reliably executed today. In light of this, the current study is primarily conducted based on hybrid computing that integrates quantum simulators and classical computers.
	
	\subsection{Architecture of Quantum Multilayer Perceptron}
\begin{figure}[H]
		\centering
		\includegraphics[scale=0.75]{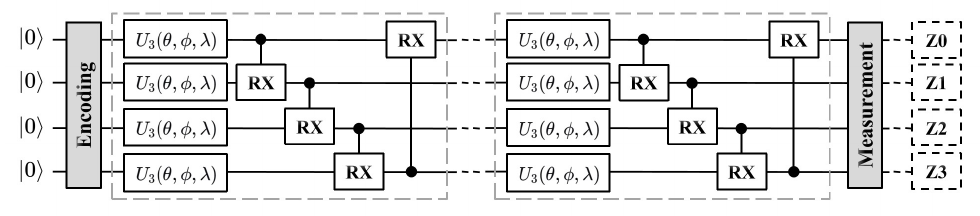}
		\caption{\small Improved architecture of quantum multilayer perceptron.}
		\label{fig_Quantum:QMLPmodel}
	\end{figure}

	\normalsize \hspace{10pt}
	As a representative network architecture, the quantum multilayer perceptron (QMLP) architecture is a versatile, parameterized variational quantum circuit designed to adapt specific computational tasks. As shown in Fig.(\ref{fig_Quantum:QMLPmodel}), we developed the improved QMLP network, which is analogous to a classical neural network. The QMLP processes transmission of information through a structured sequence of quantum layers, whose workflow begins with a linear encoding layer (amplitude state preparation QIFP) ${\rm E}({{\boldsymbol{x}}_i})$ in Eq.(\ref{eq_Quantum:bloch}), ends with either global measurement of quantum collapse or expectation value's measurement of Pauli-Z operators.
	
	The core computational body of QMLP is constructed by stacking multiple parameterized variational ansatz blocks, denoted $\Phi ({\boldsymbol{\alpha }})$, to form a deep, hierarchical structure. Each ansatz block consists of two primary components: first, a set of parameterized single-qubit rotation gates (collectively labeled as the ${U_3}(\theta ,\phi ,\lambda ) = {\rm{RZ}}(\phi ) \cdot {\rm{RY}}(\theta ) \cdot {\rm{RZ}}(\lambda )$ gate), followed by a nearest-neighbor coupling scheme implemented via parameterized two-qubit gates, specifically ${\rm{          CRX}}(\beta ) = |0\rangle \langle 0| \otimes I + |1\rangle \langle 1| \otimes {\rm{RX}}(\beta )$ in Eq.(\ref{eq:QMLP1}). The adjustable parameters within matrix circuits ${\boldsymbol{\alpha }} = \{ \theta ,\phi ,\lambda, \beta \}$ are iteratively optimized during training, functionally analogous to the weight matrices in a classical multi-layer perceptron.
	
	\begin{equation}\label{eq:QMLP1}
		\begin{array}{l}
			{U_3}(\theta ,\phi ,\lambda ) = {\rm{RZ}}(\phi ) \cdot {\rm{RY}}(\theta ) \cdot {\rm{RZ}}(\lambda ),{\rm{          CRX}}(\beta ) = |0\rangle \langle 0| \otimes I + |1\rangle \langle 1| \otimes {\rm{RX}}(\beta )\\
			{U_3}(\theta ,\phi ,\lambda ) = \left[ {\begin{array}{*{20}{c}}
					{\cos (\frac{\theta }{2})}&{ - {e^{i\lambda }}\sin (\frac{\theta }{2})}\\
					{{e^{i\phi }}\sin (\frac{\theta }{2})}&{{e^{i(\phi  + \lambda )}}\cos (\frac{\theta }{2})}
			\end{array}} \right]{\rm{,   CRX}}(\beta ) = \left[ {\begin{array}{*{20}{c}}
					1&0&0&0\\
					0&1&0&0\\
					0&0&{\cos \left( {\frac{\beta }{2}} \right)}&{ - i\sin \left( {\frac{\beta }{2}} \right)}\\
					0&0&{ - i\sin \left( {\frac{\beta }{2}} \right)}&{\cos \left( {\frac{\beta }{2}} \right)}
			\end{array}} \right]
		\end{array}
	\end{equation}
	
	Current quantum networks generally employ entanglement encoding to generate QIFPs by embedding classical data into quantum state amplitudes. This method requires only \(\log_2(n)\) qubits to represent an \(n\)-dimensional input, offering significant resource efficiency compared to alternative encoding schemes. However, the quality of QIFPs is highly susceptible to encoding inaccuracies and hardware noise in the NISQ era. Entanglement encoding relies on complex state preparation circuits composed of multiple rotation gates per qubit (e.g., \(\rm{RY}\), \(\rm{RX}\), and \(\rm{RZ}\) gates), leading to over 30\% hardware overhead and elevated error rates. Moreover, a single gate error can propagate through entanglement, corrupting the entire QIFP and degrading model performance. These limitations underscore the need for a novel quantum encoding architecture that is both hardware-efficient and robust against noise, ensuring high-fidelity input representation for reliable QNN inference\supercite{WOS:000450273400003}.
	
	Following the final improved variational layer, each qubit is measured in the Pauli-Z basis, yielding a classical expectation value in the range \([-1, 1]\). These measured values are subsequently fed into a small, classical fully-connected neural network for post-processing. This final step maps the quantum processor's output to the desired format, such as the 10-class probability distribution for classification tasks, thereby completing the hybrid quantum-classical computation pipeline.
	
	\subsection{Architecture of Quantum Convolutional Neural Network}
	\normalsize \hspace{10pt}
	Convolutional neural networks (CNNs) have established a highly successful machine learning architecture for task recognition\supercite{WOS:000402555400026,WOS:001547418900002}. A classical CNN typically comprises a sequence of alternating layers. In each layer, an intermediate two-dimensional array of pixels, known as a feature map, is generated from the previous layer. A convolutional layer produces new pixel values by computing linear combinations of adjacent pixels from the preceding feature map via a small, learned weight matrix. A pooling layer subsequently reduces the spatial dimensions of the feature map through operations like taking the maximum value of neighboring pixels, often followed by a non-linear activation function. When the feature map size becomes sufficiently small, a fully-connected layer computes the final output.
	
	\begin{figure}[H]
		\centering
		\includegraphics[scale=1.2]{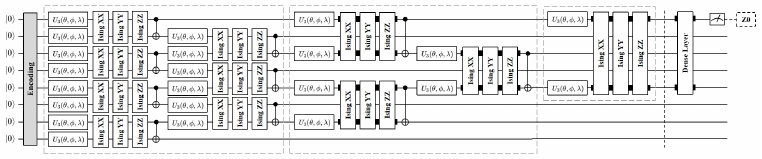}
		\caption{\small Generic architecture of quantum convolutional neural network.}
		\label{fig_Quantum:QCNNmodel}
	\end{figure}
\begin{equation}\label{eq:QCNN1}
	\begin{array}{l}
		{U_{XX}}(\theta ) = {e^{ - i\frac{\theta }{2}(X \otimes X)}},{\rm{   }}{U_{XX}}(\theta ) = \left[ {\begin{array}{*{20}{c}}
				{\cos \left( {\frac{\theta }{2}} \right)}&0&0&{ - i\sin \left( {\frac{\theta }{2}} \right)}\\
				0&{\cos \left( {\frac{\theta }{2}} \right)}&{ - i\sin \left( {\frac{\theta }{2}} \right)}&0\\
				0&{ - i\sin \left( {\frac{\theta }{2}} \right)}&{\cos \left( {\frac{\theta }{2}} \right)}&0\\
				{ - i\sin \left( {\frac{\theta }{2}} \right)}&0&0&{\cos \left( {\frac{\theta }{2}} \right)}
		\end{array}} \right]\\
		{U_{YY}}(\theta ) = {e^{ - i\frac{\theta }{2}(Y \otimes Y)}},{\rm{   }}{U_{YY}}(\theta ) = \left[ {\begin{array}{*{20}{c}}
				{\cos \left( {\frac{\theta }{2}} \right)}&0&0&{i\sin \left( {\frac{\theta }{2}} \right)}\\
				0&{\cos \left( {\frac{\theta }{2}} \right)}&{ - i\sin \left( {\frac{\theta }{2}} \right)}&0\\
				0&{ - i\sin \left( {\frac{\theta }{2}} \right)}&{\cos \left( {\frac{\theta }{2}} \right)}&0\\
				{i\sin \left( {\frac{\theta }{2}} \right)}&0&0&{\cos \left( {\frac{\theta }{2}} \right)}
		\end{array}} \right]\\
		{U_{ZZ}}(\theta ) = {e^{ - i\frac{\theta }{2}(Z \otimes Z)}},{\rm{   }}{U_{ZZ}}(\theta ) = \left[ {\begin{array}{*{20}{c}}
				{{e^{ - i\frac{\theta }{2}}}}&0&0&0\\
				0&{{e^{i\frac{\theta }{2}}}}&0&0\\
				0&0&{{e^{i\frac{\theta }{2}}}}&0\\
				0&0&0&{{e^{ - i\frac{\theta }{2}}}}
		\end{array}} \right]
	\end{array}
\end{equation}
\begin{equation}\label{eq:QCNN2}
	\begin{array}{l}
		{\cal G} = [({R_{ZZ}},0),({R_{XX}},1),({R_{YY}},2),({R_{ZX}},3),({R_{ZX}},4),({R_{XX}},5),({R_{ZX}},6),\\
		({R_{ZZ}},7),({R_{YY}},8),({R_{ZZ}},9),({R_{XX}},10),({R_{ZX}},11),({R_{ZX}},12),({R_{ZZ}},13),({R_{YY}},14)]
	\end{array}
\end{equation}
where each tuple $(R_{\alpha\beta}, i)$ denotes a two-qubit rotation gate generated by the Pauli operator $\sigma_\alpha \otimes \sigma_\beta$ applied with the $i$-th variational parameter $\theta_i$.

	Inspired by this architecture, the quantum convolutional neural network (QCNN) model was proposed that extends these principles to the quantum domain\supercite{WOS:000843205300017,WOS:000500574300022,WOS:001158795100001}. As shown in Fig.(\ref{fig_Quantum:QCNNmodel}), the quantum convolutional layer applies a single quasi-local unitary operator $U_i$ and Ising \(\rm{XX}\), \(\rm{YY}\), \(\rm{ZZ}\) gates in a translationally invariant manner with finite depth. The quantum pooling layer then performs measurements on a subset of qubits and the outcomes of these measurements deterministically control unitary rotations applied to neighboring qubits, thereby reducing the number of active degrees of freedom and introducing a form of non-linearity. This process of convolution and pooling is repeated until the system size is sufficiently small, at which point a final, fully-connected unitary transformation is applied to the remaining qubits. The quantum dense layer of current quantum neural network is implemented by sequentially applying 15 parameterized two-qubit gates to adjacent qubit pairs in a cyclic topology. The gate sequence comprises RZZ, RXX, RYY and RZX operations, with each gate indexed to a distinct variational parameter $\theta_i$ ($i = 0, \ldots, 14$) in Eq.(\ref{eq:QCNN2}). This configuration enables the circuit to capture complex quantum correlations while maintaining trainability through gradient-based optimization. The circuit's output is obtained by measuring a fixed number of output qubits. Analogous to the classical case, the QCNN's hyperparameters (e.g., the number of layers) are fixed, while the unitary transformations themselves are learned.
	
	A key advantage of this structured design is its parameter efficiency. Compared to a generic quantum circuit classifier, the QCNN achieves a doubly exponential reduction in the number of parameters required, scaling favorably as $O(\log N)$. This significantly enhances learning efficiency and reduces implementation costs. For a training set of $M$ vectors with binary labels $y_\alpha \in \{0,1\}$, the learning process involves optimizing the unitaries to minimize a cost function, such as the mean squared error between expected circuit outputs and true labels.
	
	To elucidate the QCNN's operational principles, a connection with two foundational concepts was established in quantum information: multiscale entanglement renormalization ansatz (MERA) and quantum error correction (QEC). MERA provides an efficient tensor network representation for many critical quantum states. A QCNN can be viewed as a generalization of MERA, while MERA represents a specific state via a sequence of isometries and unitaries. The QCNN processes arbitrary input states, with its intermediate measurements introducing adaptability. Specifically, the measurements in pooling layers can be interpreted as syndrome measurements in a QEC code, determining the corrective unitaries. Consequently, a QCNN with multiple pooling layers effectively combines the hierarchical structure of MERA with the error-detecting and correcting capability of nested QEC. This dual perspective positions the QCNN as a powerful architecture not only for classifying quantum states by simulating a renormalization group flow that corrects local errors, but also for designing novel, efficient quantum error-correcting codes with rich entanglement structures\supercite{WOS:001672321600001}.
	
	\subsection{Hybrid Quantum-Classical Optimization}
\begin{figure}[H]
		\centering
		\includegraphics[scale=0.62]{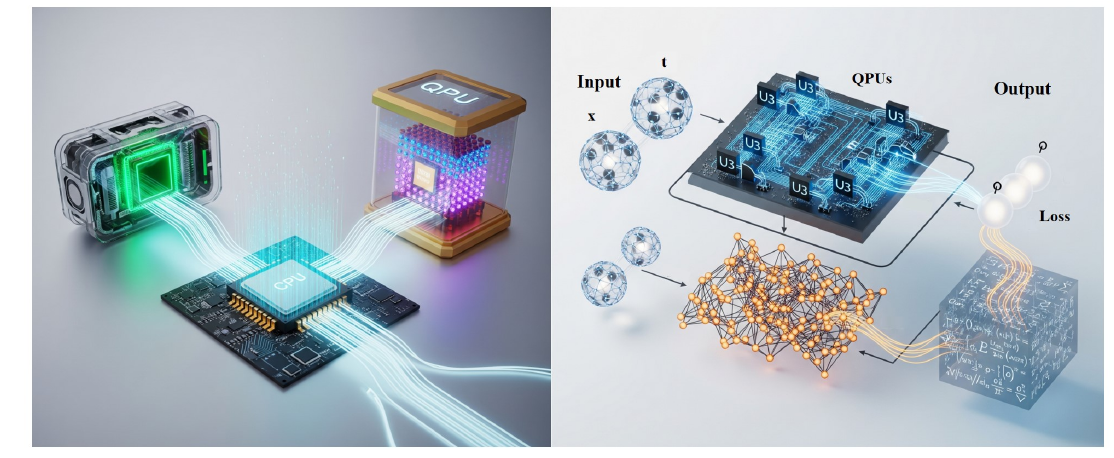}
		\caption{\small Hybrid quantum-classical pathway on quantum processor units (QPUs).}
		\label{fig_Quantum:hybridoptimization}
	\end{figure}

	\normalsize \hspace{10pt}
	Hybrid quantum-classical optimization has emerged as the dominant computing paradigm in the NISQ era, representing the most pragmatic pathway toward practical quantum advantage\supercite{WOS:000358968700012,WOS:000716870500038}. As shown in Fig.(\ref{fig_Quantum:hybridoptimization}), this architecture draws a compelling analogy with classical high-performance computing: just as GPUs serve as specialized accelerators for parallelizable tasks within a broader CPU-driven workflow, quantum processor units (QPUs) as dedicated co-processing function for classically intractable subtasks, while classical computers retain control over overall logic, data management and lightweight computations.
	
	The hybrid paradigm manifests across two principal domains. First, in the aspect of quantum artificial intelligence (AI) optimization, the VQC ansatz evaluates loss function in exponentially large feature spaces for the specified task where quantum superposition offers inherent advantages, while classical optimizers (e.g., SGD, Adam) handle parameter updates using gradient estimates obtained via techniques like the parameter-shifting rule. Second, in the aspect of quantum algorithmic acceleration, core computational bottlenecks such as matrix inversion in linear systems, phase estimation in eigenvalue problems, or Hamiltonian simulation are offloaded to quantum hardware, with classical machines managing input preprocessing, output postprocessing and iterative control flows.
	
	This division of labor mirrors the heterogeneous computing model already ubiquitous in modern supercomputing. The quantum device acts not as a standalone replacement but as a domain-specific accelerator, analogous to a GPU tailored for wavefunction manipulation rather than matrix multiplication. Only computations exhibiting quantum advantage, those with favorable scaling in Hilbert space are delegated to quantum chips, while the vast majority of operations remain on classical hardware. This co-design philosophy significantly reduces the coherence time and gate fidelity requirements for quantum processors, enabling meaningful computation within current hardware constraints.
	
	\begin{figure}[H]
		\centering
		\includegraphics[scale=0.77]{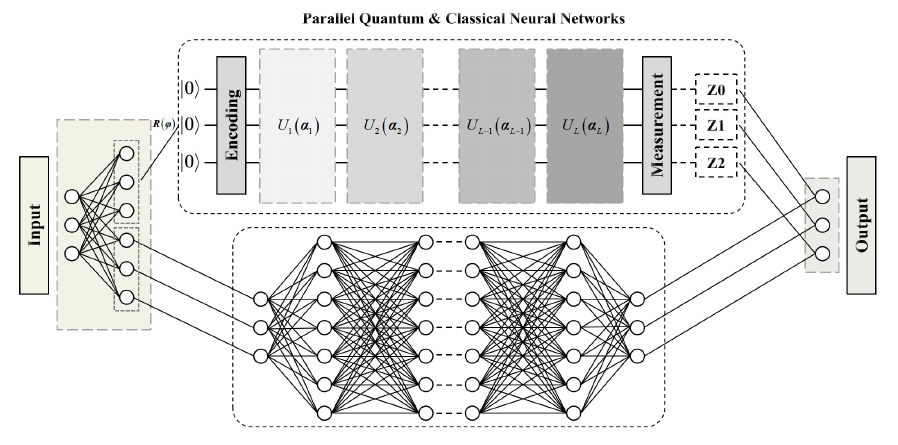}
		\caption{\small Parallel architecture of quantum-classical hybrid optimization.}
		\label{fig_Quantum:hybridoptimization3}
	\end{figure}
	\begin{figure}[H]
		\centering
		\includegraphics[scale=0.75]{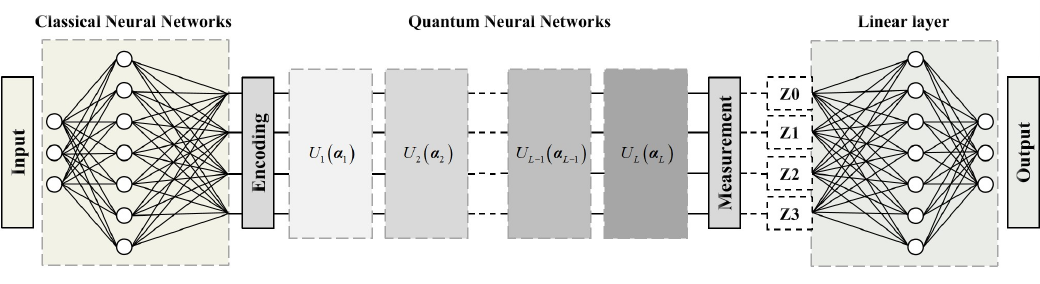}
		\caption{\small Sequential architecture of quantum-classical hybrid optimization.}
		\label{fig_Quantum:hybridoptimization2}
	\end{figure}
	
	Following these aforementioned architectures of quantum networks, classical optimizers need to be processed in a coordinated manner. As shown in Figs.(\ref{fig_Quantum:hybridoptimization3},\ref{fig_Quantum:hybridoptimization2}), the architectural topologies of hybrid quantum-classical optimization are divided into the parallel and sequential frameworks. Hybrid quantum-classical neural networks integrate parameterized quantum circuits with classical deep learning architectures to leverage quantum superposition and entanglement within the physical constraints of NISQ era.
	
	In Fig.(\ref{fig_Quantum:hybridoptimization3}), the parallel architecture bifurcates the input stream, feeding the raw data simultaneously into a classical multi-layered perceptron and a quantum neural network. The respective expectation values from the quantum measurements and the classical activations are processed concurrently and aggregated via a trainable linear combination to form the final output. Conversely, in Fig.(\ref{fig_Quantum:hybridoptimization2}), the sequential framework operates as a strictly linear pipeline, transitioning data unidirectionally between classical and quantum domains. As delineated in structural paradigms, raw input is initially processed and compressed by classical neural networks, encoded into quantum states via variable unitary transformations ($U_1 \dots U_L$), and collapsed through Pauli measurement operators ($Z_0, Z_1, Z_2, Z_3$). A terminal classical linear layer subsequently projects these expectation values into the final predictive output.
	
	It can be generalized from experience that for the parallel architecture, its circumvents information bottlenecks by maintaining direct access to uncompressed features. It can enable complementary functional fitting: the VQC maps smooth harmonic foundations (truncated Fourier series) while the classical layers fit non-harmonic discrete noise. The disadvantages have that it incurs substantial computational overhead and rapid escalation in floating-point operations (FLOPs) during classical simulation and demands complex synchronous execution between CPUs and QPUs\supercite{WOS:001513544000005,WOS:000286756800001,WOS:001406333300001}. While for the sequential architecture, it mitigates current hardware limitations by utilizing classical layers for aggressive dimensionality reduction. Allows the specialized quantum circuit to function efficiently as a highly expressive, non-linear decision boundary within a compact latent space. Its disadvantages have that it introduces severe information bottlenecks by discarding raw spatial features prior to quantum processing. Deep VQCs in this pipeline are acutely susceptible to barren plateaus (vanishing gradients) and cumulative quantum noise propagation\supercite{WOS:000361820900021,WOS:001048290700001}.
	
	Topological selection in hybrid quantum machine learning is strictly problem-dependent. Sequential models optimize resource efficiency and state preparation for high-dimensional, unstructured data, whereas parallel configurations maximize functional expressivity and gradient stability for noisy, periodic landscapes\supercite{WOS:001462597800001}.
	
	\section{Hierarchy of Quantum Kernel Networks}
	\subsection{Smoothed Particle Hydrodynamics}
	\normalsize \hspace{10pt}
	Smoothed particle hydrodynamics (SPH), as a fundamental numerical method in computational science and industrial simulation, operates without grid or volume constraints, enabling accurate tracking of spatiotemporal flows and physical analysis. Over the years, this method has garnered sustained attention from both academia and engineering communities, achieving remarkable progress in fundamental theory, algorithmic optimization and interdisciplinary applications\supercite{2019Smoothed}.
	
	\begin{figure}[H]
		\centering
		\includegraphics[scale=0.77]{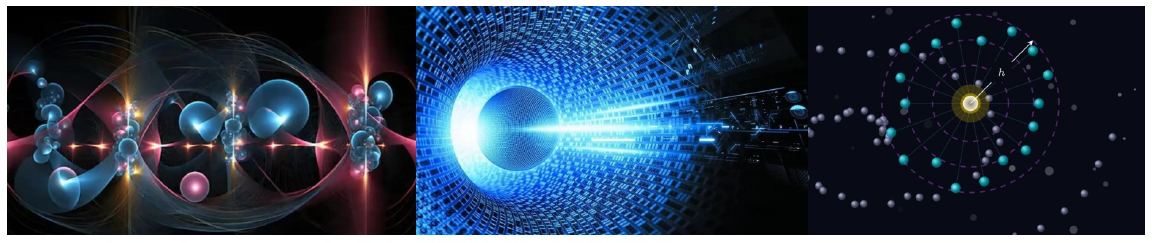}
		\caption{\small SPH discretization and computational particle physics.}
		\label{fig_Quantum:SPHmodel}
	\end{figure}
	
\begin{figure}[ht]
		\centering
		\begin{subfigure}[c]{0.4\textwidth}
			\centering
			\includegraphics[scale=0.32]{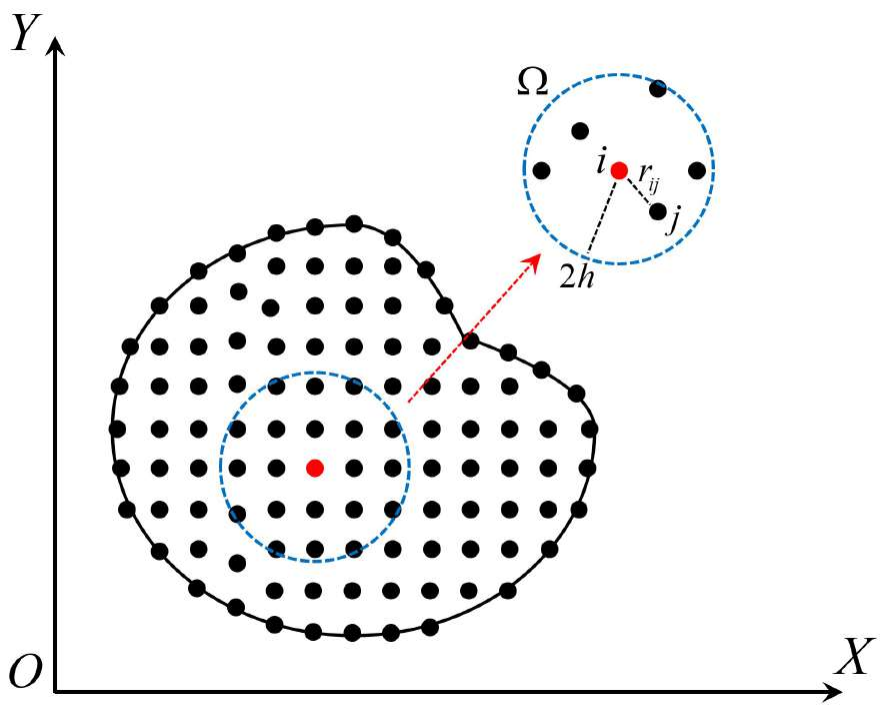}
			\caption{\footnotesize Sketch with a local support domain.}
			\label{fig:1a}
		\end{subfigure}
		\begin{subfigure}[c]{0.4\textwidth}
			\centering
			\includegraphics[scale=0.31]{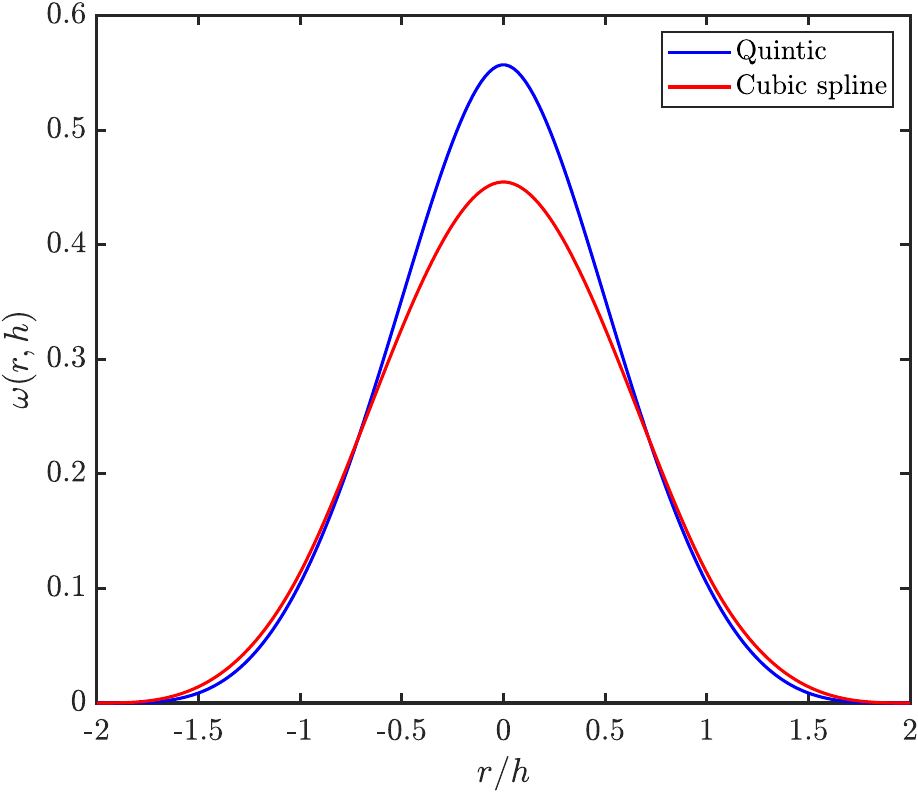}
			\caption{\centering\footnotesize Curves of common weighted kernel functions.}  
			\label{fig:1b}
		\end{subfigure}
		\caption{\footnotesize Numerical discretization of physical domain.}
		\label{fig:1}
	\end{figure}

	As a brief review, it is the rigorous mathematical foundation from Taylor series and the fundamental concept of particle-based discretization, which considers a weighted discretization of weakly integral form described in Eqs.(\ref{eq:1}-\ref{eq:5}).
	\begin{equation}\label{eq:1}
		\int_{\Omega }{f\left( \mathbf{x} \right)\Phi \left( \mathbf{x}-{{\mathbf{x}}_{i}} \right)d\mathbf{x}}={{f}_{i}}\int_{\Omega }{\Phi \left( \mathbf{x}-{{\mathbf{x}}_{i}} \right)d\mathbf{x}}+{{f}_{i,\alpha }}\int_{\Omega }{\left( {{\mathbf{x}}^{\alpha }}-\mathbf{x}_{i}^{\alpha } \right)\Phi \left( \mathbf{x}-{{\mathbf{x}}_{i}} \right)d\mathbf{x}}+O\left( {{\left( \mathbf{x}-{{\mathbf{x}}_{i}} \right)}^{2}} \right)
	\end{equation}
	\begin{equation}\label{eq:2}
		{{f}_{i}}=f\left( {{\mathbf{x}}_{i}} \right)
	\end{equation}
	\begin{equation}\label{eq:3}
		{{f}_{i,\alpha }}={{f}_{\alpha }}\left( {{\mathbf{x}}_{i}} \right)={{\left( {\partial f}/{\partial {{\mathbf{x}}^{\alpha }}}\; \right)}_{i}}
	\end{equation}
	\begin{equation}\label{eq:4}
		\left[ \begin{matrix}
			\int_{\Omega }{\omega \left( \mathbf{x}-{{\mathbf{x}}_{i}} \right)d\mathbf{x}} & \int_{\Omega }{\left( {{\mathbf{x}}^{\alpha }}-\mathbf{x}_{i}^{\alpha } \right)\omega \left( \mathbf{x}-{{\mathbf{x}}_{i}} \right)d\mathbf{x}}  \\
			\int_{\Omega }{\nabla \omega \left( \mathbf{x}-{{\mathbf{x}}_{i}} \right)d\mathbf{x}} & \int_{\Omega }{\left( {{\mathbf{x}}^{\alpha }}-\mathbf{x}_{i}^{\alpha } \right)\nabla \omega \left( \mathbf{x}-{{\mathbf{x}}_{i}} \right)d\mathbf{x}}  \\
		\end{matrix} \right]\left[ \begin{matrix}
			{{f}_{i}}  \\
			{{f}_{i,\alpha }}  \\
		\end{matrix} \right]\cong \left[ \begin{matrix}
			\int_{\Omega }{f\left( \mathbf{x} \right)\omega \left( \mathbf{x}-{{\mathbf{x}}_{i}} \right)d\mathbf{x}}  \\
			\int_{\Omega }{f\left( \mathbf{x} \right)\nabla \omega \left( \mathbf{x}-{{\mathbf{x}}_{i}} \right)d\mathbf{x}}  \\
		\end{matrix} \right]
	\end{equation}
	\begin{equation}\label{eq:5}
		\mathbf{L}=\left[ \begin{matrix}
			\int_{\Omega }{\omega \left( \mathbf{x}-{{\mathbf{x}}_{i}} \right)d\mathbf{x}} & \int_{\Omega }{\left( {{\mathbf{x}}^{\alpha }}-\mathbf{x}_{i}^{\alpha } \right)\omega \left( \mathbf{x}-{{\mathbf{x}}_{i}} \right)d\mathbf{x}}  \\
			\int_{\Omega }{\nabla \omega \left( \mathbf{x}-{{\mathbf{x}}_{i}} \right)d\mathbf{x}} & \int_{\Omega }{\left( {{\mathbf{x}}^{\alpha }}-\mathbf{x}_{i}^{\alpha } \right)\nabla \omega \left( \mathbf{x}-{{\mathbf{x}}_{i}} \right)d\mathbf{x}}  \\
		\end{matrix} \right],\mathbf{F}=\left[ \begin{matrix}
			{{f}_{i}}  \\
			{{f}_{i,\alpha }}  \\
		\end{matrix} \right],\mathbf{B}=\left[ \begin{matrix}
			\int_{\Omega }{f\left( \mathbf{x} \right)\omega \left( \mathbf{x}-{{\mathbf{x}}_{i}} \right)d\mathbf{x}}  \\
			\int_{\Omega }{f\left( \mathbf{x} \right)\nabla \omega \left( \mathbf{x}-{{\mathbf{x}}_{i}} \right)d\mathbf{x}}  \\
		\end{matrix} \right]
	\end{equation}\\
	where $\Phi \left( \mathbf{x} \right)$ denotes a basis function in the space of support domain $\Omega$. If existing a set of basis functions $\Phi (\mathbf{x})=\left\{ {{\omega }^{0}}(\mathbf{x})\text{, }{{\omega }^{1}}(\mathbf{x})\text{, }{{\omega }^{2}}(\mathbf{x}),... \right\}$ on both sides of Eq.(\ref{eq:1}) to satisfy the corresponding consistency conditions respectively shown in Fig.(\ref{fig:1b}), a linear matrix equation can be produced based on the integrated kernel forms and purely physical discretized nodes. The linear matrix equation of Eq.(\ref{eq:4}) neglecting the high-order truncational error of Talyor expansion can be produced through adopting different basis functions\supercite{2010Smoothed,WOS:001501337900001}.
	
	Considering the resolved matrix equation of Eq.(\ref{eq:4}) as $\mathbf{F}\cong {{\mathbf{L}}^{-1}}\mathbf{B}$, the improved particle method is completely introduced. If neglecting the inversion of approximated functional value ${{f}_{i}}$, it leads to the kernel gradient correction method described in Eq.(\ref{eq:6})\supercite{WOS:001501337900001}. It can be also derived based on the transformation of Eq.(\ref{eq:1}).
	\begin{equation}\label{eq:6}
		\left[ \int_{\Omega }{\left( {{\mathbf{x}}^{\alpha }}-\mathbf{x}_{i}^{\alpha } \right)\nabla \omega \left( \mathbf{x}-{{\mathbf{x}}_{i}} \right)d\mathbf{x}} \right]\left[ {{f}_{i,\alpha }} \right]\cong \left[ \int_{\Omega }{\left( f\left( \mathbf{x} \right)-{{f}_{i}} \right)\nabla \omega \left( \mathbf{x}-{{\mathbf{x}}_{i}} \right)d\mathbf{x}} \right]
	\end{equation}
	
	As for the smoothed particle hydrodynamics (SPH)\supercite{2010Smoothed}, it is obvious that neglecting the resolved matrix $\mathbf{L}$ (regarding as a unit matrix) the approximate physical variable can be directly solved in Eq.(\ref{eq:8}) as a common approach of interpolation. The symmetric form can be derived in particle-particle interactions described in Eq.(\ref{eq:13}), which possess better conservativeness of interactional movements.
	\begin{equation}\label{eq:8}
		\left[ \begin{matrix}
			{{f}_{i}}  \\
			{{f}_{i,\alpha }}  \\
		\end{matrix} \right]\cong \left[ \begin{matrix}
			\int_{\Omega }{f\left( \mathbf{x} \right)\omega \left( \mathbf{x}-{{\mathbf{x}}_{i}} \right)d\mathbf{x}}  \\
			\int_{\Omega }{f\left( \mathbf{x} \right)\nabla \omega \left( \mathbf{x}-{{\mathbf{x}}_{i}} \right)d\mathbf{x}}  \\
		\end{matrix} \right]
	\end{equation}
	\begin{equation}\label{eq:13}
		{{f}_{i}}\cong \sum{{{f}_{j}}{{\omega }_{ij}}\Delta {{V}_{j}}},\text{      }{{f}_{i,\alpha }}\cong \sum{\left( {{f}_{j}}-{{f}_{i}} \right){{\omega }_{ij,\alpha }}\Delta {{V}_{j}}}
	\end{equation}
	where the particle-particle interactional form is precisely as ${{A}_{ij}}={{A}_{i}}-{{A}_{j}}$, ${{\omega }_{ij,\beta }}$ denotes the partial derivative of basis functional value and $\Delta {{V}_{j}}$ denotes the volume of particle $j$. For instance, the Quintic kernel is employed, whose kernel curve described in Fig.(\ref{fig:1b}).$(q=\left| {{r}_{ij}} \right|/h)$ 
	\begin{equation}\label{eq:21}
		\omega (q)={{\alpha }_{D}}{{\left( 1-\frac{q}{2} \right)}^{4}}\left( 2q+1 \right),\text{      }0\le q\le 2
	\end{equation}
	where Eq.(\ref{eq:21}) represents the Quintic kernel and ${{\alpha }_{D}}$ is equal to $7/4\pi {{h}^{2}}$ in 2D and $21/16\pi {{h}^{3}}$ in 3D.
	
	\subsection{Lagrangian Quantum Network Models}
	\normalsize \hspace{10pt}
	As a purely particle-based method, SPH exhibits a strong conceptual similarity with quantum systems, rendering the pursuit of quantum SPH computations as a natural and promising research direction with inherent advantages. On the other hand, building upon this topological insight on quantum neural networks, our study identifies that sequential architectures exhibit superior performance when handling high-dimensional data and time-sensitive simulations. In the parallel architecture, where classical computers and quantum simulators operate simultaneously with concurrent data streams, the computational efficiency in time‑stepping simulations is extremely low and the implementation is not straightforward. Specially, for the demanding computational tasks inherent in SPH which involve high-dimensional state spaces and require efficient temporal evolution, a sequentially structured quantum-classical network can prove advantageous as shown in Fig.(\ref{fig_Quantum:hybridoptimization2}).
	
	In this study, we researched the hierarchy of Lagrangian quantum network models based on heuristic SPH. By integrating our improved quantum circuit design featuring optimized parameterized gates and reduced circuit depth with a classical neural network in a sequential hybrid framework, we established a novel quantum-intelligent SPH paradigm generally formulated in Eqs.(\ref{eq:Hierarchy1},\ref{eq:Hierarchy2}). To implement and validate this mathematical substitution, three progressive levels of circuit hierarchies were formulated and evaluated:
	
	\begin{figure}[H]
		\centering
		\includegraphics[scale=0.57]{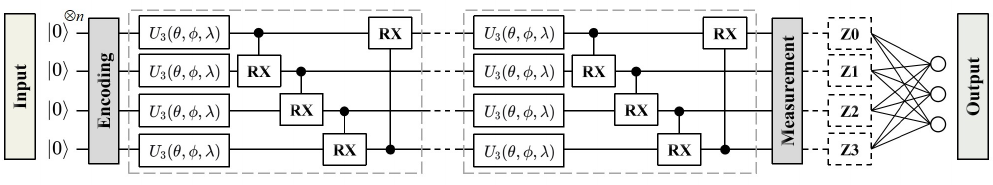}
		\caption{\small Single quantum circuit with SPH.}
		\label{fig_Hierarchy:2}
	\end{figure}
	\begin{figure}[H]
		\centering
		\begin{subfigure}[t]{1.0\textwidth}
			\centering
			\includegraphics[scale=0.57]{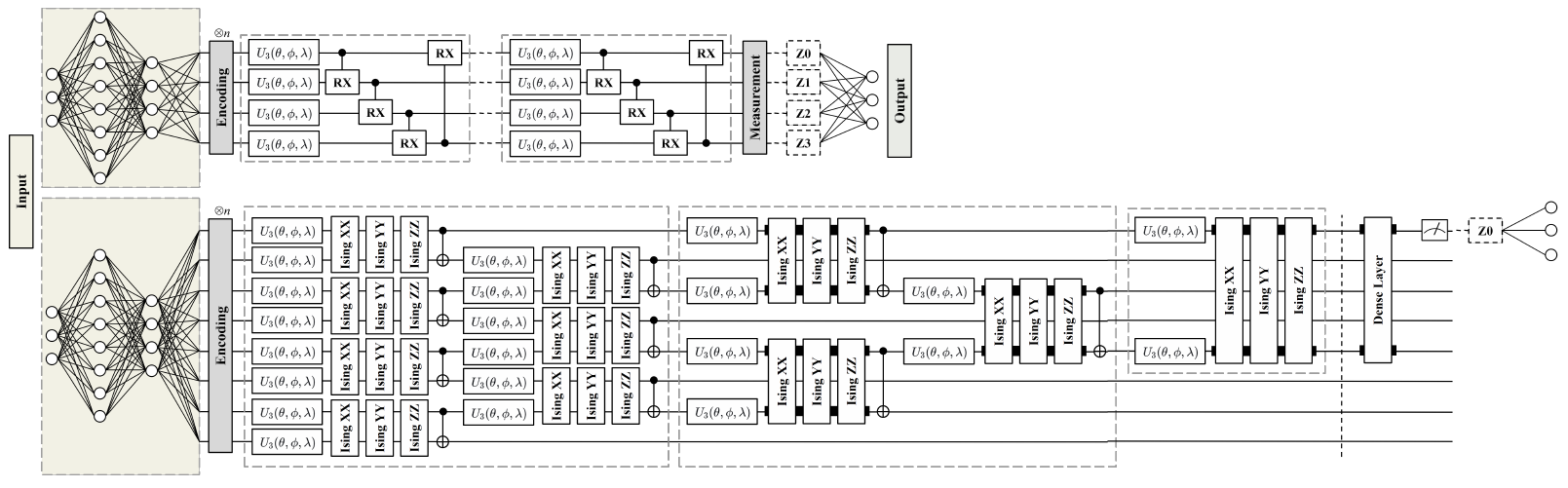}
			\caption{\centering\footnotesize Forward hierarchies of QMLP and QCNN.}  
			\label{fig_Hierarchy:1_6e}
		\end{subfigure}
		\begin{subfigure}[t]{1.0\textwidth}
			\centering
			\includegraphics[scale=0.57]{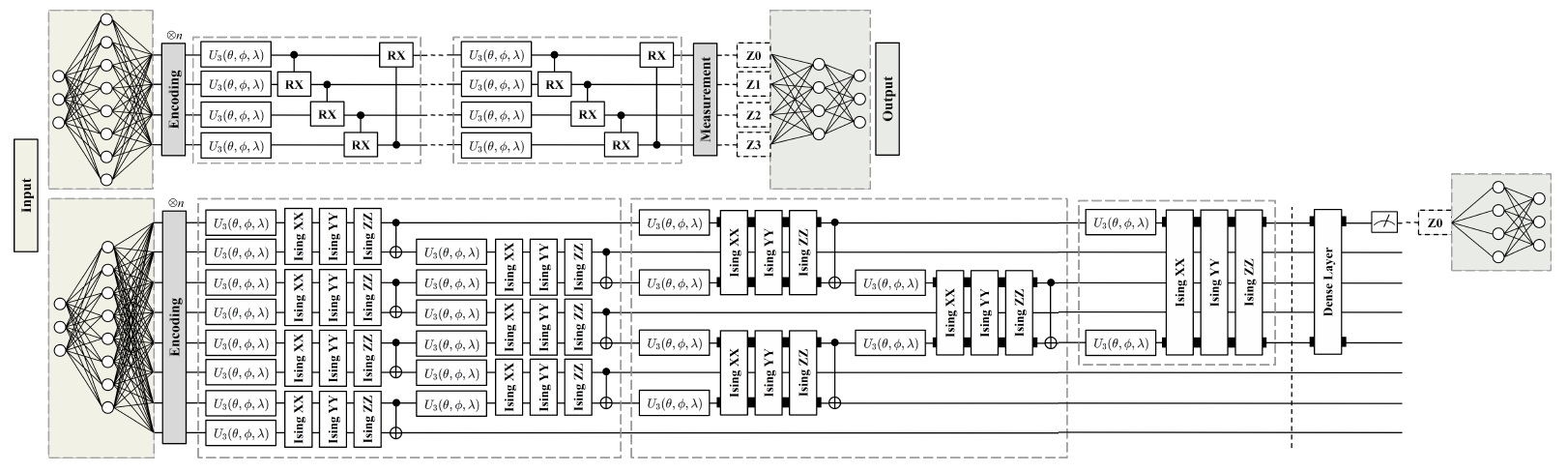}
			\caption{\centering\footnotesize Hybrid architectures of crossed-QMLP and crossed-QCNN.}  
			\label{fig_Hierarchy:1_5e}
		\end{subfigure}
		\caption{\small Hybrid quantum-classical kernel networks with SPH.}
		\label{fig_Hierarchy:1}
	\end{figure}
	
	\textbf{Single quantum circuit with SPH:} The foundational architecture in Fig.(\ref{fig_Hierarchy:2}) utilizes a straightforward quantum pipeline where classical SPH neighbor data is embedded via an encoding layer, subsequently processed through unentangled or minimally entangled parameterized rotation gates (e.g., $RX$, $U_3(\theta, \phi, \lambda)$), and ultimately collapsed via Pauli-Z measurements. While this model provides a fundamental baseline mapping from continuous SPH kernel spaces to discrete quantum states, its relatively shallow structure is often insufficient for capturing the complex, many-body nonlinear correlations intrinsic to fluid systems to an extent.
	
	\textbf{Forward hierarchies of QMLP and QCNN:} To capture more intricate hydrodynamic features, forward hierarchies for both the QMLP and QCNN were constructed in Fig.(\ref{fig_Hierarchy:1_6e}). The forward QMLP expands the expressive depth through multiple cascading layers of $U_3$ rotations. Concurrently, the forward QCNN mimics classical convolutional filters by leveraging translationally invariant Ising-type entangling gates (Ising XX, YY, ZZ) coupled with quantum pooling. While these forward topologies significantly broaden the unitary parameter space, their performance remains fundamentally bounded by the quantum noise threshold and the rigidity of predefined spatial priors.
	
	\textbf{Hybrid architectures of crossed-QMLP and crossed-QCNN:} Addressing the inherent decoherence and limited parameter trainability of pure quantum circuits, hybrid architectures were developed in Fig.(\ref{fig_Hierarchy:1_5e}). By sequentially sandwiching the deep parameterized quantum components between classical dense neural layers, this crossed-architecture effectively compresses the high-dimensional SPH neighborhood arrays into a noise-resilient latent space prior to quantum execution.
	
	\begin{equation}\label{eq:Hierarchy1}
		{f_i} \cong \sum {{\rm{QN}}{{\rm{N}}_{\boldsymbol{\theta }}}\left( {{r_{ij}},\Delta {V_j}} \right){f_j}} ,{f_{i,\alpha }} \cong \sum {{\rm{QN}}{{\rm{N}}_{\boldsymbol{\theta }}}\left( {r_{ij}^\alpha ,\Delta {V_j}} \right){f_{ji}}}
	\end{equation}
	\begin{equation}\label{eq:Hierarchy2}
		\frac{{d{{\boldsymbol{v}}_i}}}{{dt}} = \sum {{\rm{QN}}{{\rm{N}}_{\boldsymbol{\chi }}}\left( {\eta ({r_{ij}},r_{ij}^\alpha ),\Delta {V_j}} \right)\eta ({\boldsymbol{v}_{ji}})}  + {{\boldsymbol{f}}_{ext}}\left( {{\theta _i}} \right)
	\end{equation}
where $f_i$ and $f_j$ denote the physical quantities (e.g., density or pressure) associated with particles $i$ and $j$ respectively, $f_{i,\alpha}$ represents the partial derivative of the $\alpha$-direction. $r_{ij}$ is the Euclidean distance between particles $i$ and $j$, while $r_{ij}^\alpha$ denotes its component in the $\alpha$-axis, $\Delta V_j$ is the volume of particle $j$. $\boldsymbol{v}_i$ is the velocity vector of particle $i$, and $\boldsymbol{v}_{ji}$ is the velocity difference $\boldsymbol{v}_j - \boldsymbol{v}_i$. $\eta(\cdot)$ is a generic nonlinear mapping kernel function, $\boldsymbol{f}_{\text{ext}}(\theta_i)$ is an external force term depending on parameter $\theta_i$ for specified problem, $\text{QNN}_{\boldsymbol{\theta}}$ and $\text{QNN}_{\boldsymbol{\chi}}$ denote quantum kernel networks with trainable parameter sets $\boldsymbol{\theta}$ and $\boldsymbol{\chi}$, respectively, which serve as universal approximators for corresponding interaction terms in aforementioned SPH.
	
	This configuration establishes a critical foundational mapping, although there are some bottlenecks in terms of acceleration effects and the implementation of quantum hardware. By delegating the representation of complex particle interactions to the quantum unitary space and utilizing classical neural networks for adaptive error mitigation and dimensionality reduction, this sequential coupling successfully proves the feasibility of a quantum-classical SPH workflow.
	
	For the SPH-based quantum computing research proposed herein, this hybrid model provides a vital methodological bridge. While the computationally prohibitive components of SPH, such as dynamic neighbor searching and nonlinear pairwise forces, are long-term candidates for true quantum acceleration, our current framework demonstrates for the first time that unstructured Lagrangian fluid topologies can be effectively learned and simulated within a quantum intelligence paradigm. This exploratory strategy offers a realistic and highly adaptable baseline for realizing fully quantum-intelligent particle computing as hardware fidelity matures.
	
	\section{Results}
	\normalsize \hspace{10pt}
	The numerical results are presented in two main parts. The first part provides a performance comparison of quantum network architectures, evaluating the predicting accuracy, training efficiency and parameter scalability of general quantum neural networks (QNNs). The second part explores the integration of quantum kernel networks with smoothed particle hydrodynamics (SPH), focusing on quantum-enhanced kernel representations for the force computation.
	
	All numerical tests are conducted on Origin Quantum Simulators considering the constraints of current NISQ hardware, including limited coherence times and gate errors. Simulators provided a noise-free environment or manually simulated noise, enabling faithful assessment of algorithmic capabilities for the proposed models and serving as a critical benchmark for future hardware implementations\supercite{Ye2024,CHEN2024117428}.
	
	\subsection*{Performance Comparison of General QNN, Improved QMLP and QCNN Processors}
	\normalsize \hspace{10pt}
	Currently, researches on quantum network models are developing rapidly with the maturity of quantum technologies and artificial intelligence, especially for the general quantum network architectures, quantum convolutional neural network, etc\supercite{Cerezo2021Variationalquantum,WOS:000500574300022,WOS:001158795100001}. Given this necessity, the performance analysis of aforementioned network architectures (general QNN, improved QMLP and QCNN processors) is proceeded for developing an improved quantum kernel model suitable for subsequent SPH kernel computations.
	
	\begin{figure}[H]
		\centering
		\begin{subfigure}[t]{0.4\textwidth}
			\centering
			\includegraphics[scale=0.33]{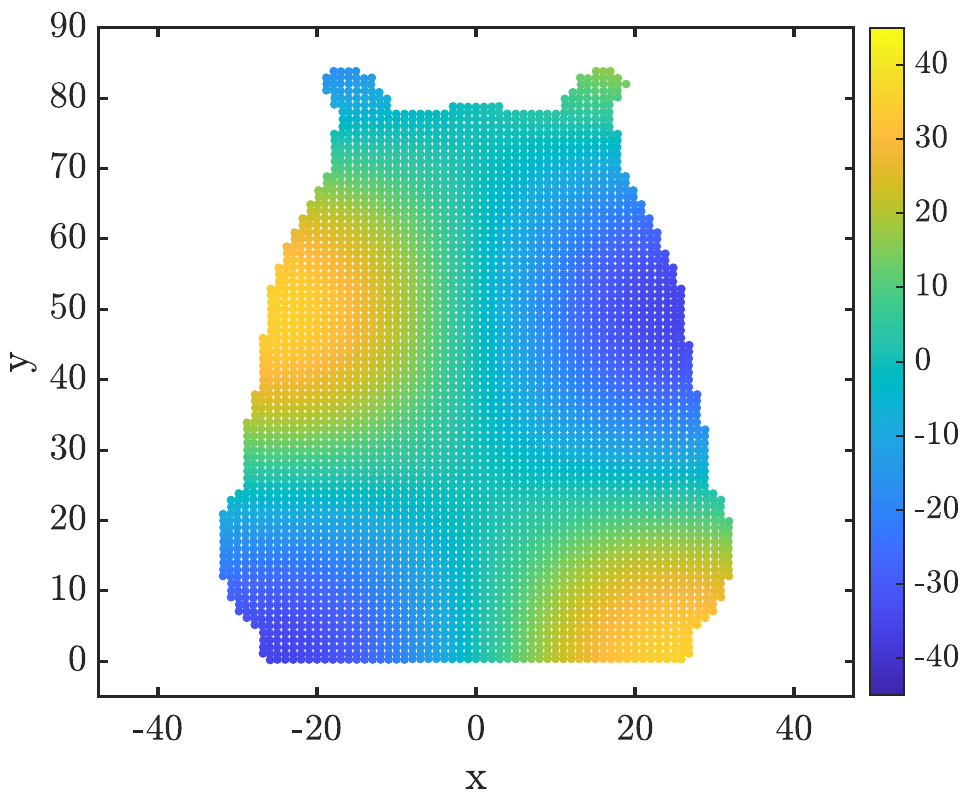 }
			\caption{\centering\footnotesize Featuring special geometries (planar unfolding).}
			\label{fig:6a}
		\end{subfigure}
		\begin{subfigure}[t]{0.4\textwidth}
			\centering
			\includegraphics[scale=0.33]{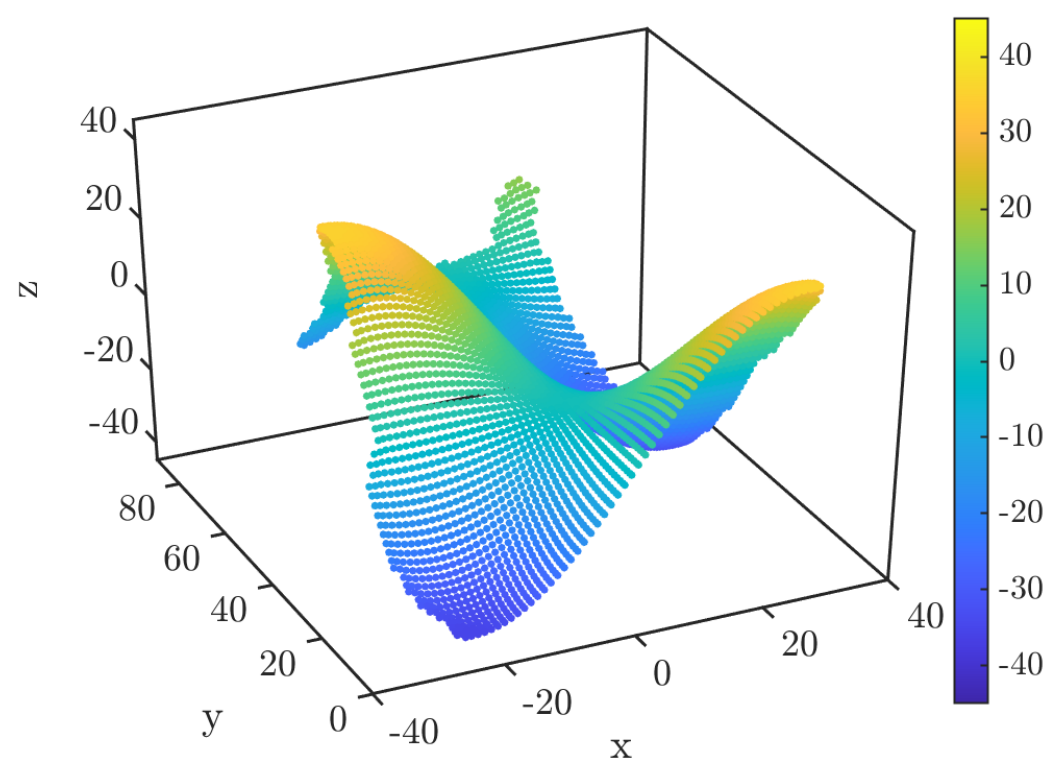 }
			\caption{\centering\footnotesize Trends on continuous function value.}  
			\label{fig:6b}
		\end{subfigure}
		\caption{\small Dedicated benchmark.}
		\label{fig_Case2:Qmodel}
	\end{figure}
\begin{figure}[H]
	\centering
	\begin{subfigure}[t]{0.4\textwidth}
		\centering
		\includegraphics[scale=0.34]{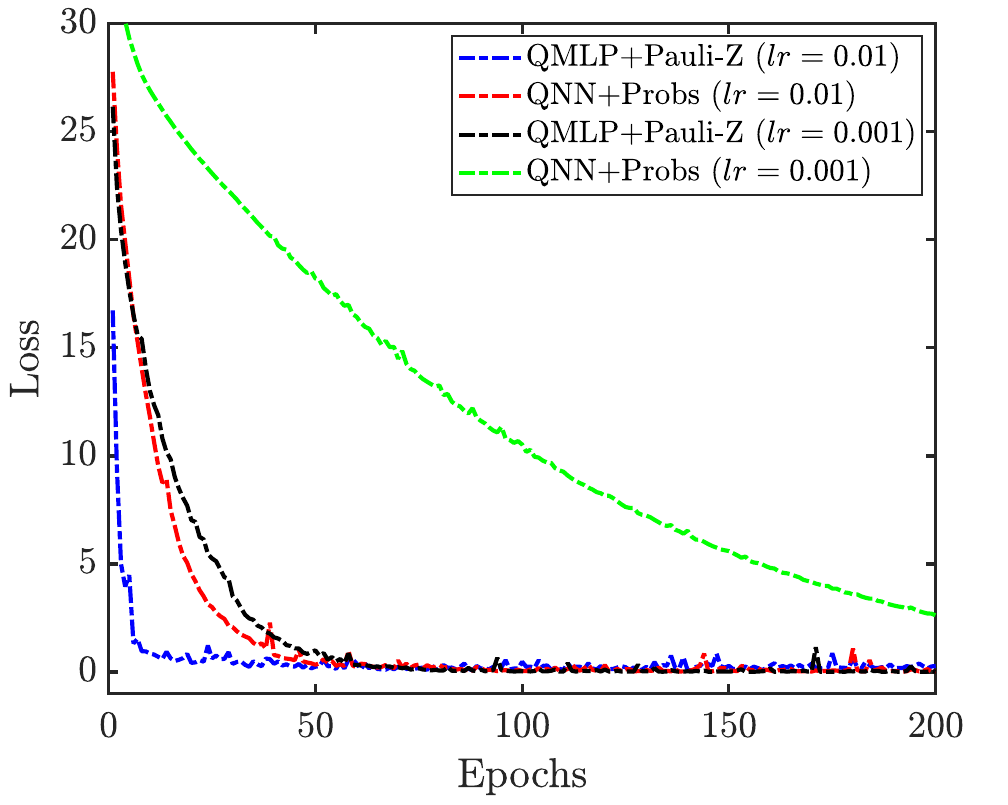}
		\caption{\centering\footnotesize Convergence of loss over epochs at different learning rates.}
		\label{fig:6a}
	\end{subfigure}
	\begin{subfigure}[t]{0.4\textwidth}
		\centering
		\includegraphics[scale=0.34]{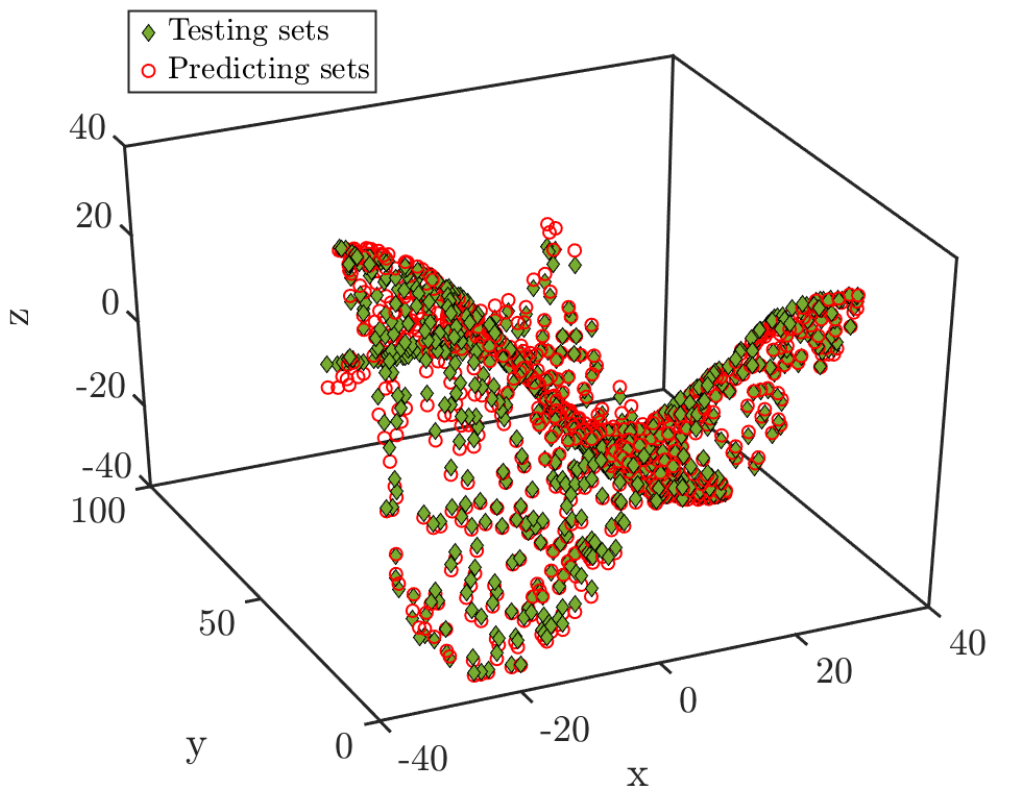}
		\caption{\centering\footnotesize Performance on quantum learning data using the improved QMLP-PauliZ.}  
		\label{fig_Case2:QMLP3_6b}
	\end{subfigure}
	\caption{\small Comparisons between the improved QMLP and QNNs with certain noise.}
	\label{fig_Case2:QMLP3}
\end{figure}
\begin{figure}[H]
	\centering
	\begin{subfigure}[t]{0.4\textwidth}
		\centering
		\includegraphics[scale=0.34]{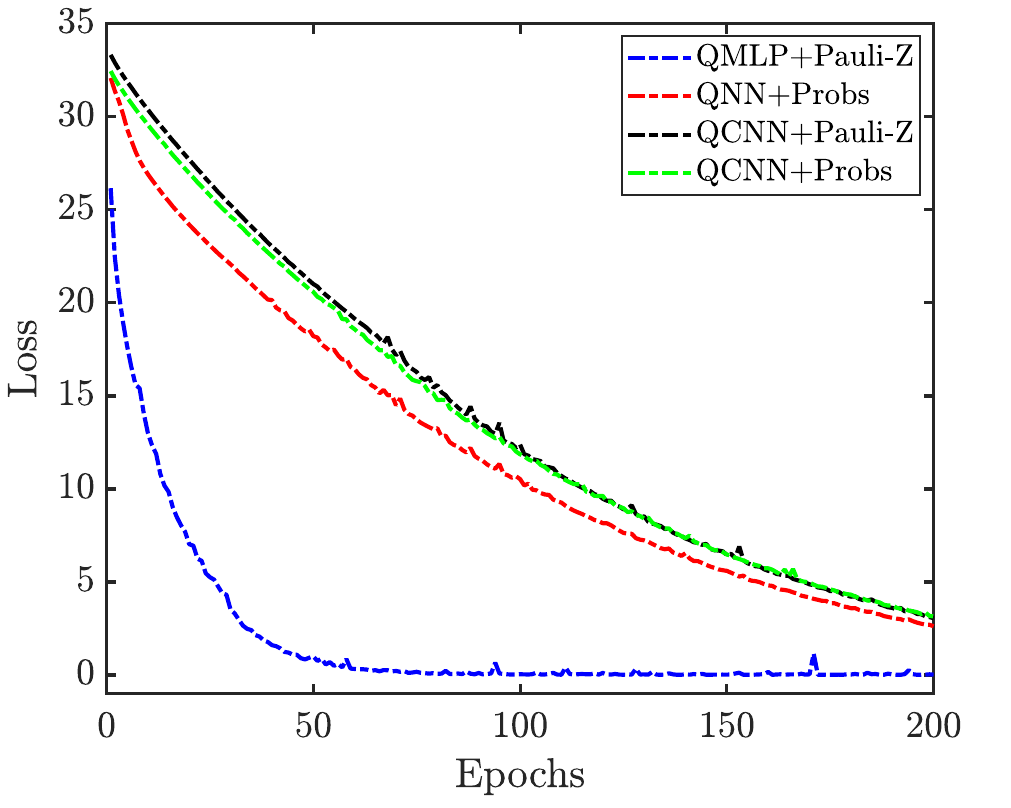}
		\caption{\centering\footnotesize Learning rate $lr = 0.001$.}
		\label{fig:6a}
	\end{subfigure}
	\begin{subfigure}[t]{0.4\textwidth}
		\centering
		\includegraphics[scale=0.34]{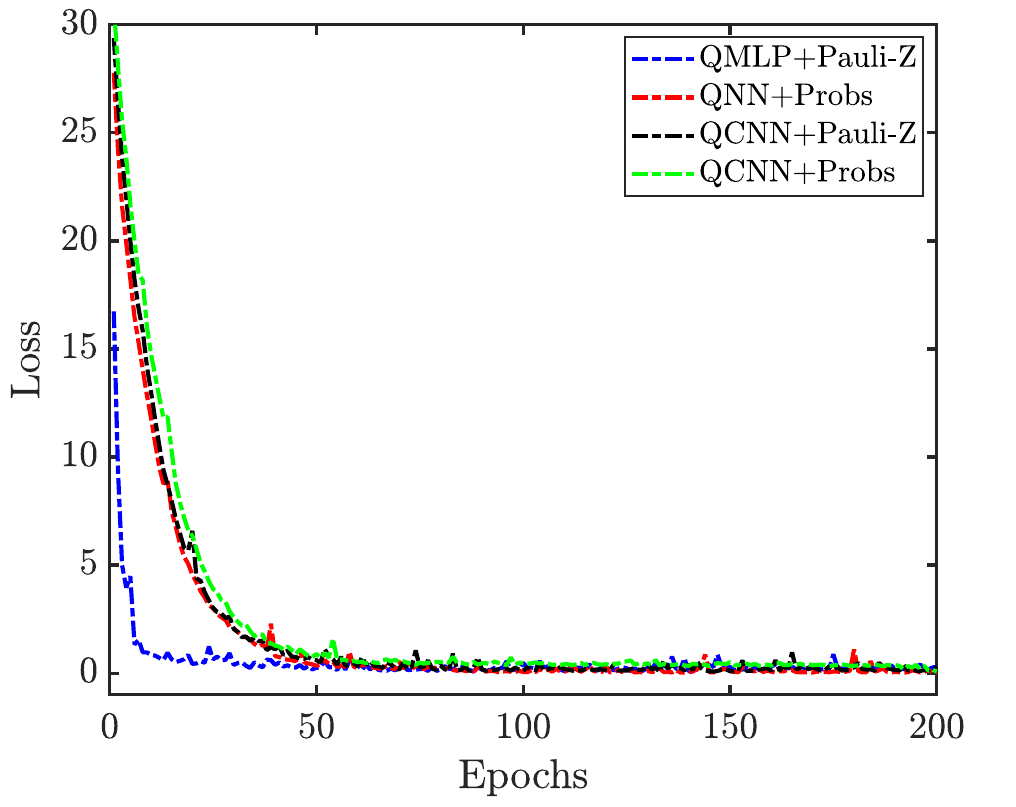}
		\caption{\centering\footnotesize Learning rate $lr = 0.01$.}  
		\label{fig:6b}
	\end{subfigure}
	\caption{\small Convergence of loss based on the aforementioned forward hierarchy of quantum networks.}
	\label{fig_Case2:QCNN3}
\end{figure}

A dedicated 2D benchmark was constructed for the present study shown in Fig.(\ref{fig_Case2:Qmodel}), featuring special geometries and continuously varying trends on function values. Our purpose is to analyze the quantum machine learning ability of aforementioned quantum architectures, in order to further improve the computational efficiency and accuracy for subsequent quantum kernel networks on SPH.
	
	First, Fig.(\ref{fig_Case2:QMLP3}) depicts the computational performance between the improved QMLP architecture and the general QNN architecture at different learning rates. We also investigated different output-forms on collapsing quantum states. It can be observed that the improved QMLP with PauliZ output in Fig.(\ref{fig_Hierarchy:1_6e}) possesses the superior convergence rate. The total loss function consists of a data fitting term, defined as the mean squared error between the network predictions and the reference data. Additionally, physical constraints, such as residuals of governing equations, can be incorporated as auxiliary loss terms to enhance the physical consistency of indicated model within subsequent SPH framework\supercite{WOS:001584061000001}. Fig.(\ref{fig_Case2:QMLP3_6b}) presents the numerical performance of developed quantum network on the training and test sets, demonstrating that the multi-parameter quantum learning integrated with a feed forward architecture achieves strong generalization in the quantum unitary space. 
	
	Subsequently, we further discussed the concerned QCNN processors shown in Fig.(\ref{fig_Hierarchy:1_6e}) at different learning rates. It also demonstrated that the improved QMLP architecture with PauliZ output exhibits significant advantages to an extent. It notes that the advantage of QCNN is not remarkable in this case, which may be attributed to the fact that such networks typically excel in tasks involving spatial structure, such as image recognition, feature pooling and dimensionality reduction\supercite{WOS:000500574300022}. For the general-purpose data learning task considered in this work, these advantages may not be fully leveraged observed from Fig.(\ref{fig_Case2:QCNN3}). On the other hand, three types of quantum neural networks were compared in this study due to the current scarcity of comparative research on quantum neural networks. Future work still holds potential for further exploration in areas such as feature pooling and dimensionality reduction.

This design of improved QMLP model not only enhances the model expressivity, but also offers the flexibility for seamless extension to cross-coupled forward and backward neural configurations from Figs.(\ref{fig_Hierarchy:1},\ref{fig_Case2:QCNN3}). It is worth noting that a comprehensive comparison between forward hierarchy and hybrid architectures is not explicitly presented, as the differences among these configurations are marginal. In practice, the hierarchy segment is typically task-dependent and tailored to specific application requirements, whereas the forward neural component plays a more fundamental role in determining algorithmic behavior. Therefore, from the perspective of core algorithmic performance, focusing on the forward neural architecture is both sufficient and more instructive, ensuring a fair and meaningful comparison across different network designs.

Second, the preceding analysis primarily focuses on the computational efficiency of compared network architectures, while the following discussion addresses their accuracy. In practice, the accuracy differences among the evaluated networks are marginal, with efficiency serving as the dominant distinguishing factor. Nevertheless, certain configurations particularly those incorporating both forward and backward neural components (hybrid crossed-hierarchies), exhibit modest advantages, providing valuable numerical benchmarks for hybrid quantum network models. These observations also serve to highlight the inherent advantages of improved quantum network architecture developed in this work, offering insight into its optimization potential. This comparative study aims to establish a reference point for various quantum network models, facilitating further exploration by subsequent researchers. The contribution of particular significance was given in the current scarcity of studies addressing the fitting problem in this domain.
	
	\begin{figure}[H]
		\centering
		\begin{subfigure}[t]{0.4\textwidth}
			\centering
			\includegraphics[scale=0.34]{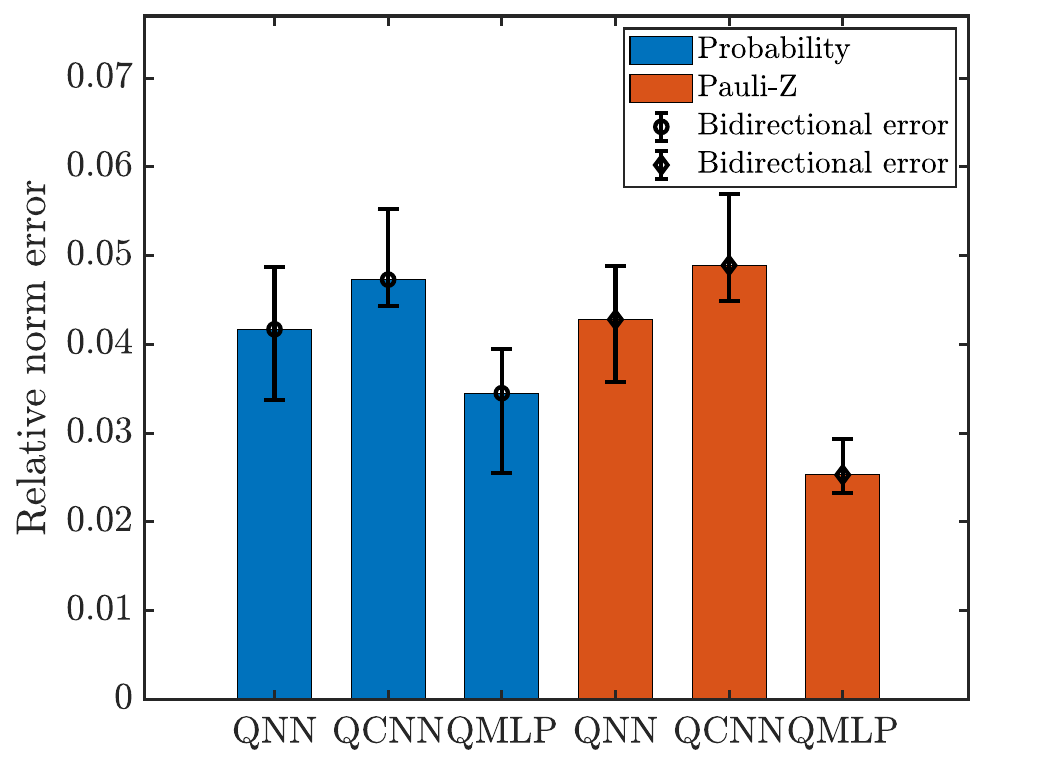}
			\caption{\centering\footnotesize Counted errors based on the forward hierarchy.}
			\label{fig:6a}
		\end{subfigure}
		\begin{subfigure}[t]{0.4\textwidth}
			\centering
			\includegraphics[scale=0.34]{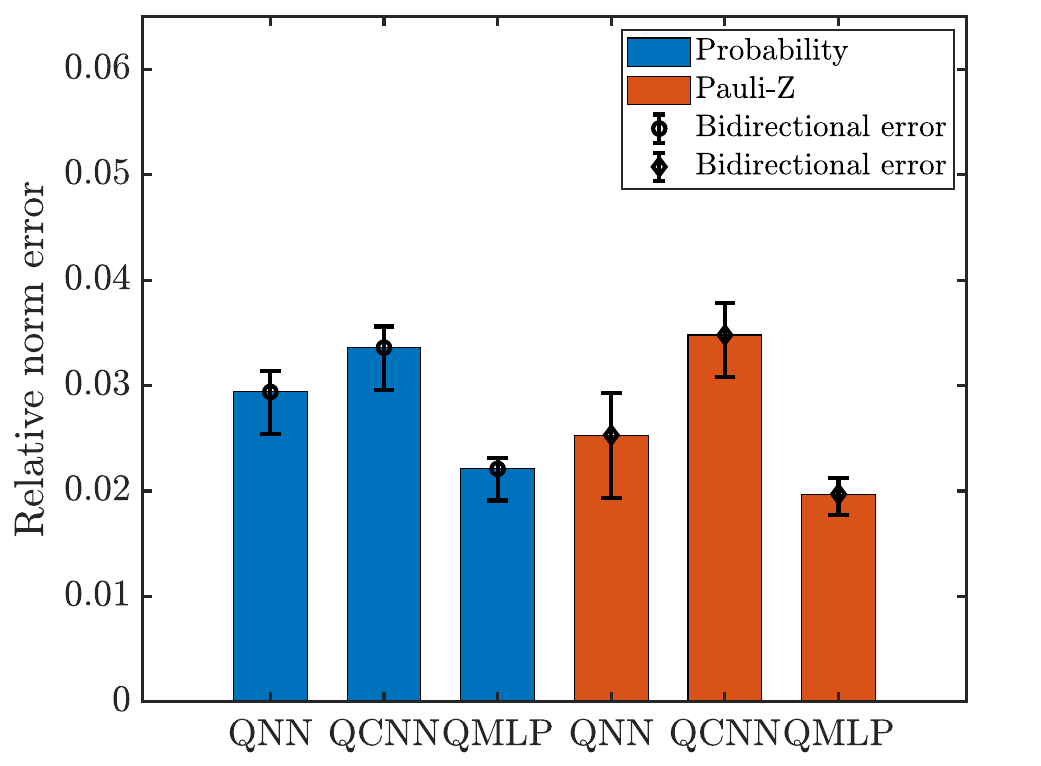}
			\caption{\centering\footnotesize Counted errors based on the hybrid architectures.}  
			\label{fig:6b}
		\end{subfigure}
		\caption{\small Computational accuracy of three quantum networks based on different quantum architectures.}
		\label{fig_Case2:Accuracy1}
	\end{figure}
	\begin{figure}[H]
		\centering
		\begin{subfigure}[t]{0.4\textwidth}
			\centering
			\includegraphics[scale=0.34]{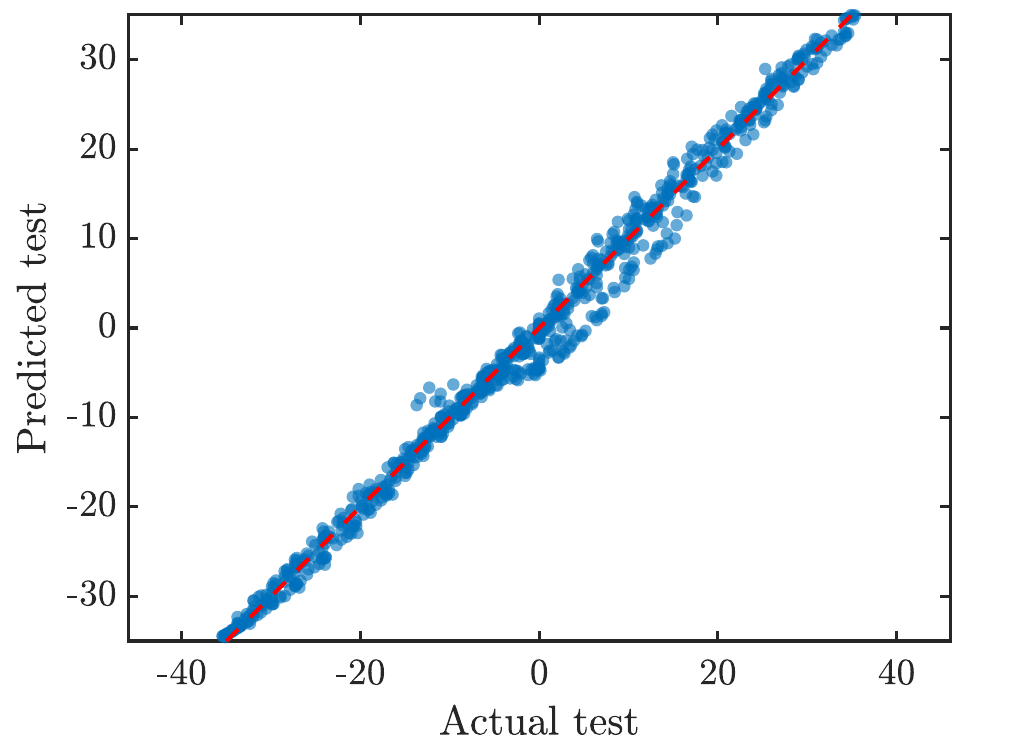}
			\caption{\centering\footnotesize QNN-Probability in forward hierarchy.}
			\label{fig:6a}
		\end{subfigure}
		\begin{subfigure}[t]{0.4\textwidth}
			\centering
			\includegraphics[scale=0.34]{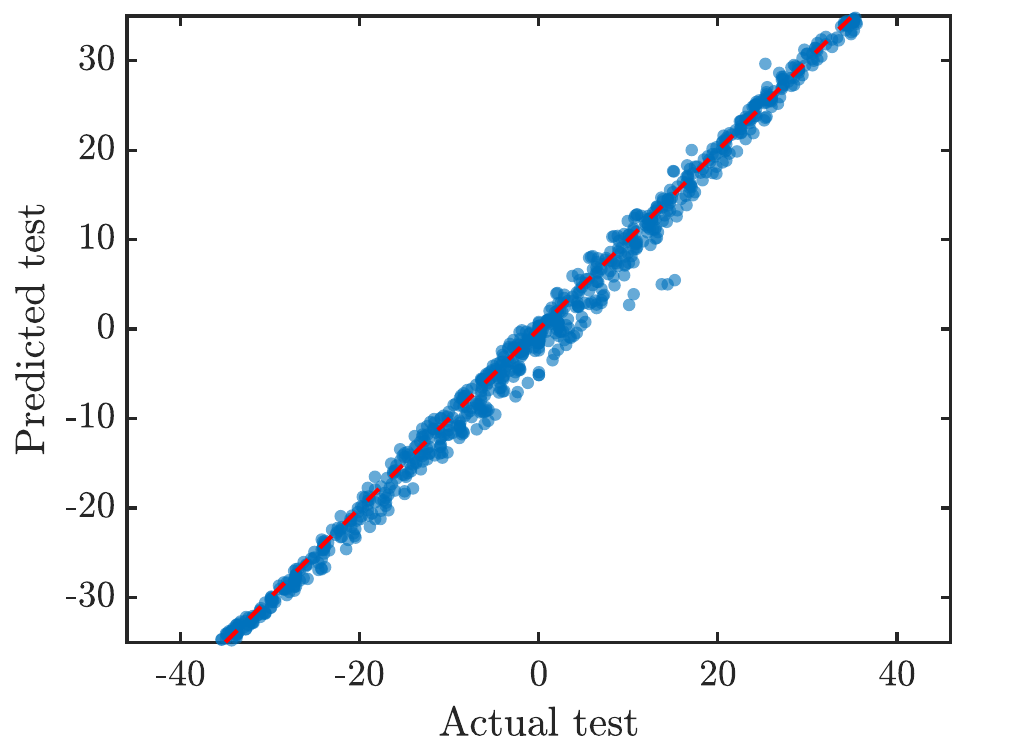}
			\caption{\centering\footnotesize QCNN-PauliZ in forward hierarchy.}  
			\label{fig:6b}
		\end{subfigure}
		\begin{subfigure}[t]{0.4\textwidth}
			\centering
			\includegraphics[scale=0.34]{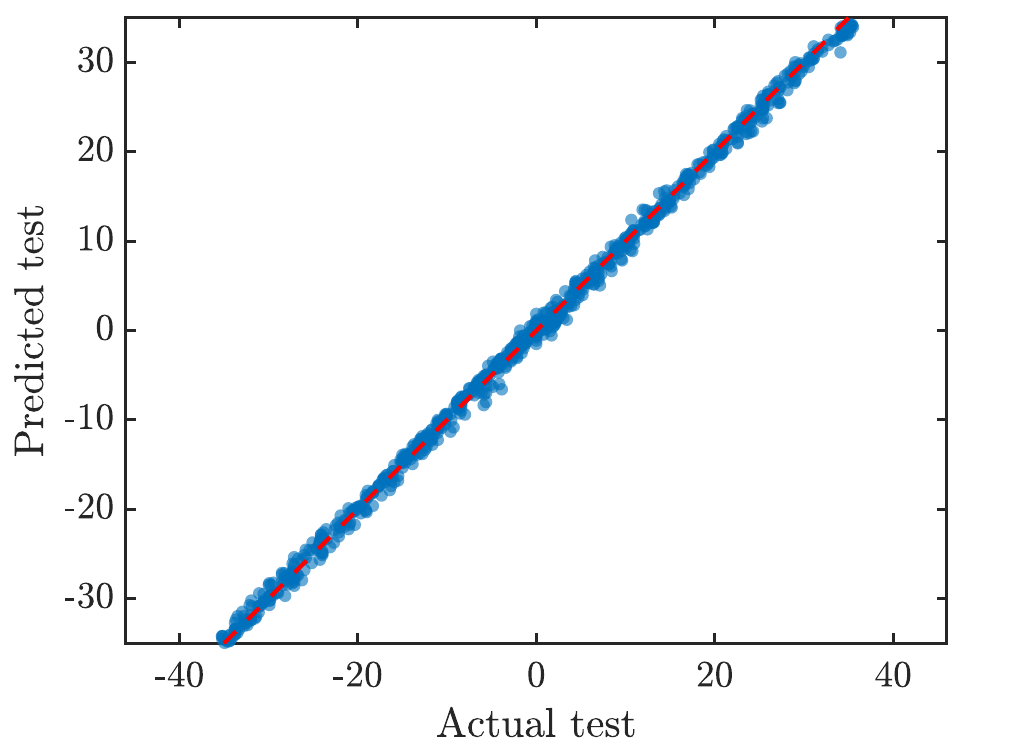}
			\caption{\centering\footnotesize QMLP-PauliZ in forward hierarchy.}
			\label{fig:6a}
		\end{subfigure}
		\begin{subfigure}[t]{0.4\textwidth}
			\centering
			\includegraphics[scale=0.34]{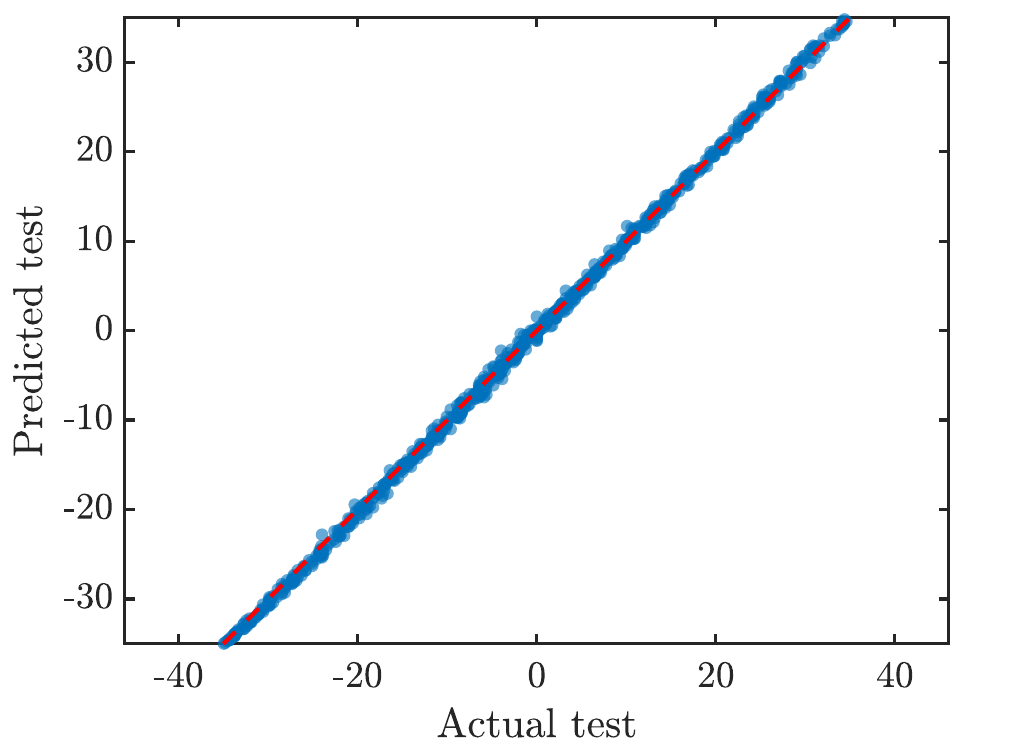}
			\caption{\centering\footnotesize QMLP-PauliZ in hybrid architectures.}  
			\label{fig:6b}
		\end{subfigure}
		\caption{\small Analysis of test results obtained from quantum neural network models.}
		\label{fig_Case2:Accuracy2}
	\end{figure}

We explored the computational accuracy on the aforementioned three quantum networks based on the forward hierarchy and hybrid architectures respectively shown in Fig.(\ref{fig_Hierarchy:1}). The relative norm errors of current three quantum models are presented in Fig.(\ref{fig_Case2:Accuracy1}), where the error bars are arranged in a tree-like structure to illustrate the distribution. It can be observed that the hybrid architectures of Fig.(\ref{fig_Hierarchy:1_5e}) generally have more optimal accuracy than the forward hierarchy of Fig.(\ref{fig_Hierarchy:1_6e}). The rationale lies in fact that the incorporation of a backward neural network not only increases the number of optimizable parameters but also further expands the trainability of quantum parameters within unitary space. However, the primary capability originates from the strong generalization of forward neural network, which further unlocks the optimization mechanism of the quantum circuit's unitary parameter space. The backward network, in turn, needs to be tailored in conjunction with the specific problem to achieve the robust optimization capability characteristic of quantum-intelligent circuits. 

On the other hand, the improved QMLP model exhibits moderately superior accuracy performance among these quantum networks, demonstrating the same pattern of advantage as observed in the earlier comparison of computational efficiency. Related results can be also analyzed from Fig.(\ref{fig_Case2:Accuracy2}). It notes that the output of current quantum networks can be realized either as total probability observations (Probability) or as the expectation value of PauliZ. Overall, the PauliZ expectation yields superior results, as it captures the expected probabilistic information within the unitary space. This is because the expectation value retains phase coherence and provides a continuous, differentiable signal that is better suited for gradient-based optimization, whereas total probability observations offer only discrete measurement snapshots and discard phase information, limiting their representational capacity to an extent. 
		
	\begin{figure}[H]
		\centering
		\begin{subfigure}[t]{0.4\textwidth}
			\centering
			\includegraphics[scale=0.34]{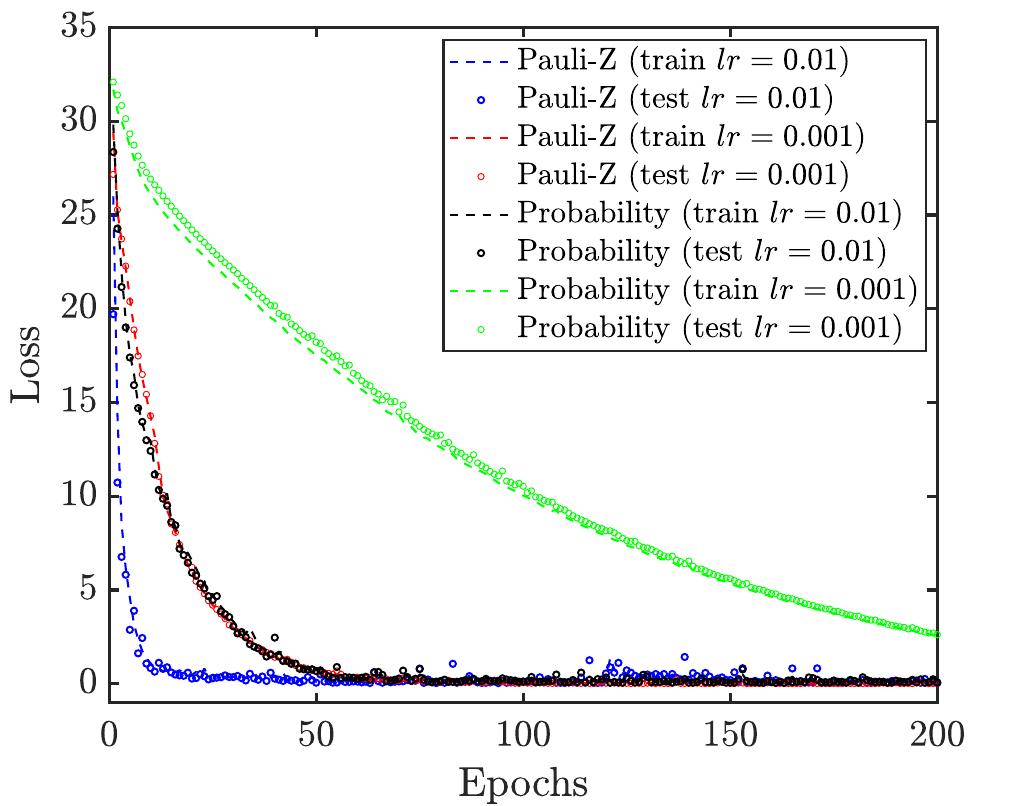}
			\caption{\centering\footnotesize Forward hierarchy.}
			\label{fig:6a}
		\end{subfigure}
		\begin{subfigure}[t]{0.4\textwidth}
			\centering
			\includegraphics[scale=0.34]{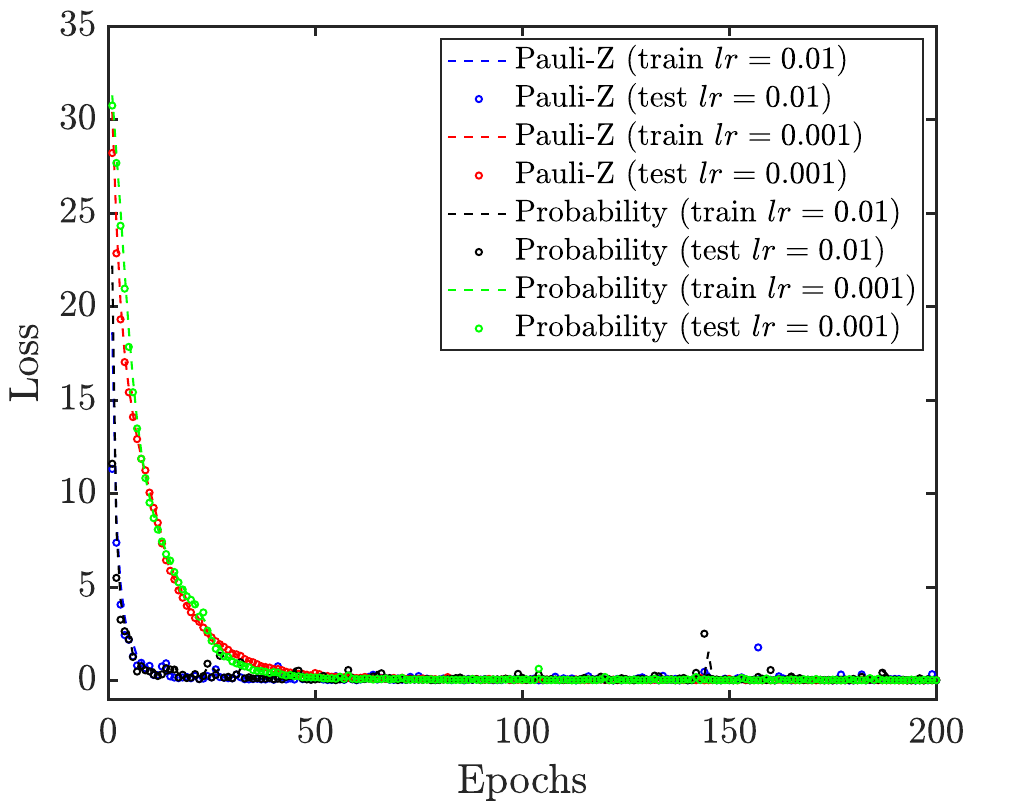}
			\caption{\centering\footnotesize Hybrid architectures.}  
			\label{fig:6b}
		\end{subfigure}
		\caption{\small Numerical performance of QNN at different learning rates and outputs.}
		\label{fig_Case2:QNN1}
	\end{figure}
	\begin{figure}[H]
		\centering
		\begin{subfigure}[t]{0.4\textwidth}
			\centering
			\includegraphics[scale=0.34]{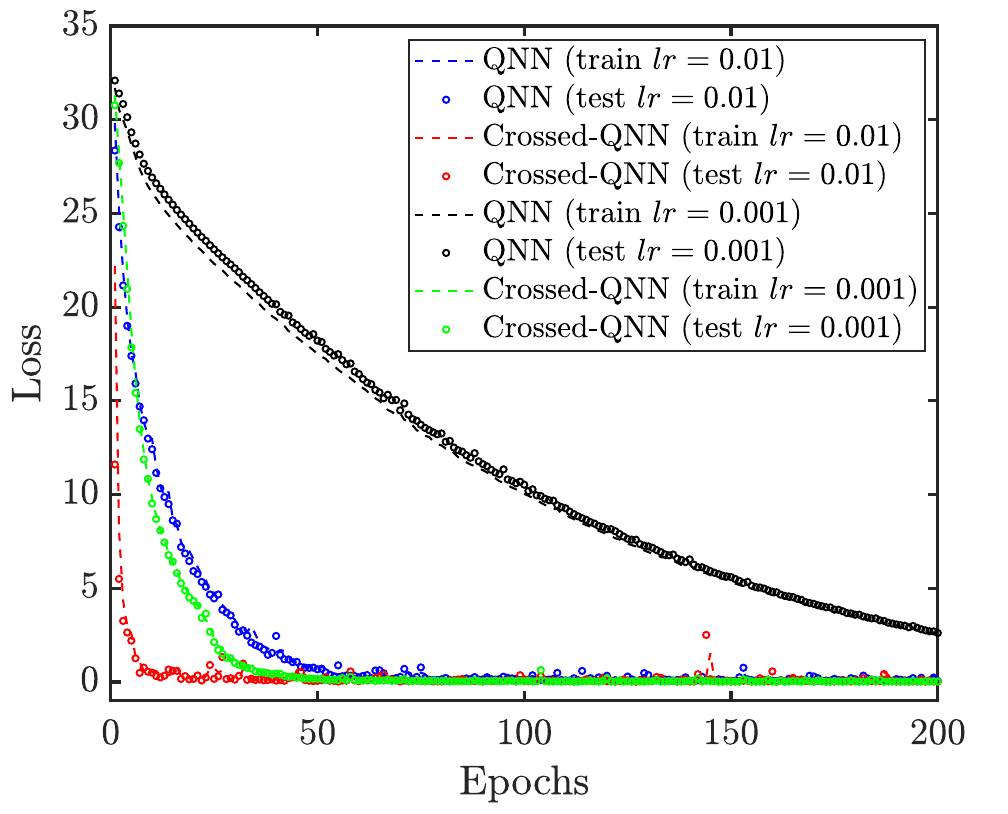}
			\caption{\centering\footnotesize Probability output.}
			\label{fig:6a}
		\end{subfigure}
		\begin{subfigure}[t]{0.4\textwidth}
			\centering
			\includegraphics[scale=0.34]{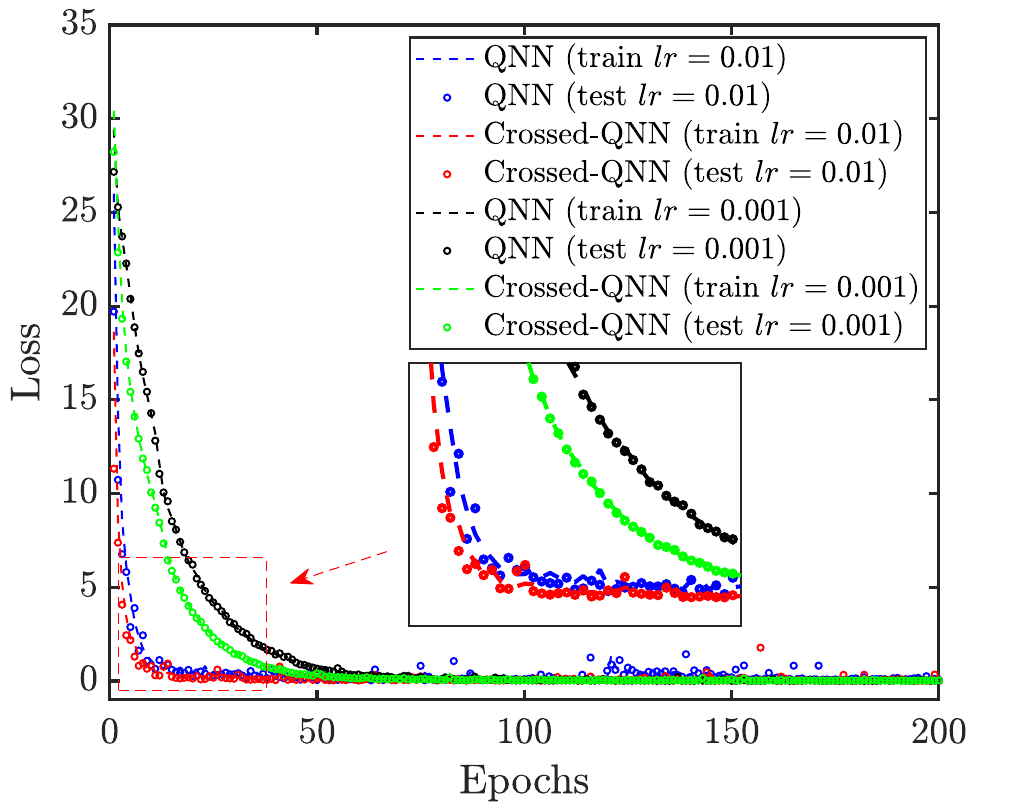}
			\caption{\centering\footnotesize PauliZ output.}  
			\label{fig:6b}
		\end{subfigure}
		\caption{\small Comparisons on QNN between forward hierarchy and hybrid architectures at different learning rates and outputs.}
		\label{fig_Case2:QNN2}
	\end{figure}

\begin{figure}[H]
		\centering
		\begin{subfigure}[t]{0.4\textwidth}
			\centering
			\includegraphics[scale=0.34]{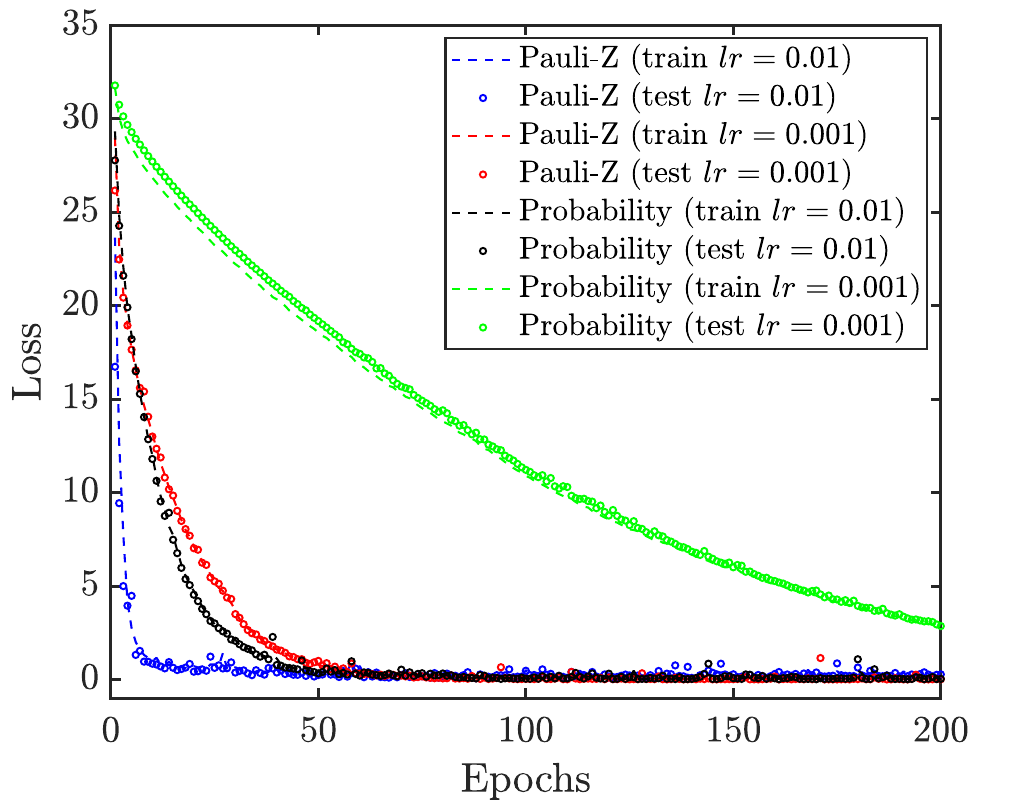}
			\caption{\centering\footnotesize Forward hierarchy.}
			\label{fig:6a}
		\end{subfigure}
		\begin{subfigure}[t]{0.4\textwidth}
			\centering
			\includegraphics[scale=0.34]{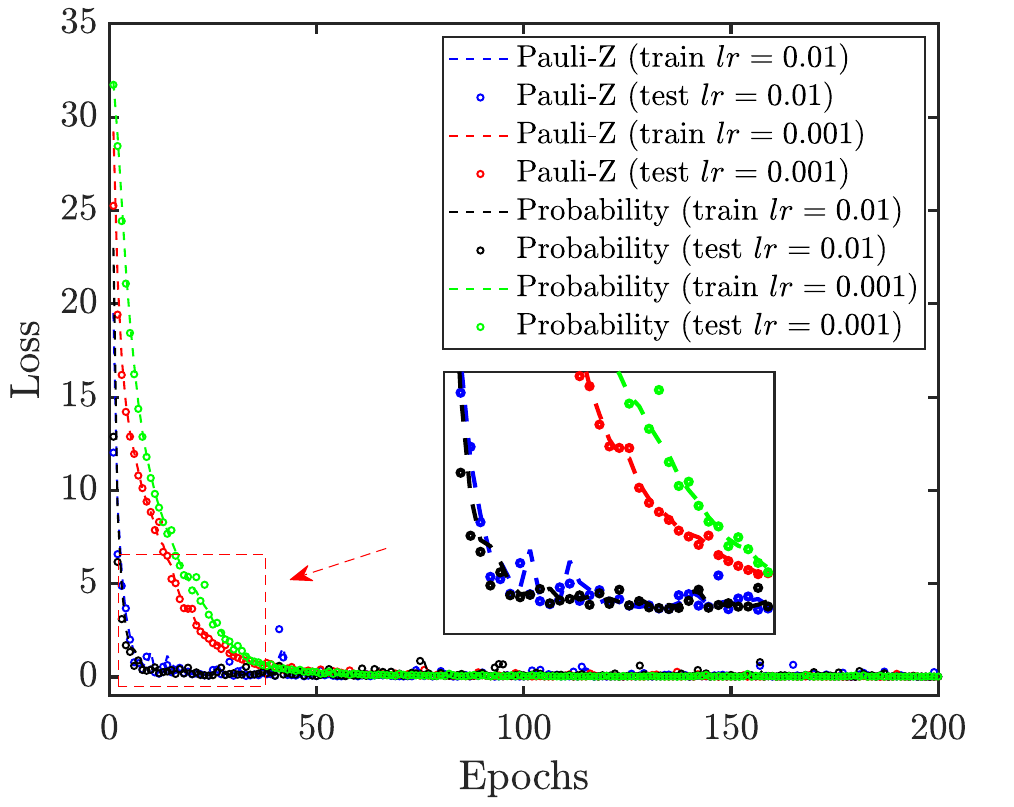}
			\caption{\centering\footnotesize Hybrid architectures.}  
			\label{fig:6b}
		\end{subfigure}
		\caption{\small Numerical performance of QMLP at different learning rates and outputs.}
		\label{fig_Case2:QMLP1}
	\end{figure}

\begin{figure}[H]
		\centering
		\begin{subfigure}[t]{0.4\textwidth}
			\centering
			\includegraphics[scale=0.34]{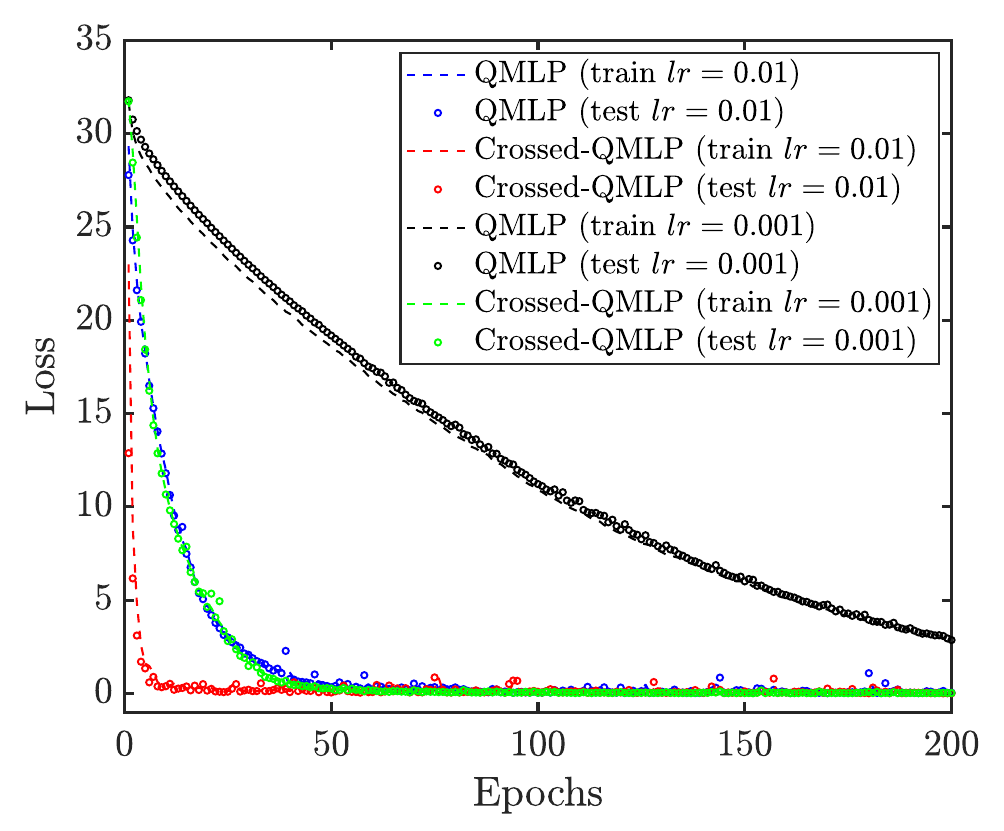}
			\caption{\centering\footnotesize Probability output.}
			\label{fig:6a}
		\end{subfigure}
		\begin{subfigure}[t]{0.4\textwidth}
			\centering
			\includegraphics[scale=0.34]{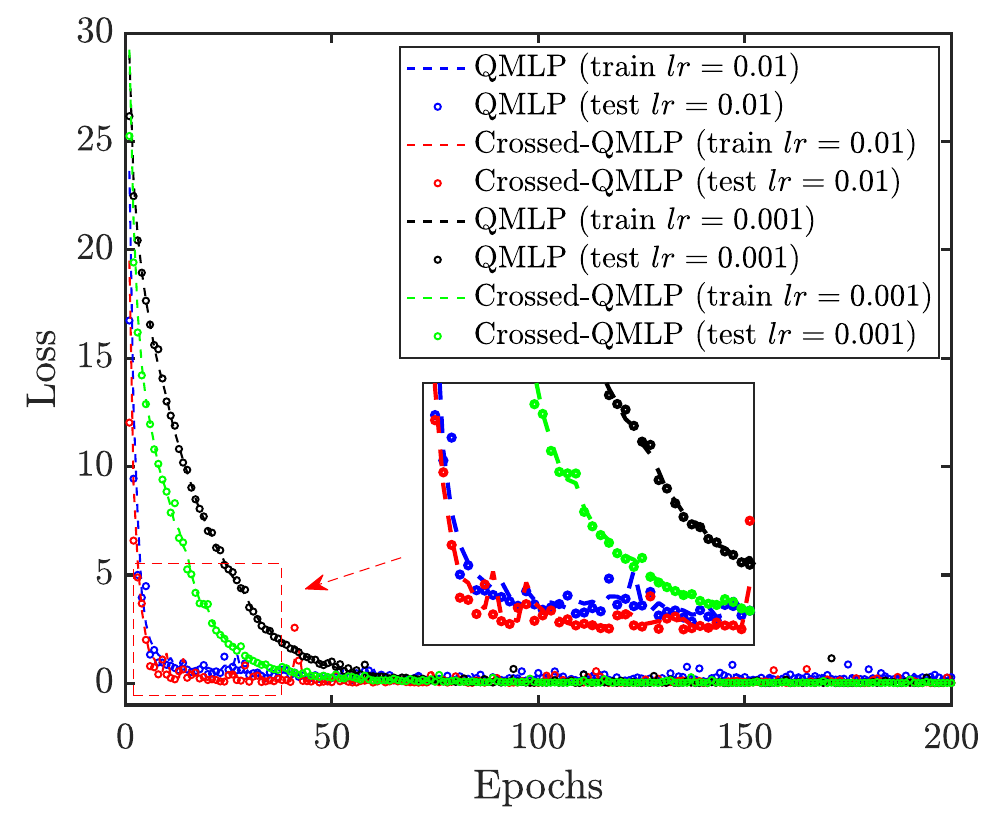}
			\caption{\centering\footnotesize PauliZ output.}  
			\label{fig:6b}
		\end{subfigure}
		\caption{\small Comparisons on QMLP between forward hierarchy and hybrid architectures at different learning rates and outputs.}
		\label{fig_Case2:QMLP2}
	\end{figure}

\begin{figure}[H]
		\centering
		\begin{subfigure}[t]{0.4\textwidth}
			\centering
			\includegraphics[scale=0.34]{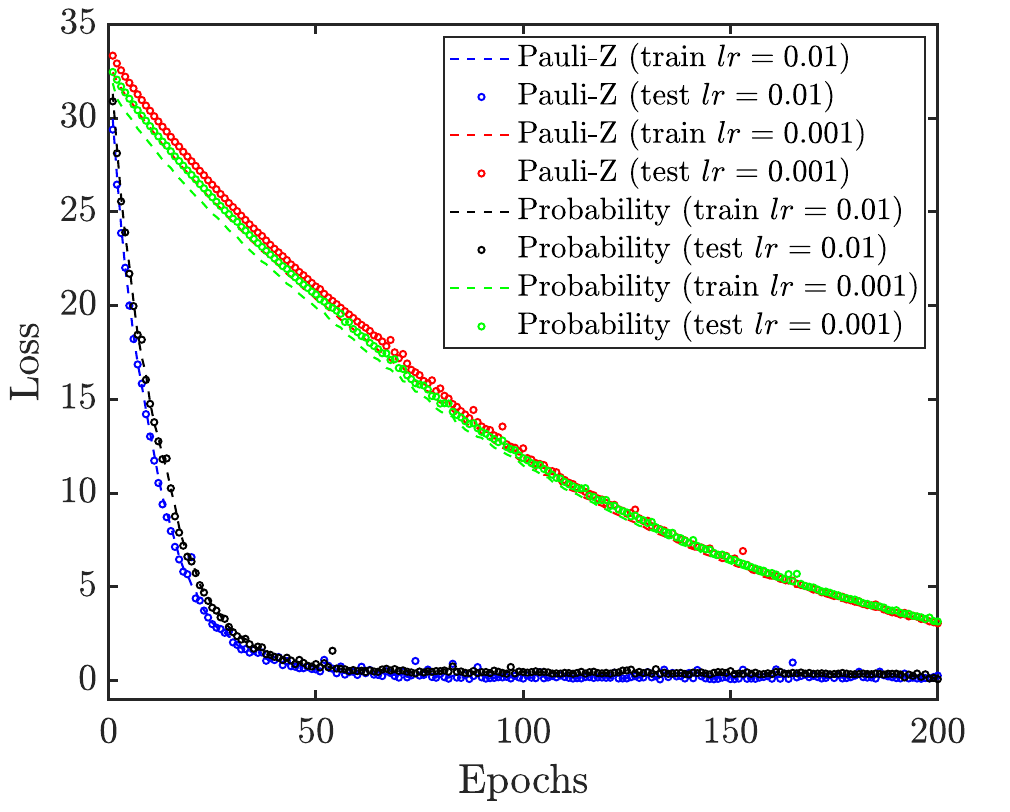}
			\caption{\centering\footnotesize Forward hierarchy.}
			\label{fig:6a}
		\end{subfigure}
		\begin{subfigure}[t]{0.4\textwidth}
			\centering
			\includegraphics[scale=0.34]{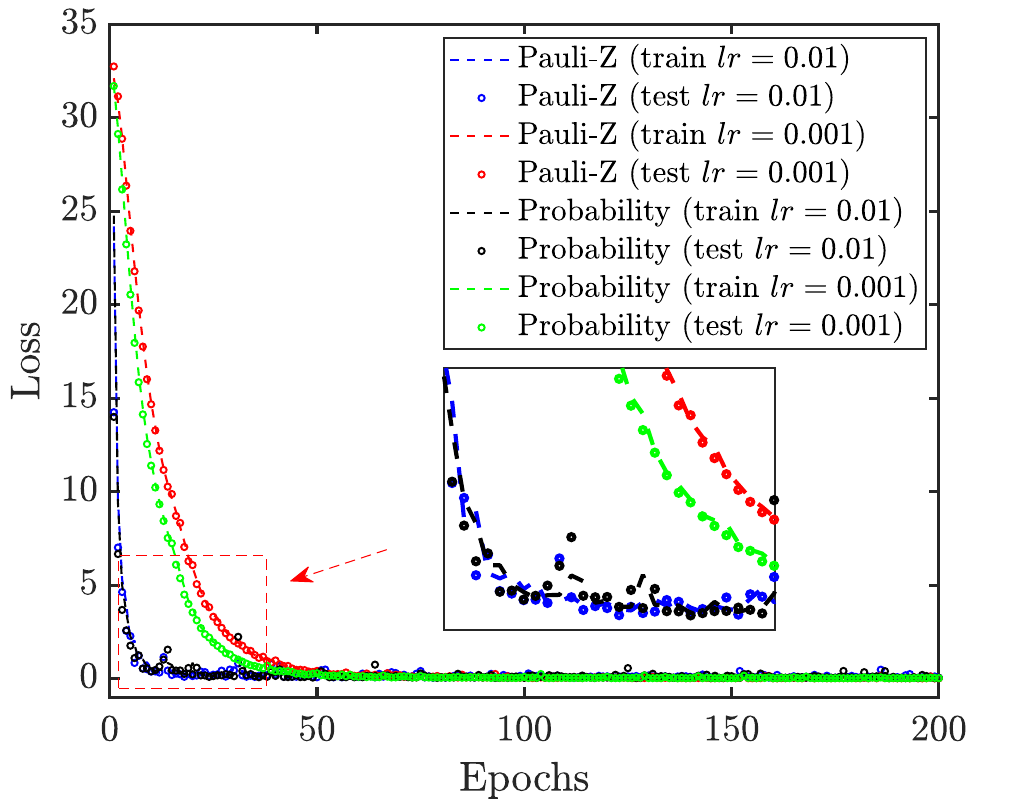}
			\caption{\centering\footnotesize Hybrid architectures.}  
			\label{fig:6b}
		\end{subfigure}
		\caption{\small Numerical performance of QCNN at different learning rates and outputs.}
		\label{fig_Case2:QCNN1}
	\end{figure}

\begin{figure}[H]
		\centering
		\begin{subfigure}[t]{0.4\textwidth}
			\centering
			\includegraphics[scale=0.34]{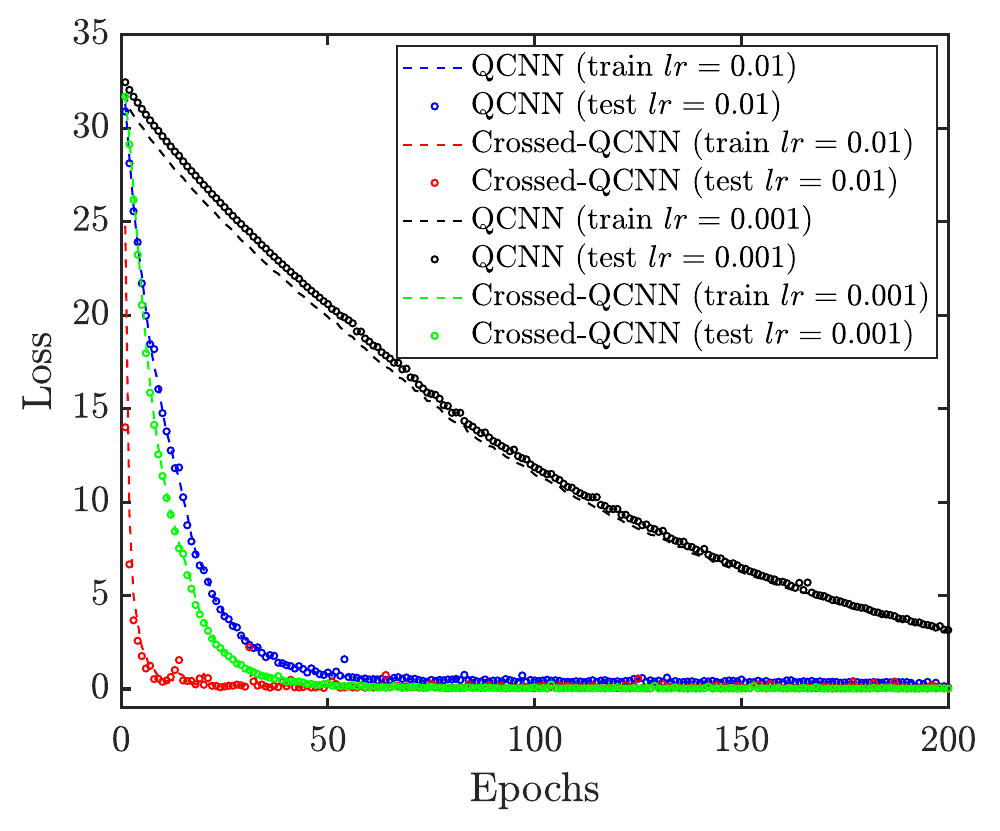}
			\caption{\centering\footnotesize Probability output.}
			\label{fig:6a}
		\end{subfigure}
		\begin{subfigure}[t]{0.4\textwidth}
			\centering
			\includegraphics[scale=0.34]{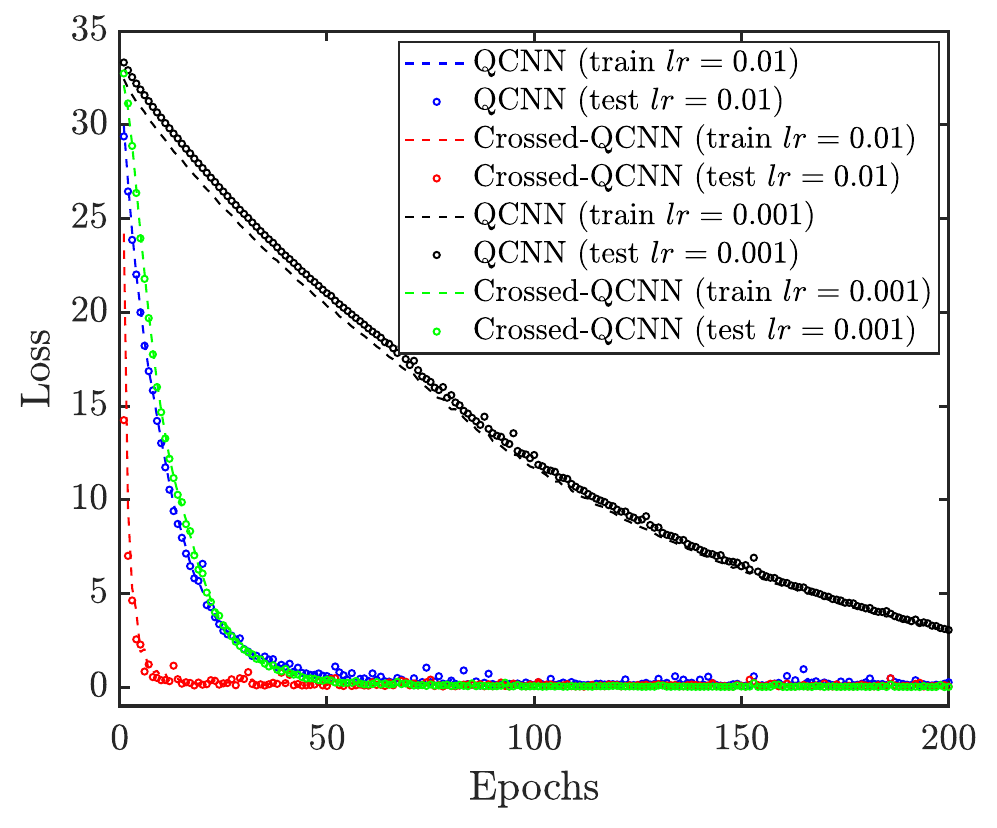}
			\caption{\centering\footnotesize PauliZ output.}  
			\label{fig:6b}
		\end{subfigure}
		\caption{\small Comparisons on QCNN between forward hierarchy and hybrid architectures at different learning rates and outputs.}
		\label{fig_Case2:QCNN2}
	\end{figure}

Third, for indicated three quantum networks in this study, we further evaluated the performance separately, and the specific results can be obtained from Figs.(\ref{fig_Case2:QNN1}-\ref{fig_Case2:QCNN2}). Fig.(\ref{fig_Case2:QNN1}) compares the numerical performance of general QNN under forward hierarchy and hybrid architectures, respectively. The QNN with Pauli‑Z output converges faster than its probability‑based counterpart. Fig.(\ref{fig_Case2:QNN2}) further contrasts the forward and hybrid QNNs at two learning rates, confirming that the hybrid architecture yields slightly improved stability. Fig.(\ref{fig_Case2:QMLP1}) and Fig.(\ref{fig_Case2:QMLP2}) present analogous results for the improved QMLP networks. The QMLP with Pauli‑Z expectation consistently outperforms the probability output, achieving lower loss and tighter error bounds. Moreover, the hybrid crossed‑QMLP architecture in Fig.(\ref{fig_Case2:QMLP2}) exhibits superior convergence and reduced variance compared to the forward hierarchy, underscoring the benefit of classical forward and backward processing layers. Fig.(\ref{fig_Case2:QCNN1}) and Fig.(\ref{fig_Case2:QCNN2}) show the computing performance on QCNN. While the QCNN with Pauli‑Z output behaves similarly to the QMLP for the forward hierarchy, its hybrid architecture does not show a significant improvement over the forward one from Fig.(\ref{fig_Case2:QCNN2})), likely because the convolutional structure already imposes strong spatial priors, leaving less room for additional classical layers to enhance generalization.


Overall, these figures collectively demonstrate that the improved QMLP with Pauli‑Z output and hybrid crossed architecture offer the most balanced trade‑off between accuracy, convergence speed and architectural flexibility among aforementioned three evaluated quantum network models.
	
	\subsection*{Static Quantum Kernel on Smoothed Particle Hydrodynamics}
	\normalsize \hspace{10pt}
	The developed hierarchy of Lagrangian quantum networks aims to proceed the computational integration of quantum intelligence and SPH kernel space. Through the aforementioned performance comparison of general QNN, improved QMLP and QCNN processors, we primarily employed the improved QMLP‑PauliZ model for SPH kernel studies\supercite{WOS:000989244100003}. The performance of other neural architectures exhibited qualitatively similar behavior and is therefore not elaborated here, as the methodological rationale remains consistent.
	
	The static quantum kernel on SPH is calculated and the benchmark of a single vortex component \(i\) is purposefully given as described in Eq.(\ref{StaticQuantumKernel_1}). It generates two‑dimensional scalar field distributions with nebulae. It constructs complex interference patterns through the linear superposition of multiple rotating vortices, governed by carefully designed mathematical expressions. The core idea is to validate the computational effectiveness of quantum SPH kernel on multi-complex static problems.
	\begin{equation}\label{StaticQuantumKernel_1}
		f_i(x,y,t) = A_i \cdot \underbrace{\exp\left(-\frac{r_i^2}{2\sigma_i^2}\right)}_{\text{Gaussian envelope}} 
		\cdot \underbrace{\sin(\omega_i t + k_i r_i + m_i \theta_i)}_{\text{primary wave term}}
		\cdot \underbrace{[1 + \alpha_i \cos(\beta_i \theta_i)]}_{\text{azimuthal modulation}}
		\cdot \underbrace{\tanh\left(\frac{r_i}{\sigma_i}\right)}_{\text{radial modulation}}
	\end{equation}
	where \(r_i = \sqrt{(x-c_{x,i})^2 + (y-c_{y,i})^2}\) is the radial distance to vortex center, \(\theta_i = \arctan\left(\frac{y-c_{y,i}}{x-c_{x,i}}\right)\) is the azimuthal angle, \(A_i\) is the amplitude controlling vortex intensity, \(\sigma_i\) is the characteristic size defining the vortex influence range, \(\omega_i\) is the angular frequency governing temporal evolution, \(k_i\) is the radial wavenumber determining the radial fringe density, \(m_i\) is the azimuthal mode number controlling the number of lobes (\(m=3\) yields three lobes, \(m=5\) yields five lobes), \(\alpha_i, \beta_i\) are azimuthal modulation parameters enhancing geometric complexity.
	
	The total field is composed of three main superimposed layers: the primary vortex layer, the fine‑structure layer and the background fluctuation layer. 
	
	\textbf{Primary vortex layer:} three vortex components with distinct parameters are linearly superposed,
	\begin{equation}
		Z_{\text{vortex}}(x,y,t) = \sum_{i=1}^{3} f_i(x,y,t).
	\end{equation}
	Each vortex has its own spatial position \((c_x,c_y)\) and physical parameters, producing complex interference patterns. The adopted parameter configuration is
	\begin{align*}
		&\text{Vortex 1: } (0.4, 0.6, 1.2, 0.25, 1.5, 15.0, 5, 0.3, 2.0) \\
		&\text{Vortex 2: } (0.6, 0.4, 1.0, 0.20, 2.0, 12.0, 3, 0.4, 2.5) \\
		&\text{Vortex 3: } (0.5, 0.5, 0.8, 0.35, 0.8, 8.0, 7, 0.2, 1.5).
	\end{align*}
	
	\textbf{Fine‑structure layer:} to simulate the fine-structures in nebulae, high‑frequency noise components are added,
	\begin{equation}
		Z_{\text{fine}}(x,y,t) = \sum_{j=1}^{5} 0.1 \cdot \sin(\kappa_{x,j}x + \kappa_{y,j}y + \phi_j + t)
	\end{equation}
	with \(\kappa_{x,j}=20+5j\), \(\kappa_{y,j}=15+3j\) and \(\phi_j \sim U(0,2\pi)\) providing multi‑scale textural details.
	
	\textbf{Background fluctuation layer:} a large‑scale low‑frequency background field enhances overall smoothness,
	\begin{align}
		Z_{\text{bg}}(x,y,t) = 0.15 \big[&\sin(3\pi x)\cos(2\pi y)\cos(0.3t) \nonumber \\
		&+ \sin(2\pi(x+y))\cos(0.5t)\big].
	\end{align}
	
	In a result, the post‑processing and nonlinear transformation with hyperbolic tangent compression function means that the linear superposition of original fields yields,
	\begin{equation}
		{Z_{{\rm{total}}}}(x,y,t) = \tanh (1.5 \cdot ({Z_{{\rm{vortex}}}} + {Z_{{\rm{fine}}}} + {Z_{{\rm{bg}}}})).
	\end{equation}
	This transformation restricts the field values to \([-1,1]\) while enhancing contrast in the medium‑intensity region.
	
	\begin{figure}[H]
		\centering
		\begin{subfigure}[t]{0.4\textwidth}
			\centering
			\includegraphics[scale=0.345]{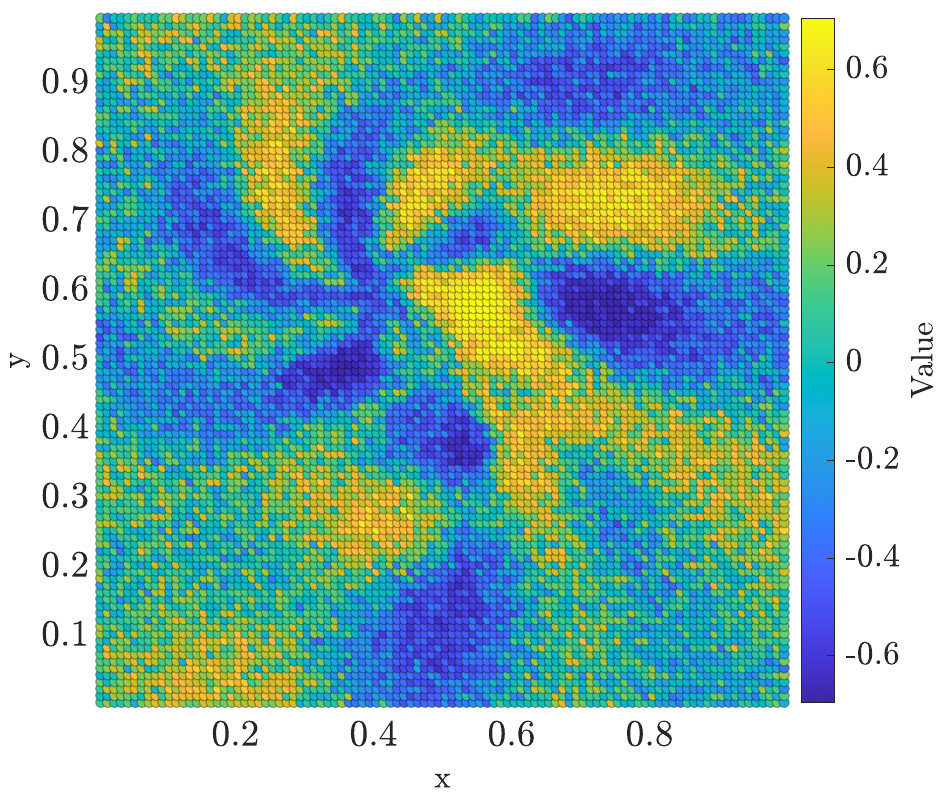}
			\caption{\centering\footnotesize Origin distribution.}
			\label{fig_Case3:Learning1_6a}
		\end{subfigure}
		\begin{subfigure}[t]{0.4\textwidth}
			\centering
			\includegraphics[scale=0.345]{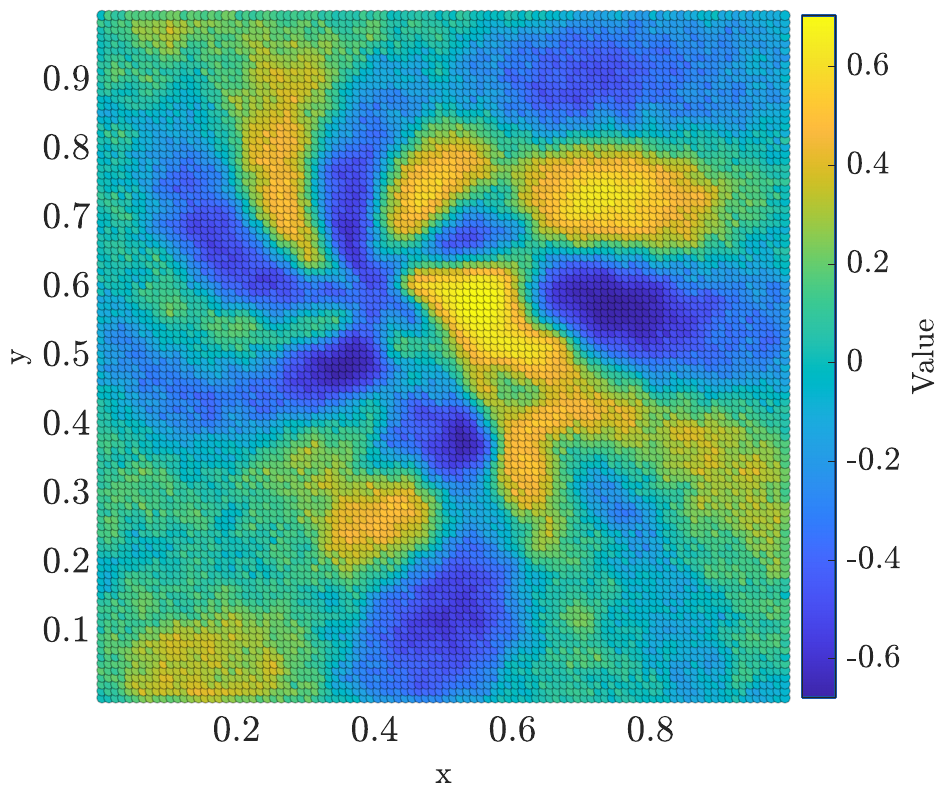}
			\caption{\centering\footnotesize Conventional SPH computing.}  
			\label{fig_Case3:Learning1_6ab}
		\end{subfigure}
		\begin{subfigure}[t]{0.4\textwidth}
			\centering
			\includegraphics[scale=0.345]{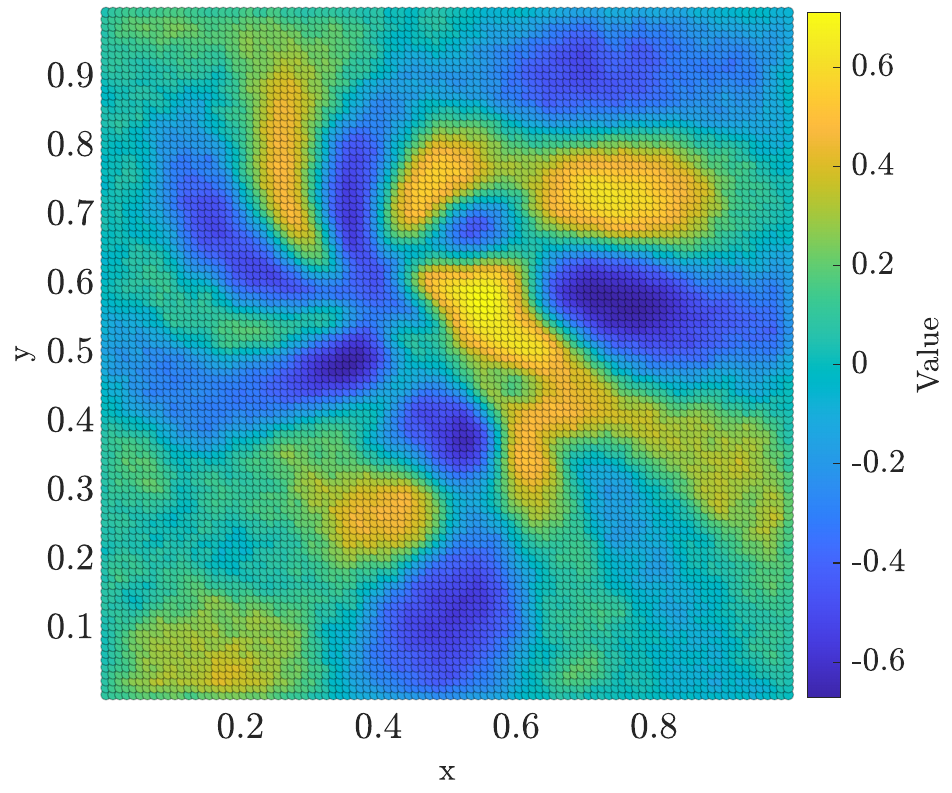}
			\caption{\centering\footnotesize Crossed-QMLP architecture with PauliZ.}
			\label{fig_Case3:Learning1_6ac}
		\end{subfigure}
		\begin{subfigure}[t]{0.4\textwidth}
			\centering
			\includegraphics[scale=0.345]{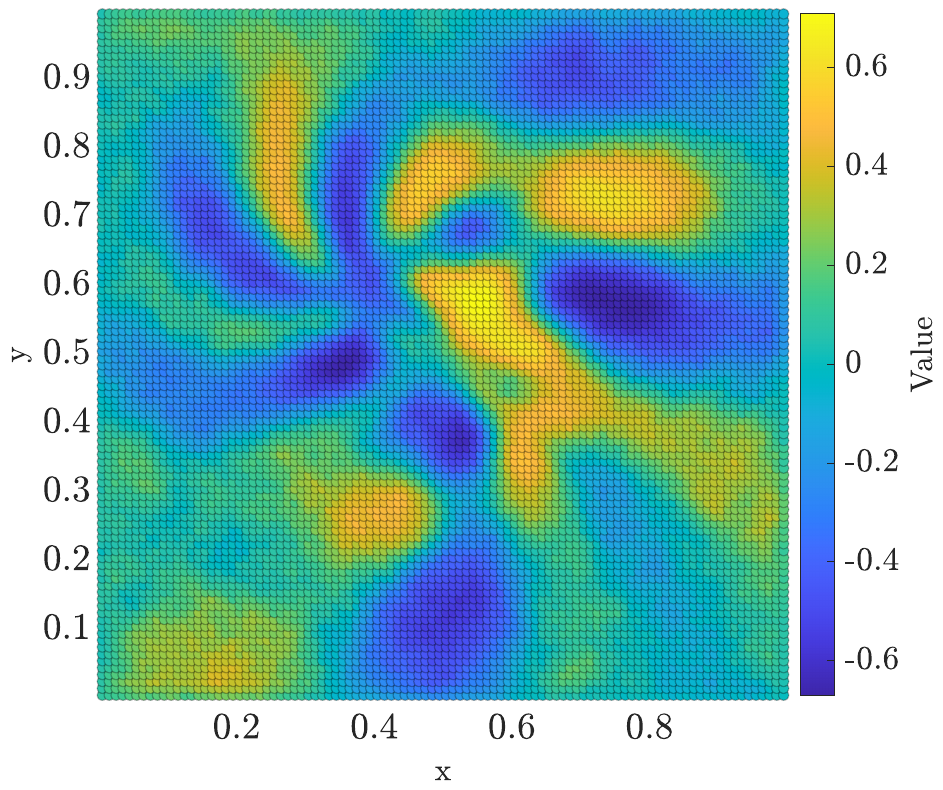}
			\caption{\centering\footnotesize Single quantum circuit with PauliZ.}  
			\label{fig_Case3:Learning1_6ad}
		\end{subfigure}
		\caption{\small Numerical performance on different computing models.}
		\label{fig_Case3:Learning1}
	\end{figure}
	
	As shown in Fig.(\ref{fig_Case3:Learning1_6a}), the regular point sampling is performed for present computational validation. From a fluid dynamics perspective, this model can be regarded as a simplified representation of velocity potential induced by multiple point vortices. Each vortex component resembles a Gaussian vortex core with vorticity distribution \(\omega_i \propto \exp(-r_i^2/\sigma_i^2)\). The linear superposition property corresponds to the linear superposition principle of vorticity fields, while the azimuthal modulation term models possible azimuthal inhomogeneities in realistic vortices.
	
	First, based on the complicated vorticity distribution, we adopted the static quantum kernel on SPH to demonstrate the validity of current study. Fig.(\ref{fig_Case3:Learning1_6ab}) gives the conventional computing results using SPH and Figs.(\ref{fig_Case3:Learning1_6ac}, \ref{fig_Case3:Learning1_6ad}) depict the similar distributions by the crossed-QMLP architecture and single quantum circuit, respectively. It can be observed that the current Lagrangian quantum networks with SPH can be effectively approximated similarly as the conventional SPH method for the von Neumann computational paradigm. 
	
	\begin{figure}[H]
		\centering
		\begin{subfigure}[t]{0.4\textwidth}
			\centering
			\includegraphics[scale=0.31]{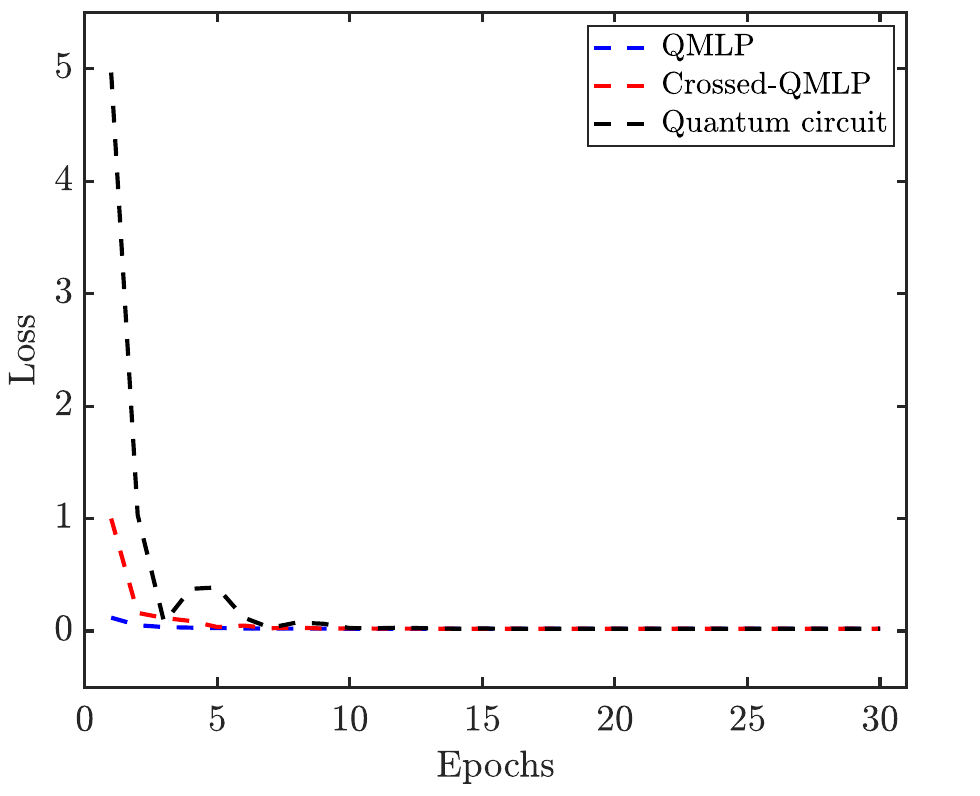}
			\caption{\centering\footnotesize Training procedure on different hierarchies of QMLP.}
			\label{fig:6a}
		\end{subfigure}
		\begin{subfigure}[t]{0.4\textwidth}
			\centering
			\includegraphics[scale=0.31]{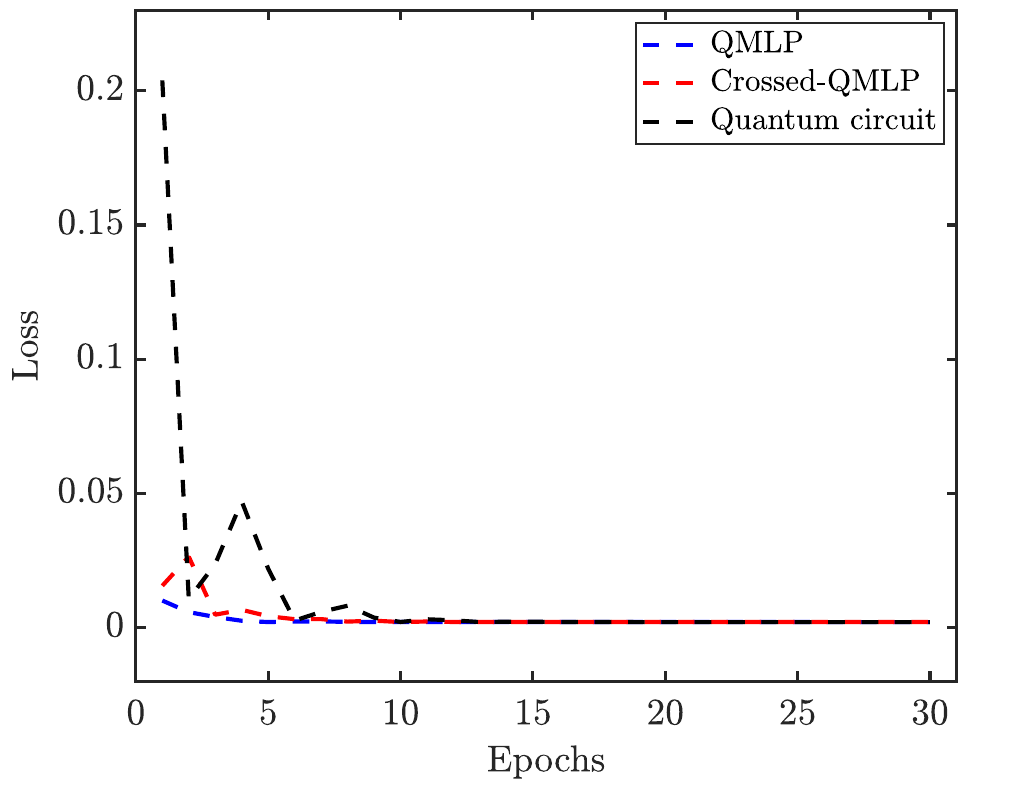}
			\caption{\centering\footnotesize Predicting procedure on different hierarchies of QMLP.}  
			\label{fig:6b}
		\end{subfigure}
		\begin{subfigure}[t]{0.4\textwidth}
			\centering
			\includegraphics[scale=0.345]{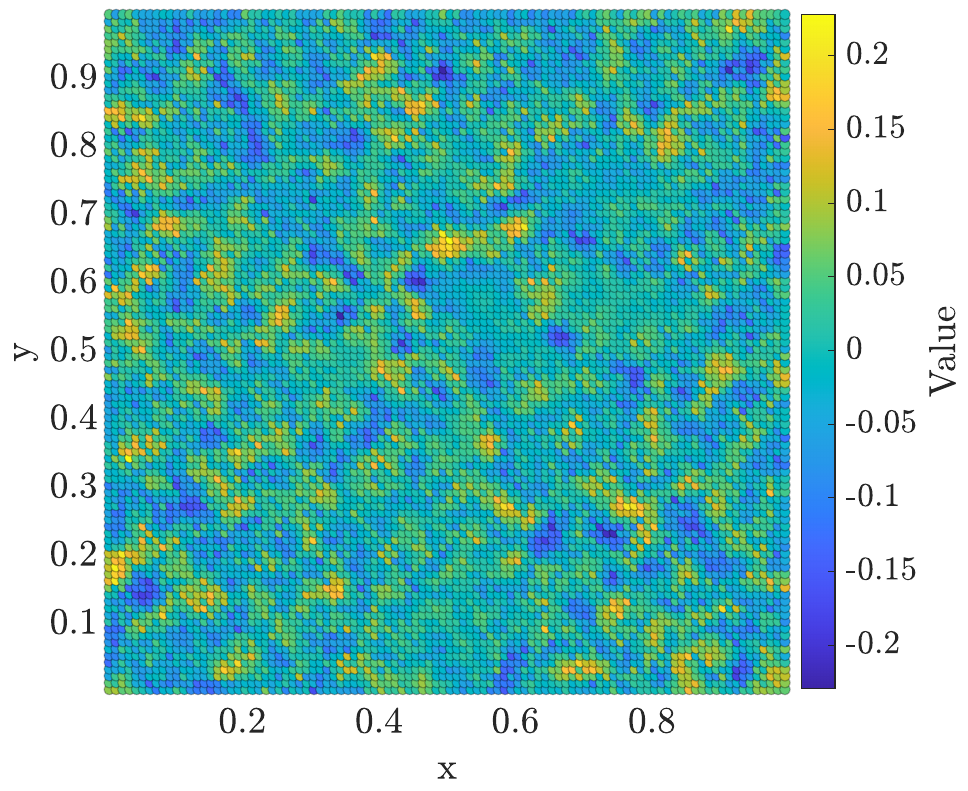}
			\caption{\centering\footnotesize Relative errors on smoothing characteristic between single quantum circuit and original SPH computing.}
			\label{fig:6a}
		\end{subfigure}
		\begin{subfigure}[t]{0.4\textwidth}
			\centering
			\includegraphics[scale=0.345]{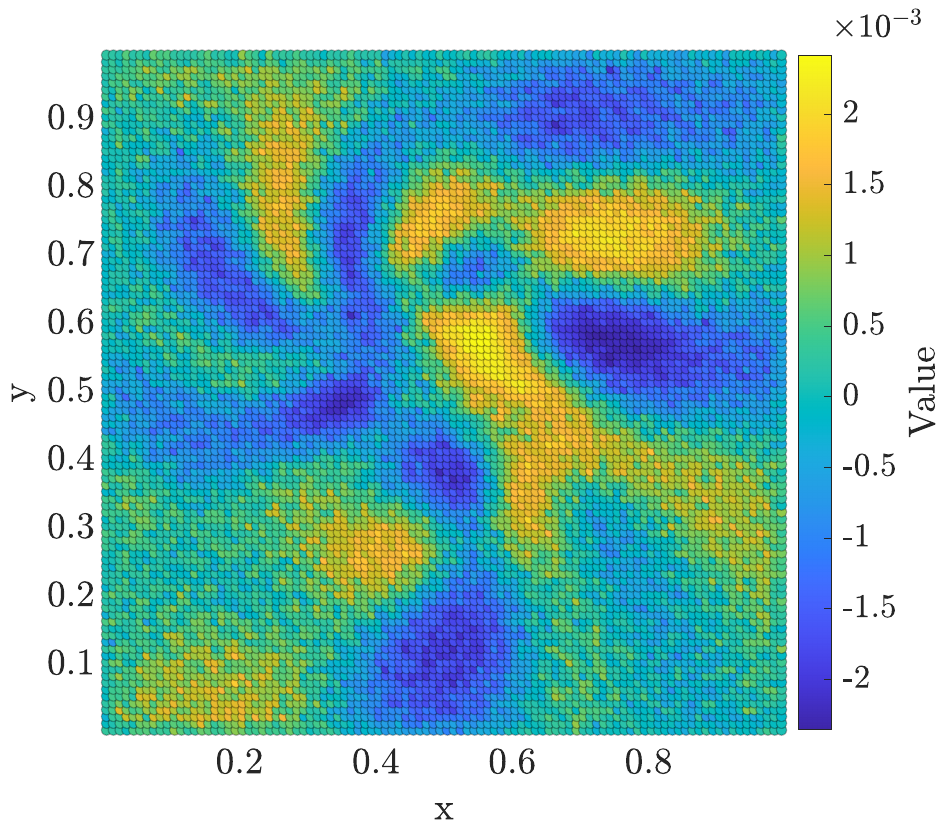}
			\caption{\centering\footnotesize Relative errors on smoothing characteristic between single quantum circuit and crossed QMLP-PauliZ architecture.}  
			\label{fig:6b}
		\end{subfigure}
		\caption{\small Performance comparisons on different hierarchies of quantum networks.}
		\label{fig_Case3:Learning2}
	\end{figure}
	\begin{figure}[H]
		\centering
		\begin{subfigure}[t]{0.4\textwidth}
			\centering
			\includegraphics[scale=0.345]{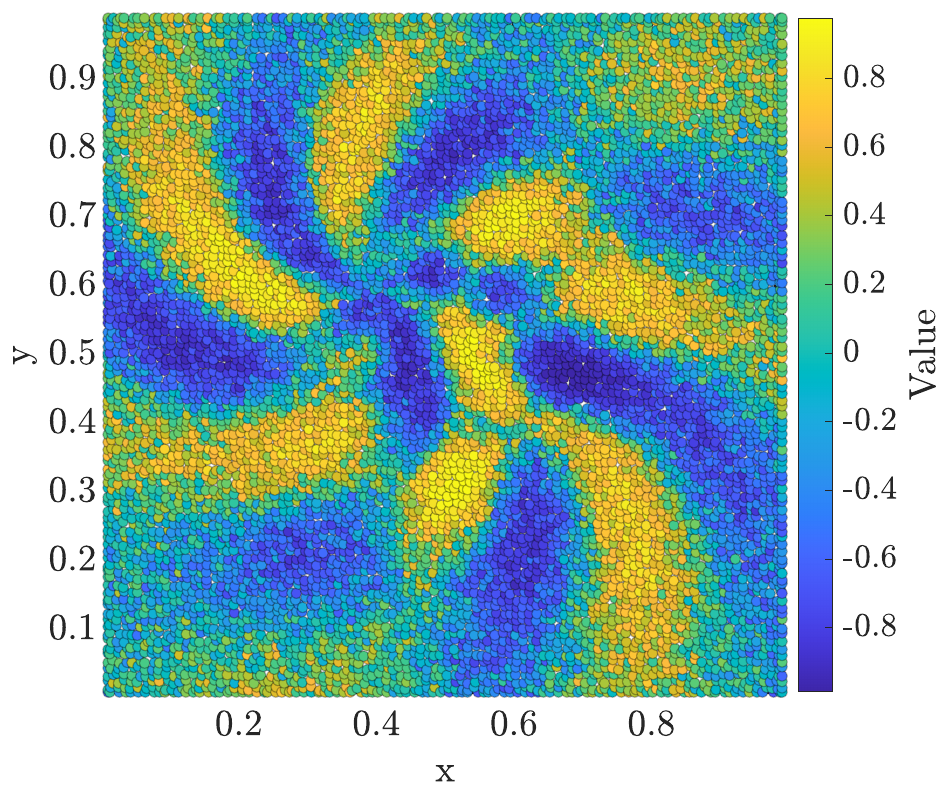}
			\caption{\centering\footnotesize Origin distribution.}
			\label{fig:6a}
		\end{subfigure}
		\begin{subfigure}[t]{0.4\textwidth}
			\centering
			\includegraphics[scale=0.345]{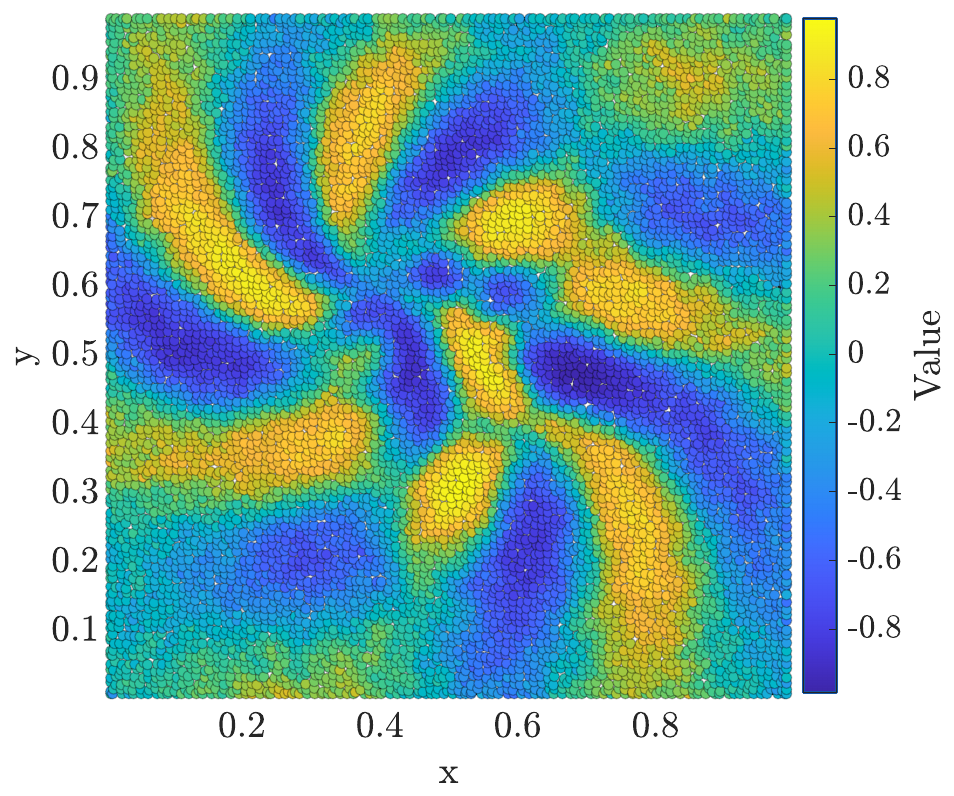}
			\caption{\centering\footnotesize SPH computing with corrected kernel.}  
			\label{fig:6b}
		\end{subfigure}
		\begin{subfigure}[t]{0.4\textwidth}
			\centering
			\includegraphics[scale=0.345]{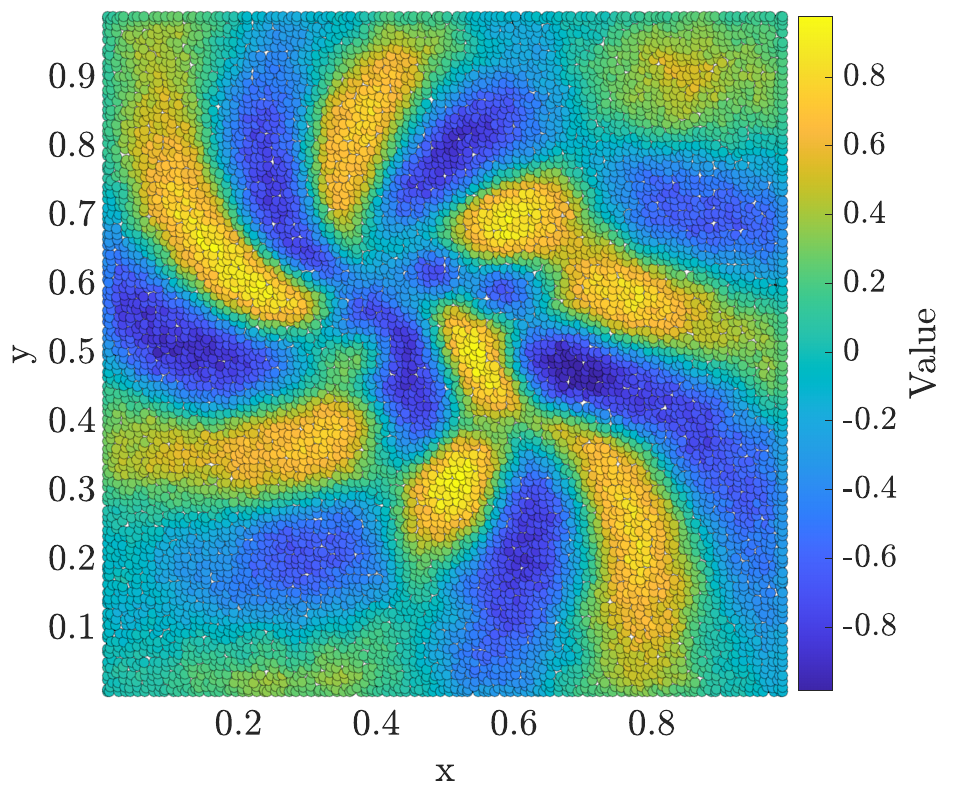}
			\caption{\centering\footnotesize QMLP-PauliZ architecture.}
			\label{fig:6a}
		\end{subfigure}
		\begin{subfigure}[t]{0.4\textwidth}
			\centering
			\includegraphics[scale=0.345]{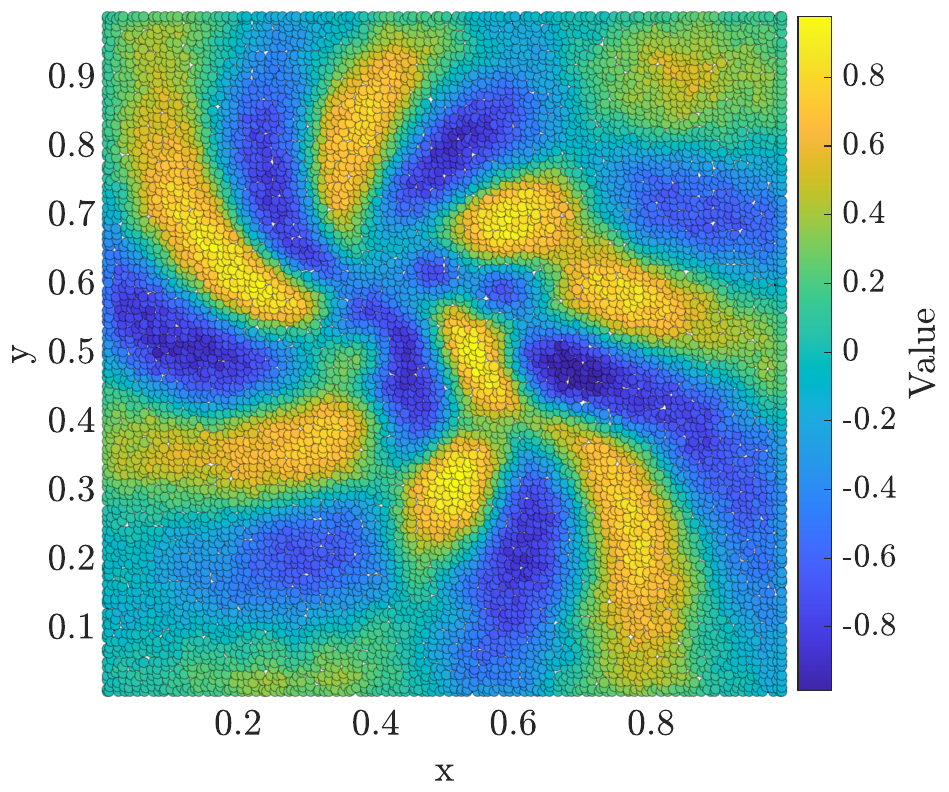}
			\caption{\centering\footnotesize Single quantum circuit with PauliZ.}  
			\label{fig:6b}
		\end{subfigure}
		\caption{\small Numerical performance on different computing models.}
		\label{fig_Case3:Learning3}
	\end{figure}

Accordingly, Fig.(\ref{fig_Case3:Learning2}) described the performance comparisons on different hierarchies of quantum networks, which can be observed that the sequential hybrid architecture (crossed-QMLP) exhibits a profound advantage over the elementary single quantum circuit in both the training and prediction phases. As illustrated by the loss trajectories over consecutive epochs, the crossed-QMLP model achieves rapid convergence, with its loss plummeting to near-zero values. Furthermore, the spatial relative error distributions on smoothing characteristic corroborate these dynamic findings. The error landscape between single quantum circuit and classical SPH baseline displays pronounced deviations, fluctuating broadly between -0.2 and 0.2. This is because that the smoothing characteristic from SPH resembles a moderately fitted state in Fig.(\ref{fig_Case3:Learning1_6a}) and it is unavoidable that some local points exhibit relatively large gradient variations for quantum kernel networks. Such deviations are confined to a small subset of grid points, while the relative error statistics over the majority of physical domain remain around 0.05. Meanwhile, the relative errors between single quantum circuit and crossed-QMLP architecture are shown to be tightly constrained within an order of magnitude of $10^{-3}$. These comparisons compellingly demonstrate that integrating classical neural layers as forward compression and hybrid processing modules is indispensable to unlock a robust, high-fidelity mapping of unstructured fluid topological features within quantum unitary space.
	
	\begin{figure}[H]
		\centering
		\begin{subfigure}[t]{0.4\textwidth}
			\centering
			\includegraphics[scale=0.33]{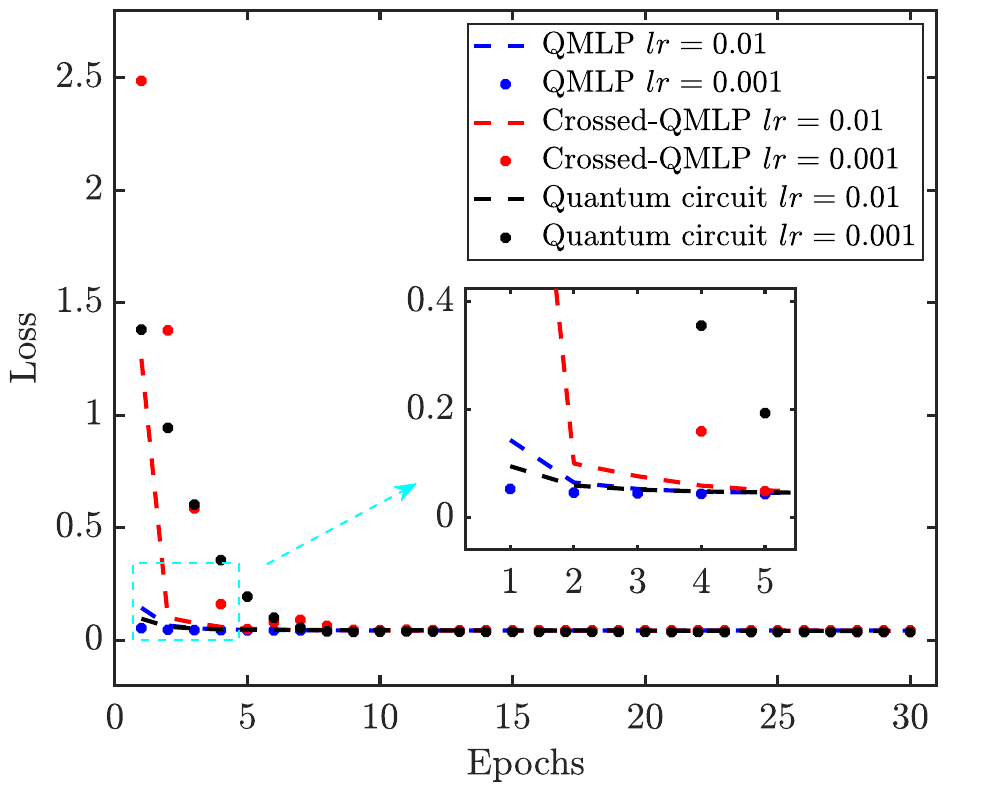}
			\caption{\centering\footnotesize Training procedure on different hierarchies of QMLP.}
			\label{fig:6a}
		\end{subfigure}
		\begin{subfigure}[t]{0.4\textwidth}
			\centering
			\includegraphics[scale=0.33]{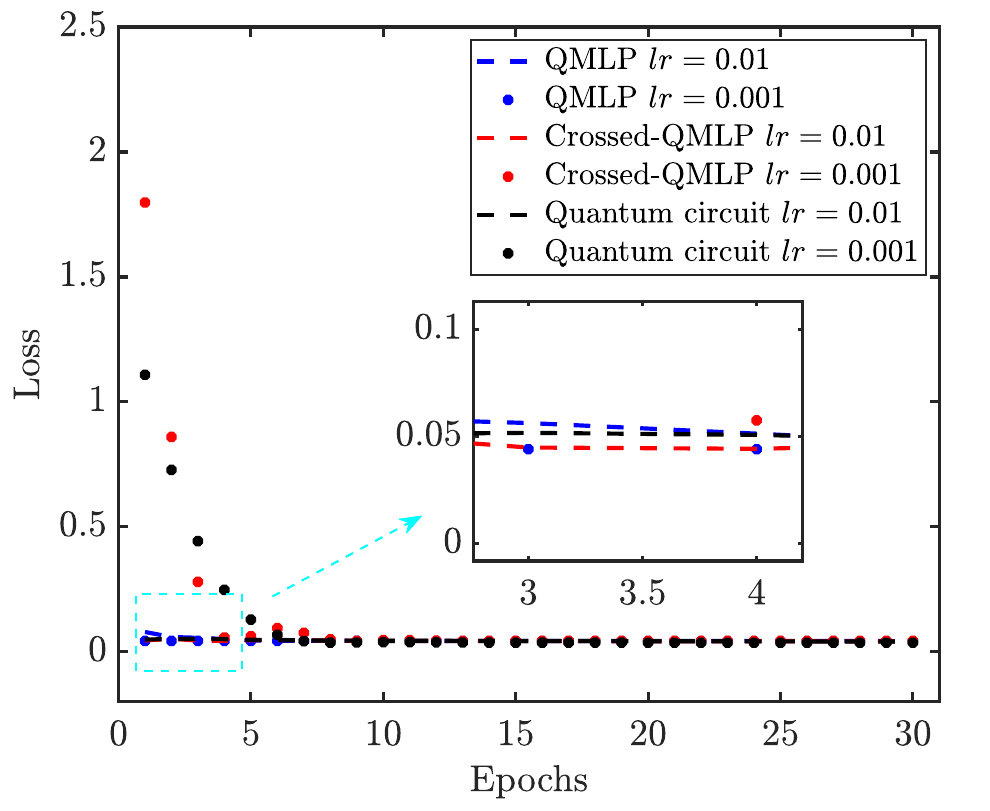}
			\caption{\centering\footnotesize Predicting procedure on different hierarchies of QMLP.}  
			\label{fig:6b}
		\end{subfigure}
		\caption{\small Learning epochs on current quantum networks of SPH.}
		\label{fig_Case3:Learning4}
	\end{figure}
	\begin{figure}[H]
		\centering
		\includegraphics[scale=0.35]{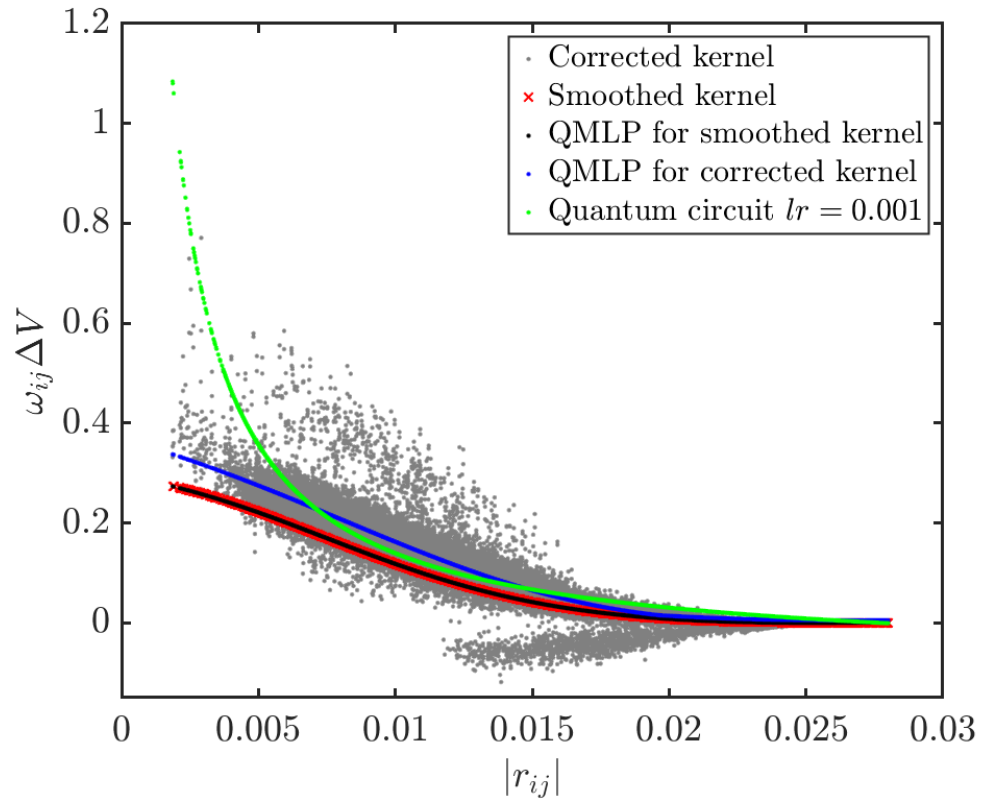}
		\caption{\small Quantized kernel space between standard smoothing kernel and irregular particle distributions at $lr=0.001$.}
		\label{fig_Case3:Learning5}
	\end{figure}

\begin{figure}[H]
	\centering
	\begin{subfigure}[t]{0.4\textwidth}
		\centering
		\includegraphics[scale=0.38]{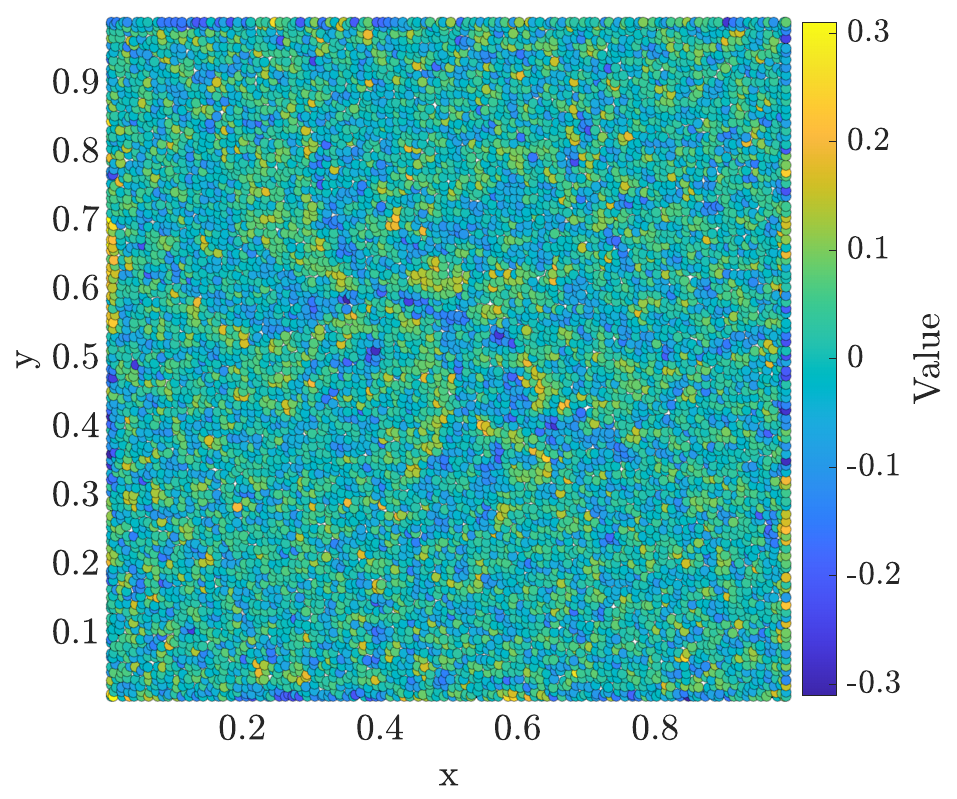}
		\caption{\centering\footnotesize Relative errors on smoothing characteristic between QMLP-PauliZ and corrected SPH computing.}
		\label{fig:6a}
	\end{subfigure}
	\begin{subfigure}[t]{0.4\textwidth}
		\centering
		\includegraphics[scale=0.38]{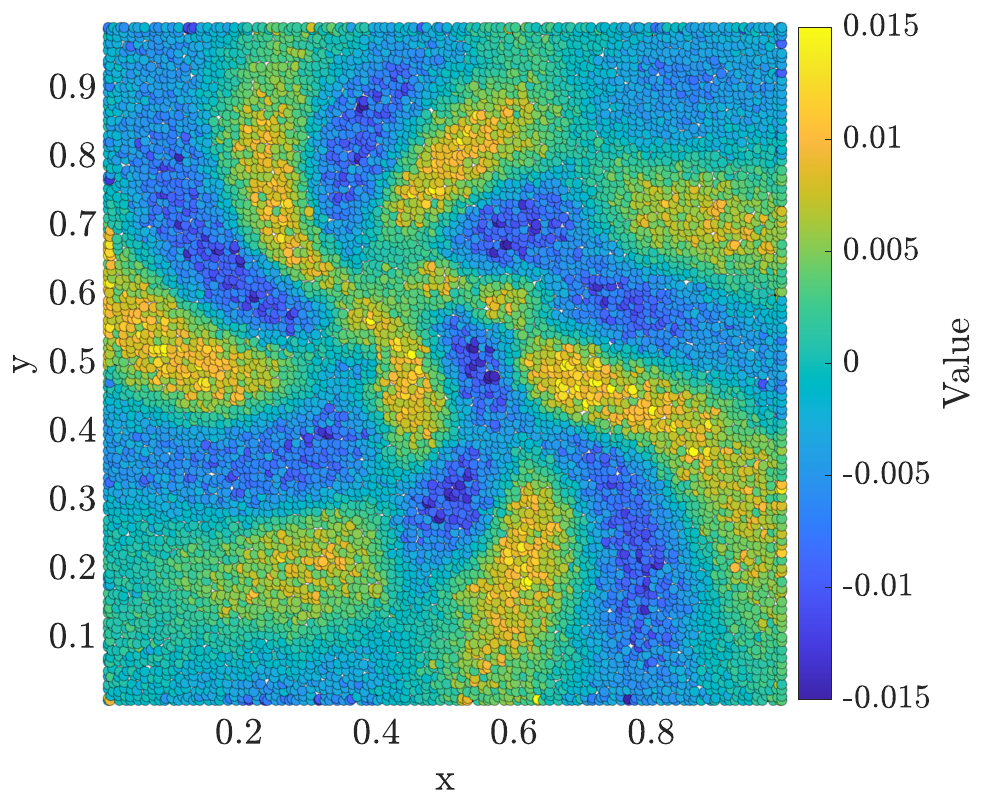}
		\caption{\centering\footnotesize Relative errors on smoothing characteristic between single quantum circuit and crossed-QMLP architecture with PauliZ.}  
		\label{fig:6b}
	\end{subfigure}
	\caption{\small Performance comparisons on different hierarchies of quantum networks.}
	\label{fig_Case3:Learning6}
\end{figure}

Second, it addresses the further case of irregular particle distributions combing with the corrected kernel SPH method. This case adopted the aforementioned vorticity distribution of Eq.(\ref{StaticQuantumKernel_1}) at $t=500$. Fig.(\ref{fig_Case3:Learning3}) described the numerical performance on different computing models. Fig.(\ref{fig_Case3:Learning4}) captured the learning epochs on current quantum networks of SPH, which can be observed that these approaches exhibit fast convergence rates. Notably, the quantum-classical hybrid circuit (QMLP and crossed-QMLP) integrated with a neural network achieves even faster convergence, which is consistent with the findings from the first case study presented earlier. Subsequently, statistical analysis of Fig.(\ref{fig_Case3:Learning5}) was conducted on the kernel learning results for both the regular particle distribution using the standard smoothing kernel and the irregular particle distribution employing the corrected kernel, with a learning rate of 0.001. Fig.(\ref{fig_Case3:Learning6}) similarly computed the quantified relative errors on different hierarchies of quantum networks. Although minor variations exist among the quantum circuits employed on different kernels, the results remain acceptable, which further supports the observation that elementary quantum gates lack strong generalization capability. Moreover, the performance of quantum AI is inherently linked to its training data and circuit architecture, leading to certain correlations in the observed behaviors.

In summary, through a series of static quantum kernel tests on SPH, elementary quantum gates inherently lack strong generalization capability. To overcome this limitation, a forward network (i.e., hidden layers) is introduced to provide the necessary generalization ability, thereby establishing a pathway that adapts the parameter‑specific generalization of quantum gates and enhances quantum learning performance in data fitting (generalization) tasks. Consequently, while single‑qubit circuits may suffice for simple fitting problems, their generalization capacity becomes insufficient for moderately complex fitting tasks, necessitating the incorporation of a forward neural or hybrid crossed‑neural architecture. Certainly, this insight is derived solely from the present study, and further improvements require additional extensions and in-depth considerations.
	
	\begin{figure}[H]
		\centering
		\begin{subfigure}[t]{0.4\textwidth}
			\centering
			\includegraphics[scale=0.345]{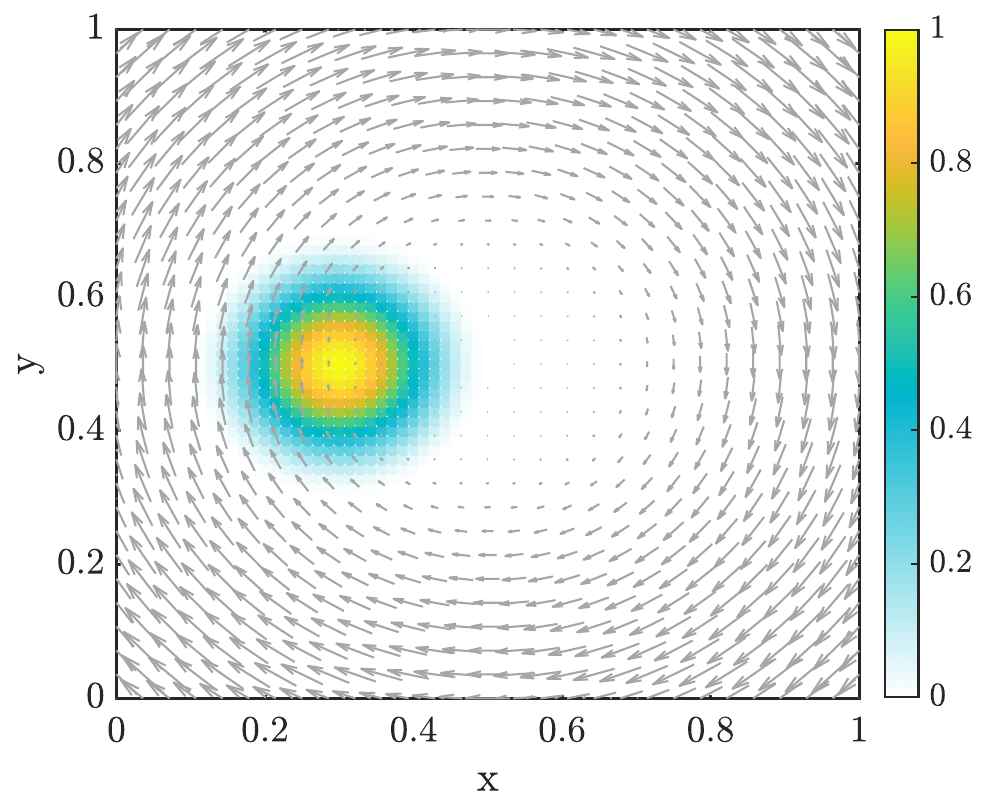}
			\caption{\centering\footnotesize $t$=0 s.}
			\label{fig_Case4:Learning1a}
		\end{subfigure}
		\begin{subfigure}[t]{0.4\textwidth}
			\centering
			\includegraphics[scale=0.345]{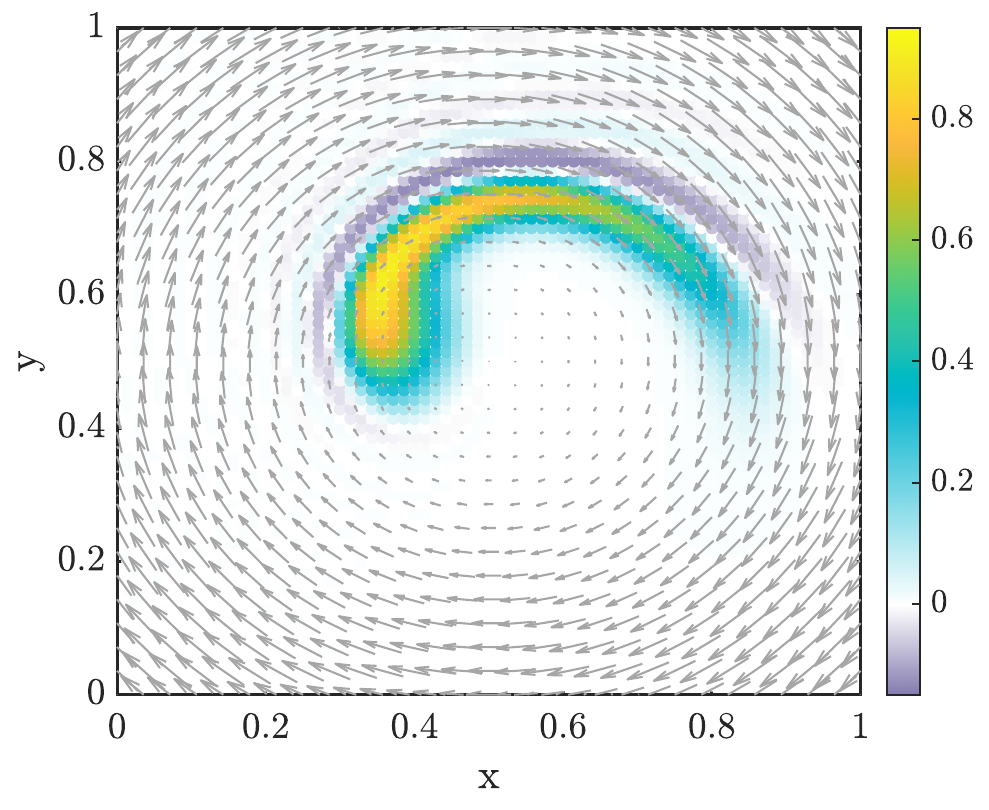}
			\caption{\centering\footnotesize $t$=0.15 s.}  
			\label{fig:6b}
		\end{subfigure}
		\begin{subfigure}[t]{0.4\textwidth}
			\centering
			\includegraphics[scale=0.345]{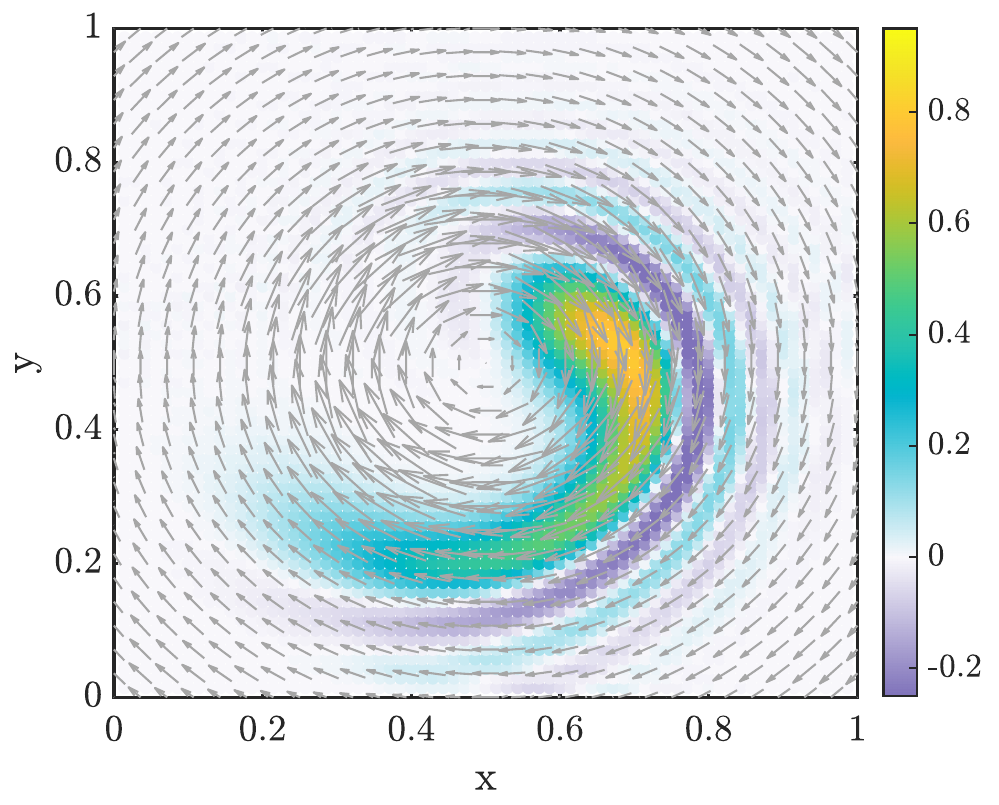}
			\caption{\centering\footnotesize $t$=0.35 s.}
			\label{fig:6a}
		\end{subfigure}
		\begin{subfigure}[t]{0.4\textwidth}
			\centering
			\includegraphics[scale=0.345]{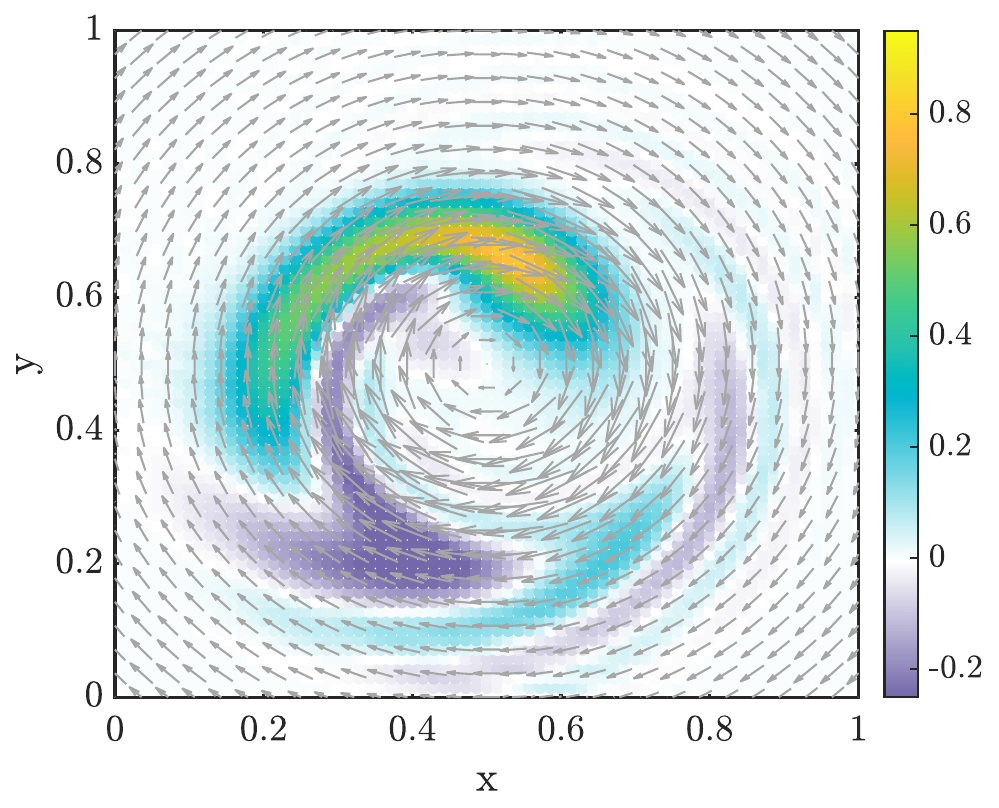}
			\caption{\centering\footnotesize $t$=0.60 s.}  
			\label{fig:6b}
		\end{subfigure}
		\caption{\small Original results with spacing $\Delta d$=0.02 m.}
		\label{fig_Case4:Learning1}
	\end{figure}
	
	\subsection*{Transient Quantum Kernel on Smoothed Particle Hydrodynamics}
	\normalsize \hspace{10pt}
	In this subsection, a circular scalar field with advective transport is usually stretched and deformed into a crescent and then it returns to the original position and height. The purpose is to promote the further tests for present transient quantum kernel on SPH with evolutional Euler field distributions\supercite{flyer2016enhancing[35]}, whose mode of particle morphology is quasi-uniform. The corresponding governing equation is defined on computational domain [0 m, 1 m]×[0 m, 1 m] and given by
	\begin{equation}\label{eq:39}
		\frac{\partial \psi }{\partial t}=-\frac{\partial (u\psi )}{\partial x}-\frac{\partial (v\psi )}{\partial y}
	\end{equation}
	\begin{equation}\label{eq:40}
		\left\{ \begin{aligned}
			& u(x,y,t)={{u}_{\theta }}(r,t)\sin \theta  \\ 
			& v(x,y,t)=-{{u}_{\theta }}(r,t)\cos \theta  \\ 
		\end{aligned} \right.
	\end{equation}
	\begin{equation}\label{eq:41}
		{{u}_{\theta }}(r,t)=\frac{4\pi r}{T}\left[ 1-\cos \left( \frac{2\pi t}{T} \right)\frac{1-{{\left( 4r \right)}^{6}}}{1+{{\left( 4r \right)}^{6}}} \right]
	\end{equation}
	\begin{equation}\label{eq:42}
		r=\sqrt{{{\left( x-0.5 \right)}^{2}}+{{\left( y-0.5 \right)}^{2}}},\text{      }\theta ={{\tan }^{-1}}\left( \frac{y-0.5}{x-0.5} \right)
	\end{equation}
	\begin{equation}\label{eq:43}
		{{\left. \psi  \right|}_{t=0}}=\left\{ \begin{matrix}
			0.5+0.5\cos \left( \pi \hat{r} \right),\text{  }\hat{r}\le 1  \\
			0,\text{                           }\hat{r}>1  \\
		\end{matrix} \right.
	\end{equation}
	\begin{equation}\label{eq:44}
		\hat{r}=5\sqrt{{{\left( x-0.3 \right)}^{2}}+{{\left( y-0.5 \right)}^{2}}}
	\end{equation}
where $T$ denotes the kinetic period of divergence-free velocity field $(u,v)$. The rotary velocity field with transport scalar $\psi$ is centered at node (0.5 m, 0.5 m). We promote the initial condition of transport scalar $\psi$ in Eq.(\ref{eq:43}). As shown in Fig.(\ref{fig_Case4:Learning1a}), the field on transport scalar $\psi $ is centered at node (0.3 m, 0.5 m) and enforced to generate an advective transport by the rotary divergence-free velocity field. In this procedure, the period $T$ is 1.00 s and after one period it should have ideally returned to its original height and position. It is worth noting that the nodes near physical boundary are extrapolated to 2-3 layers in order to completely eliminate the effect on boundary truncation.

\begin{figure}[H]
	\centering
	\begin{subfigure}[t]{0.4\textwidth}
		\centering
		\includegraphics[scale=0.33]{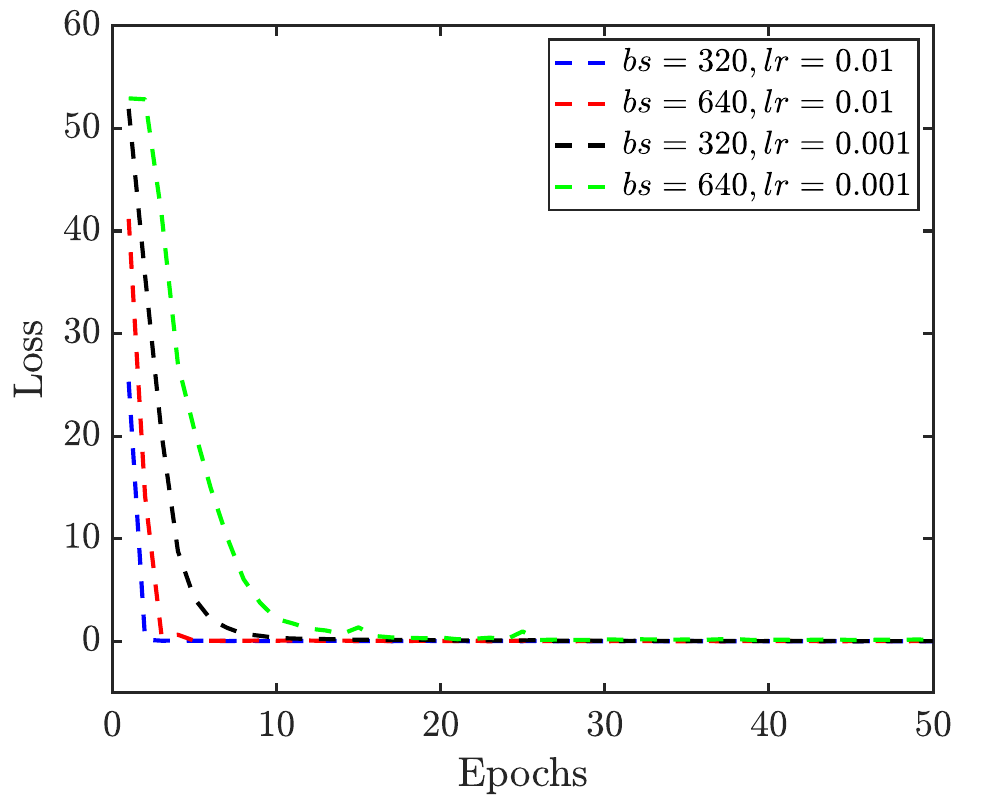}
		\caption{\centering\footnotesize Training procedure.}
		\label{fig:6a}
	\end{subfigure}
	\begin{subfigure}[t]{0.4\textwidth}
		\centering
		\includegraphics[scale=0.33]{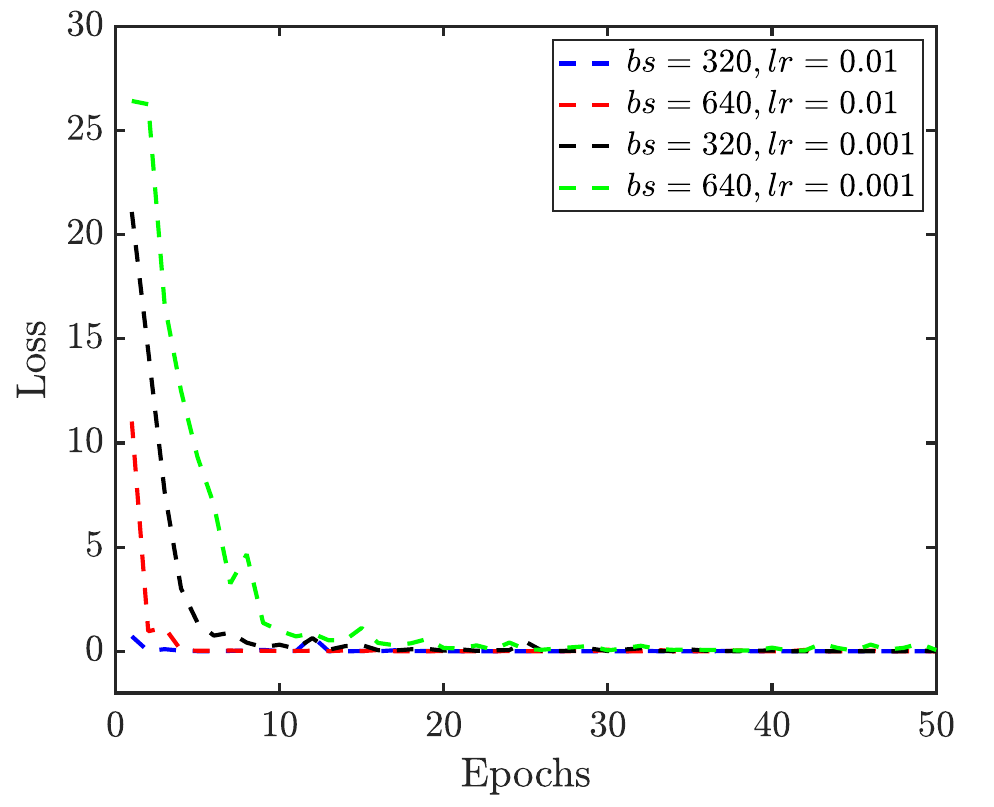}
		\caption{\centering\footnotesize Testing procedure.}  
		\label{fig:6b}
	\end{subfigure}
	\caption{\small Learning epochs on hybrid crossed-QMLP architecture.}
	\label{fig_Case4:Learning2}
\end{figure}

In this procedure on advective transport of a scalar $\psi$, an explicit time step is fixed to be $\Delta t=1\times {{10}^{-4}}\text{ s}$ and Fig.(\ref{fig_Case4:Learning1}) describes the flow regimes in a period. In a fact, with the enforcement of rotary velocity field, we discovered that after a period the circular scalar field cannot be returned to its original position completely for aforementioned SPH theme. The main distinguish and reason are the selection of explicit time evolutional scheme differently from the implicit iterative solutions. The purpose on this paper is to investigate the numerical performance on corresponding quantum schemes so it is not studied about the time integration scheme detailedly\supercite{flyer2016enhancing[35],WOS:001501337900001}. 

\begin{figure}[H]
	\centering
	\begin{subfigure}[t]{0.4\textwidth}
		\centering
		\includegraphics[scale=0.33]{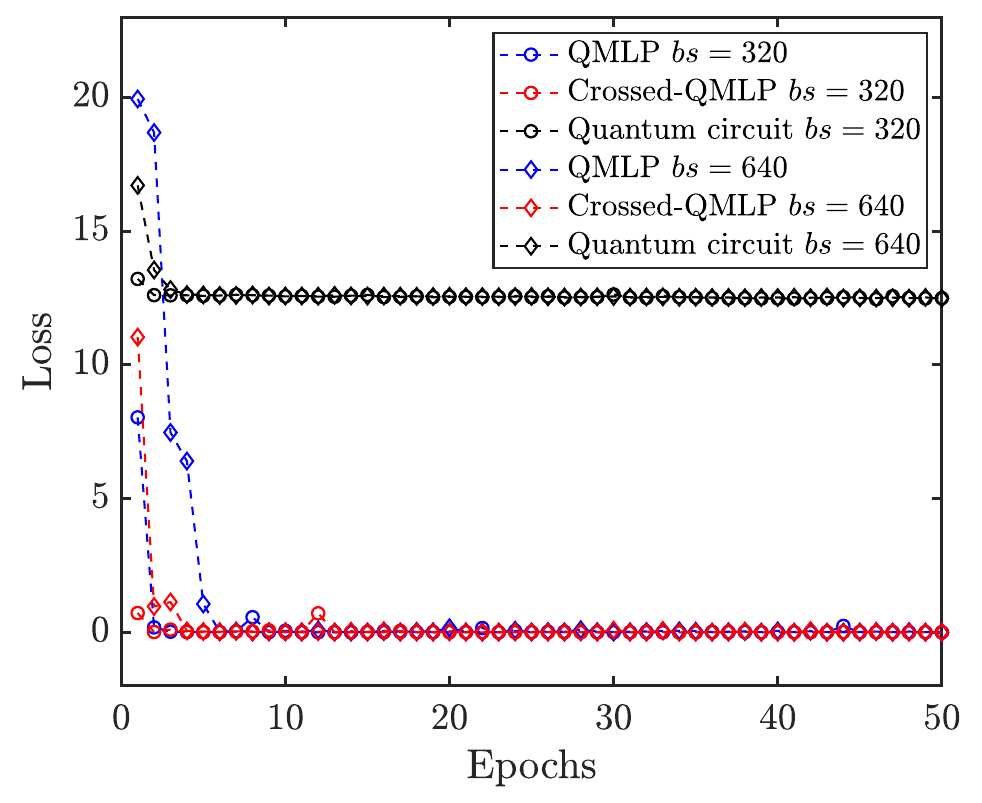}
		\caption{\centering\footnotesize With certain noise at learning rate $lr = 0.01$.}
		\label{fig:6a}
	\end{subfigure}
	\begin{subfigure}[t]{0.4\textwidth}
		\centering
		\includegraphics[scale=0.33]{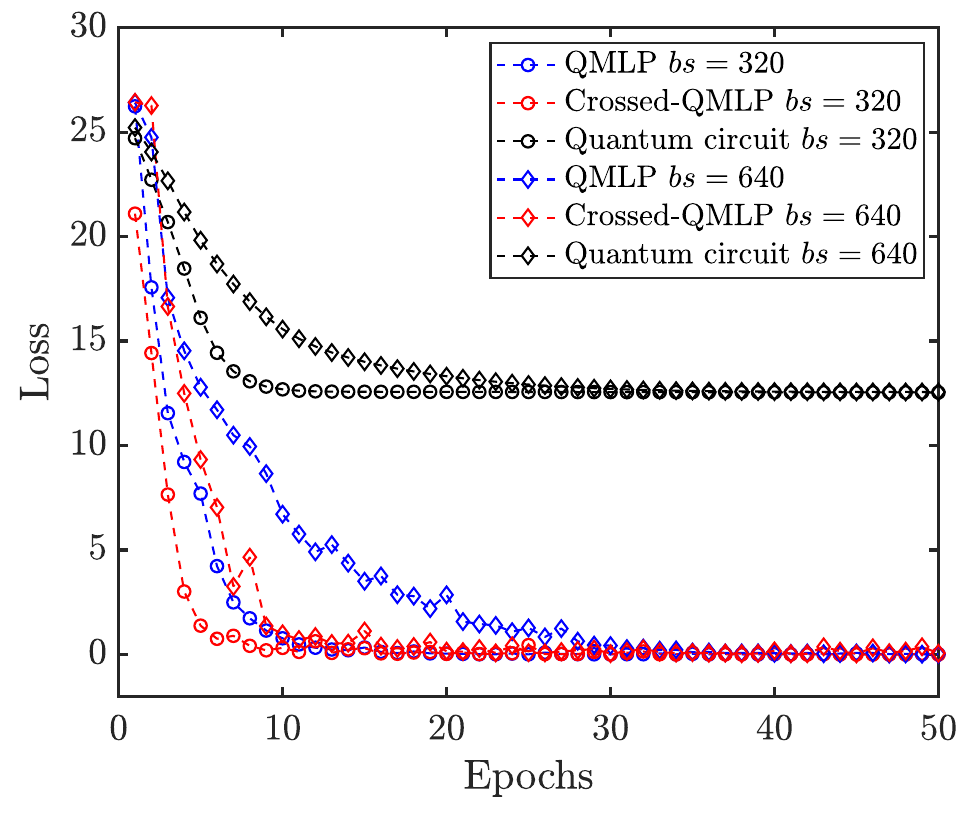}
		\caption{\centering\footnotesize With certain noise at learning rate $lr = 0.001$.}  
		\label{fig:6b}
	\end{subfigure}
	\caption{\small Learning epochs on different hierarchies of quantum networks.}
	\label{fig_Case4:Learning22}
\end{figure}

The convergence behaviors of tested quantum architectures during this transient process are systematically presented in Fig.(\ref{fig_Case4:Learning2}) and Fig.(\ref{fig_Case4:Learning22}). Different batch size $bs$ and learning rate $lr$ of quantum machine learning process are captured by using present hybrid crossed-QMLP architecture. During both the training and testing phases across varying batch sizes, the elementary single quantum circuit struggles to minimize the loss, quickly plateauing at an unacceptably high error margin. Conversely, the quantum-classical hybrid circuits (QMLP and particularly the crossed-QMLP) exhibit remarkably fast and stable convergence rates.

\begin{figure}[H]
	\centering
	\begin{subfigure}[t]{0.4\textwidth}
		\centering
		\includegraphics[scale=0.33]{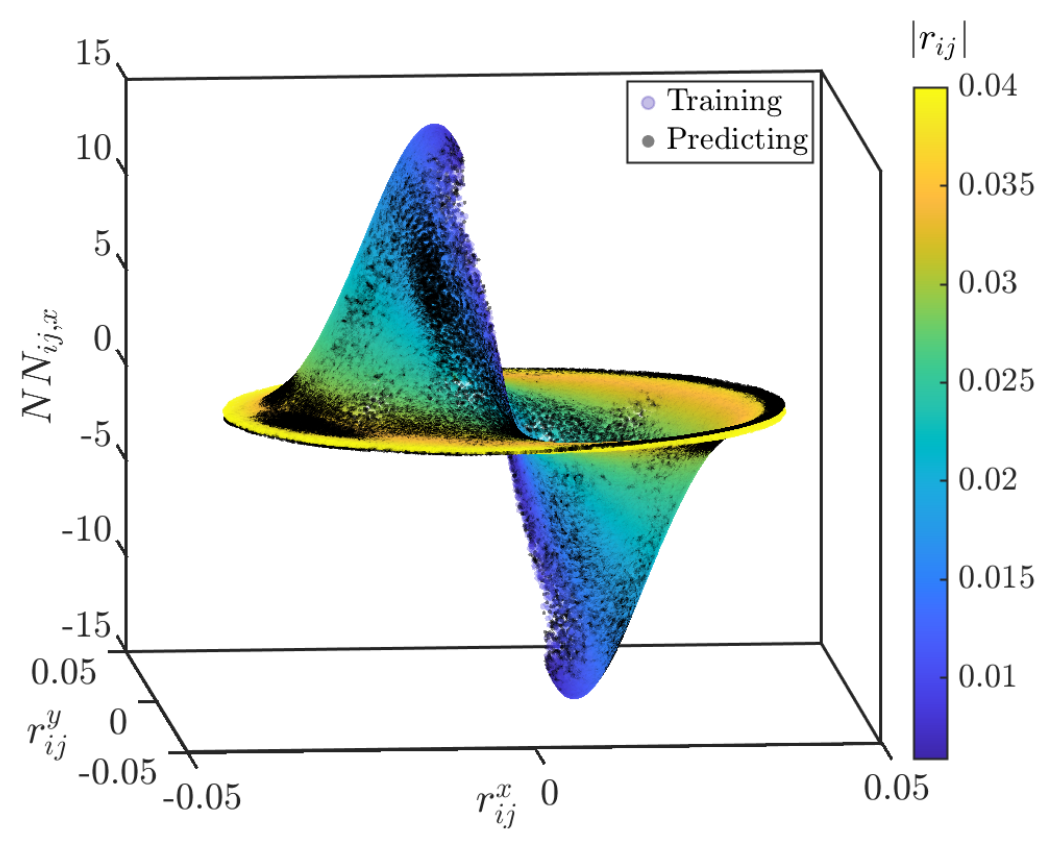}
		\caption{\centering\footnotesize Gradient along X-direction for hybrid crossed-QMLP architecture.}
		\label{fig:6a}
	\end{subfigure}
	\begin{subfigure}[t]{0.4\textwidth}
		\centering
		\includegraphics[scale=0.33]{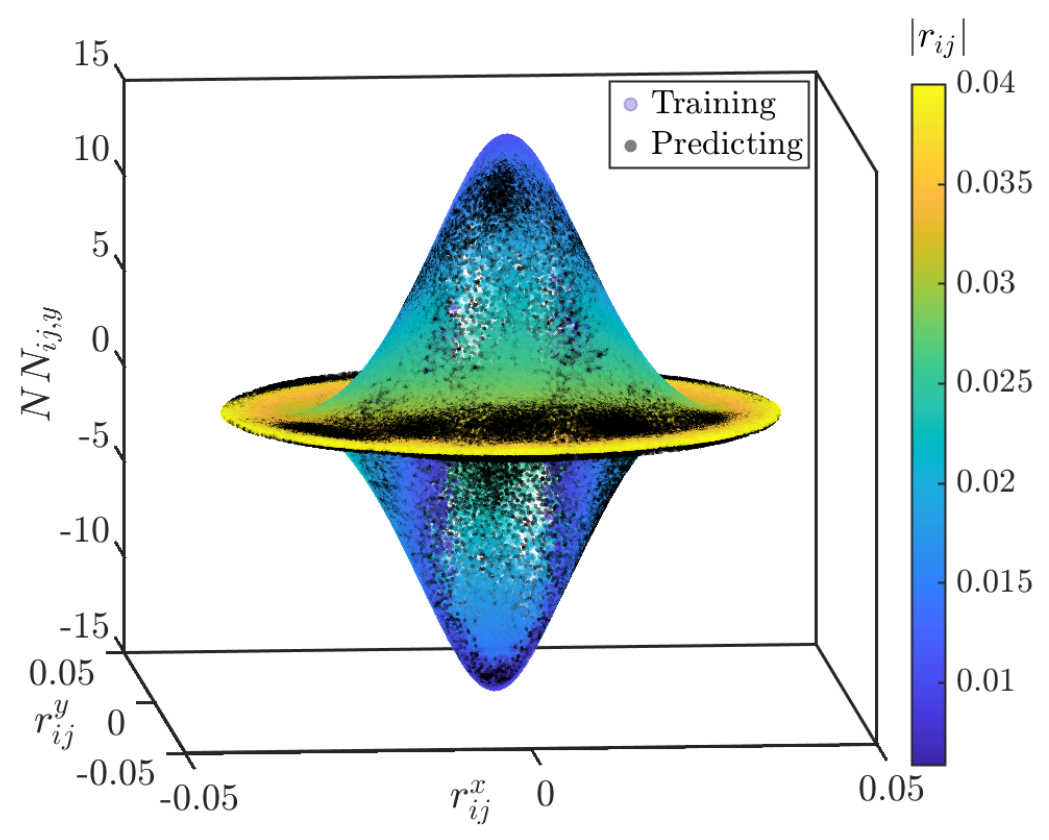}
		\caption{\centering\footnotesize Gradient along Y-direction for hybrid crossed-QMLP architecture.}  
		\label{fig:6b}
	\end{subfigure}
	\caption{\small Kernel distributions on quantum kernel networks for SPH ($bs$=640, $lr$=0.01).}
	\label{fig_Case4:smoothingKernel2}
\end{figure}

\begin{figure}[H]
		\centering
		\begin{subfigure}[t]{0.4\textwidth}
			\centering
			\includegraphics[scale=0.33]{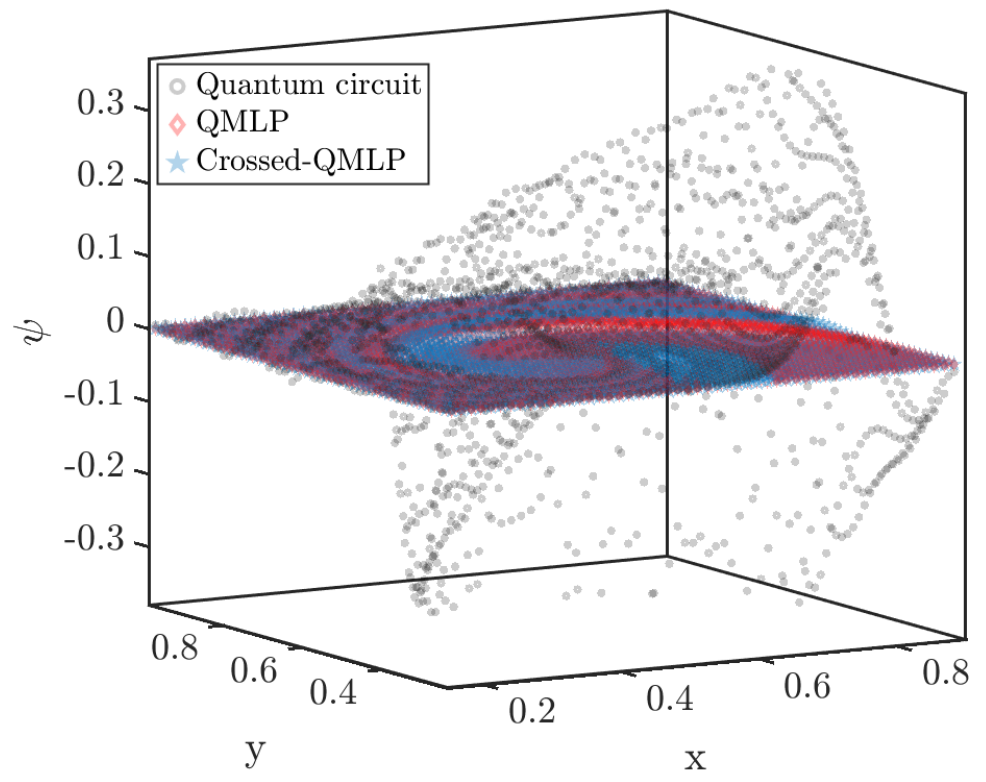}
			\caption{\centering\footnotesize Three quantum models at $t$=0.15 s.}
			\label{fig:6a}
		\end{subfigure}
		\begin{subfigure}[t]{0.4\textwidth}
			\centering
			\includegraphics[scale=0.33]{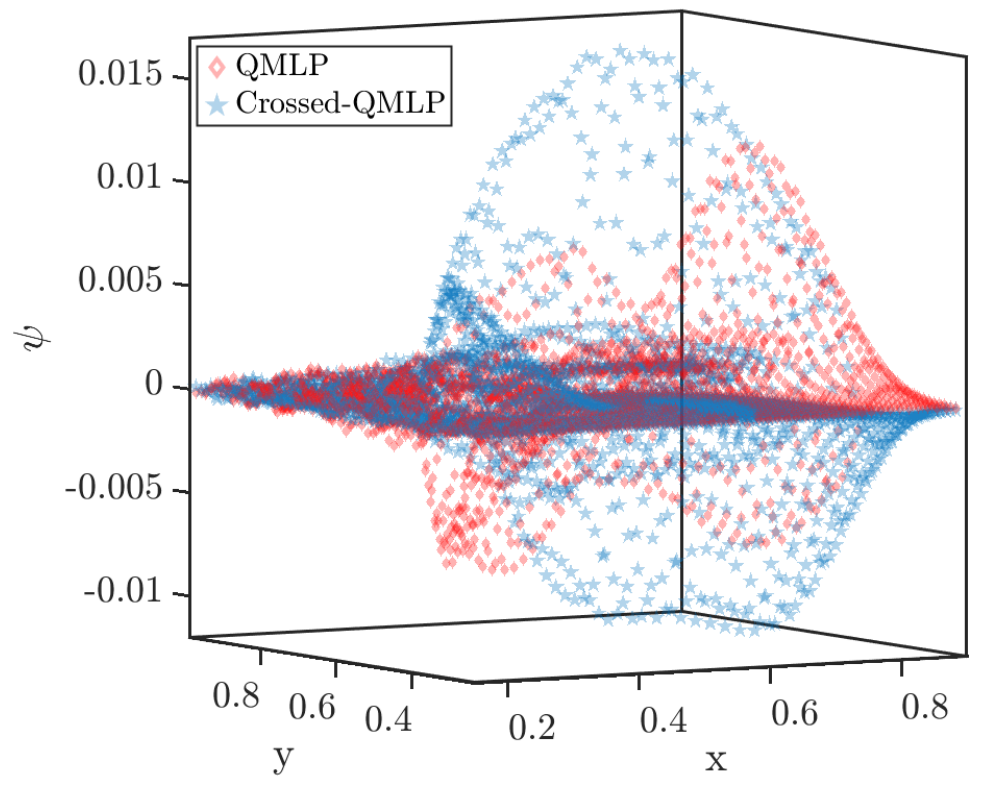}
			\caption{\centering\footnotesize Magnifying results between forward hierarchy and hybrid architecture of crossed-QMLP at $t$=0.15 s.}  
			\label{fig:6b}
		\end{subfigure}
		\begin{subfigure}[t]{0.4\textwidth}
			\centering
			\includegraphics[scale=0.33]{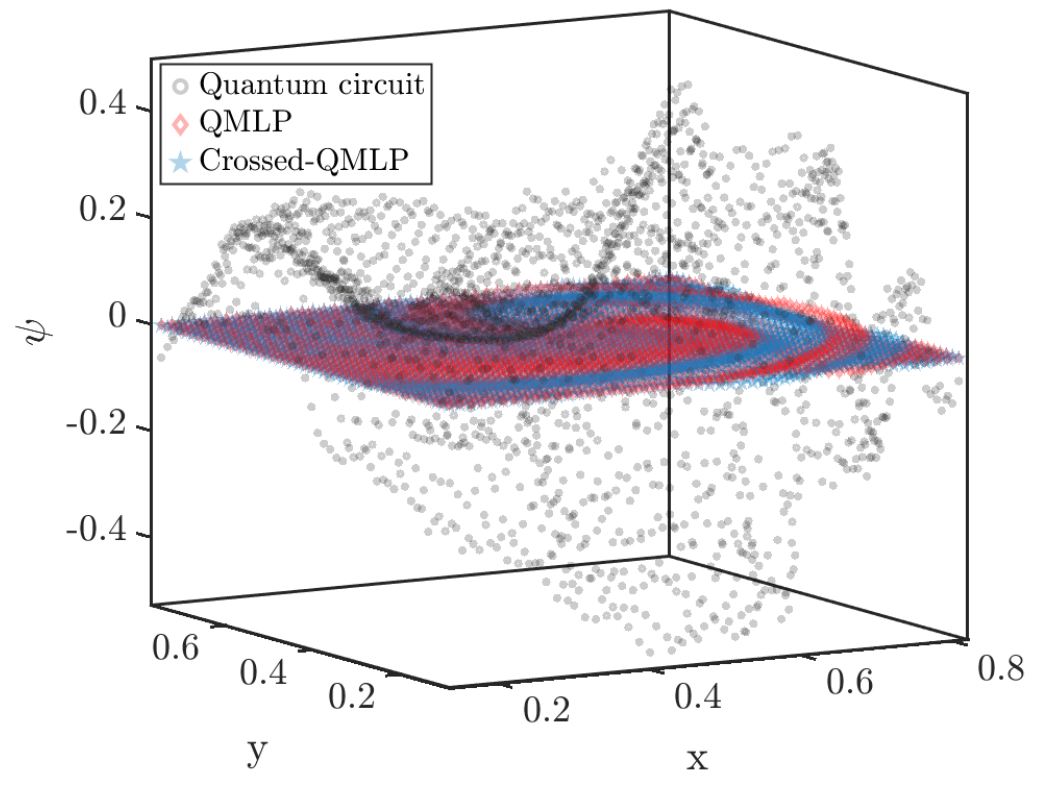}
			\caption{\centering\footnotesize Three quantum models at $t$=0.35 s.}
			\label{fig:6a}
		\end{subfigure}
		\begin{subfigure}[t]{0.4\textwidth}
			\centering
			\includegraphics[scale=0.33]{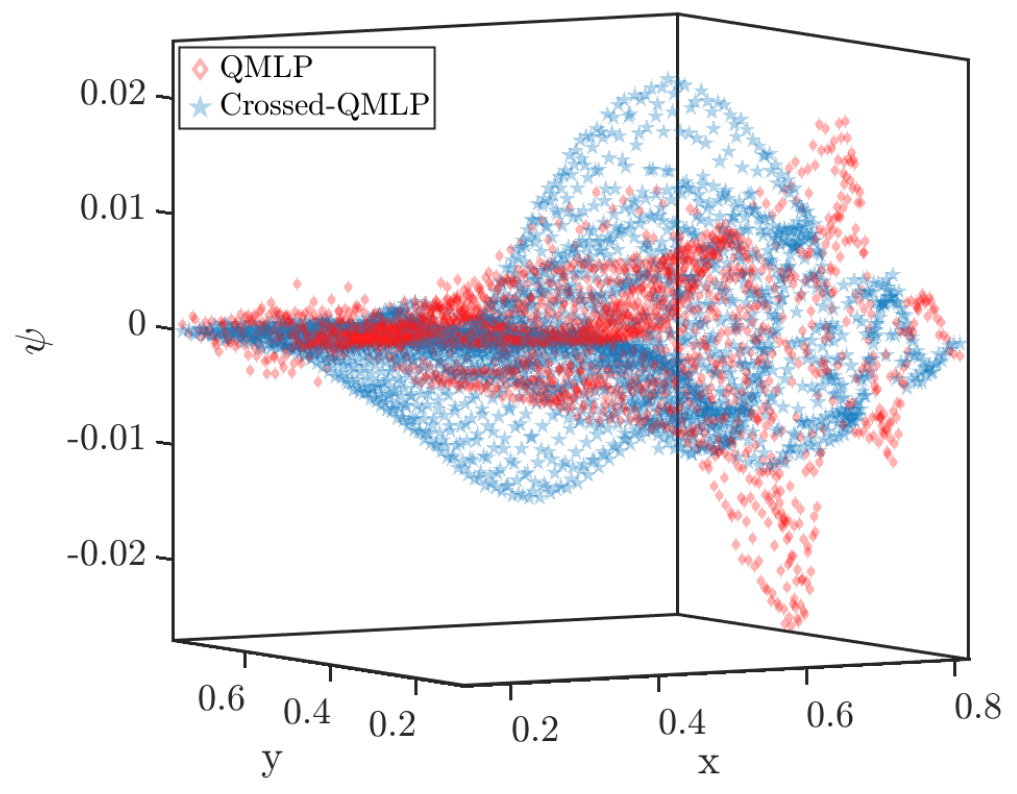}
			\caption{\centering\footnotesize Magnifying results between forward hierarchy and hybrid architecture of crossed-QMLP at $t$=0.35 s.}  
			\label{fig:6b}
		\end{subfigure}
		\begin{subfigure}[t]{0.4\textwidth}
			\centering
			\includegraphics[scale=0.33]{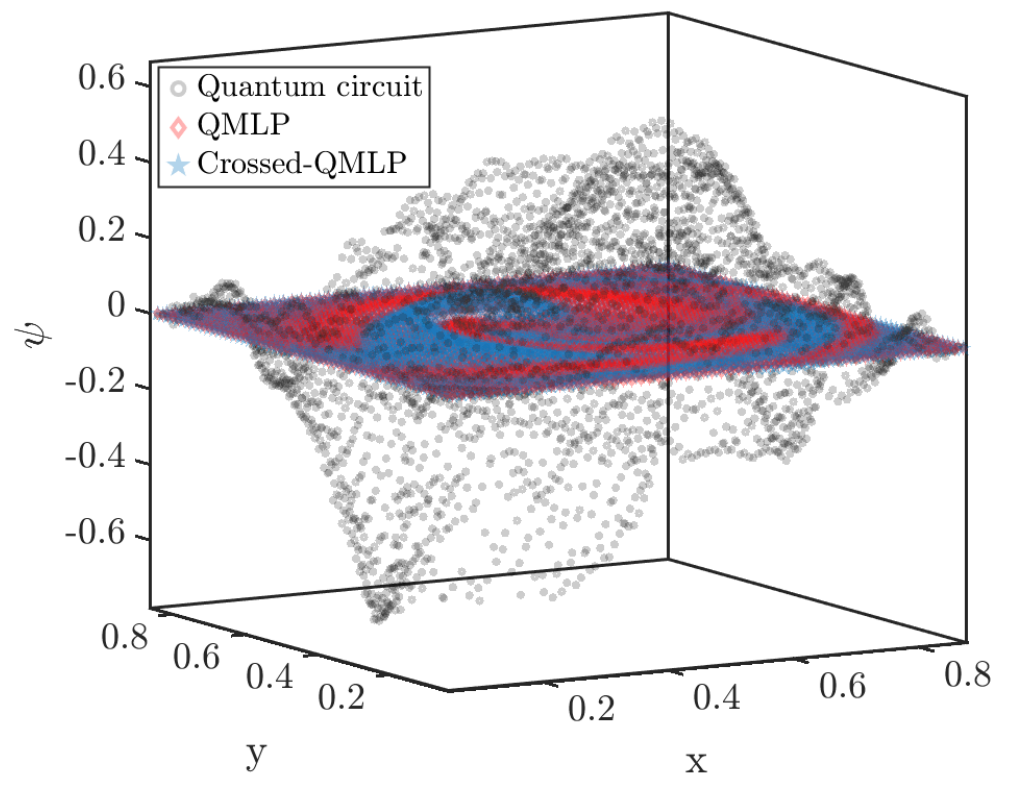}
			\caption{\centering\footnotesize Three quantum models at $t$=0.60 s.}
			\label{fig:6a}
		\end{subfigure}
		\begin{subfigure}[t]{0.4\textwidth}
			\centering
			\includegraphics[scale=0.33]{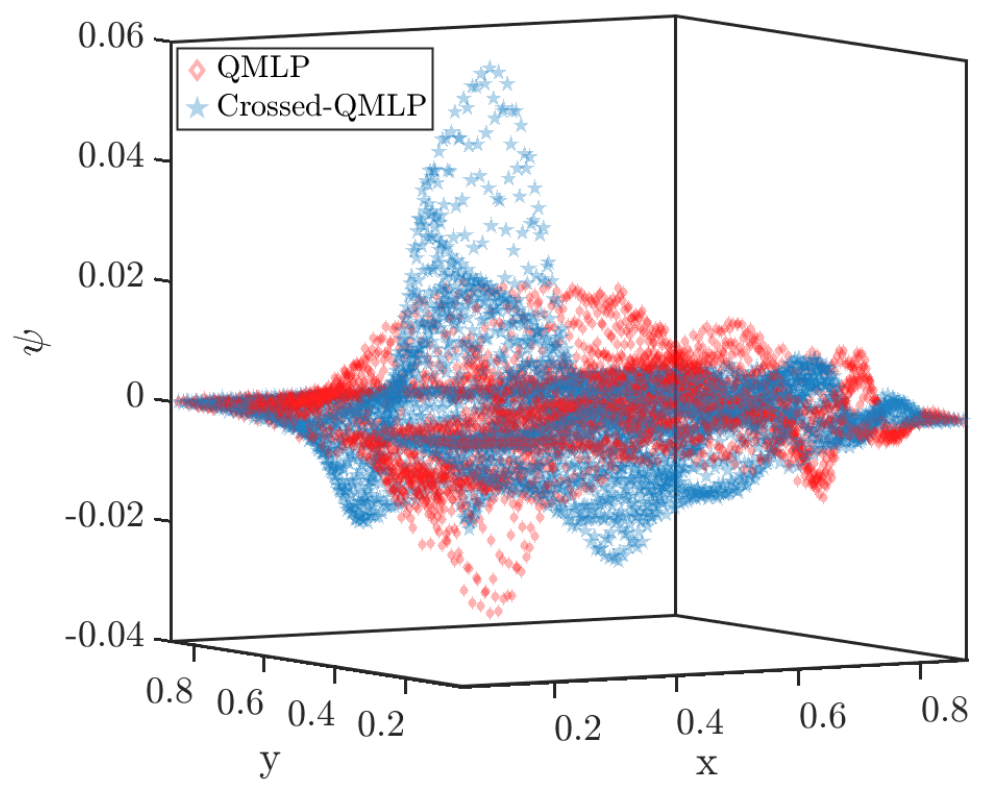}
			\caption{\centering\footnotesize Magnifying results between forward hierarchy and hybrid architecture of crossed-QMLP at $t$=0.60 s.}  
			\label{fig:6b}
		\end{subfigure}
		\caption{\small Counted errors of different quantum kernel networks compared with SPH.}
		\label{fig_Case4:Learning3}
	\end{figure}

This dynamic errors of present transient quantum SPH kernel network models are visually corroborated by the spatiotemporal field reconstructions at intermediate kinetic stages. Fig.(\ref{fig_Case4:Learning3}) gives these errors' distributions for peak deformation points ($t$=0.15, 0.35, 0.60 s). It can be observed that the elementary quantum circuit experiences severe morphological distortion and a complete generalization collapse, failing to track the crescent-shaped topology. In stark contrast, the sequentially coupled crossed-QMLP model preserves the structural integrity of scalar fields perfectly, tracking the extreme topological deformations with precision that identically mirrors the baseline set by the classical SPH. The magnitude error maps confirm that the hybrid architecture confines numerical deviations to negligible levels (the quantified relative norm errors are all around 3.43\%).

\begin{figure}[H]
	\centering
	\begin{subfigure}[t]{0.4\textwidth}
		\centering
		\includegraphics[scale=0.33]{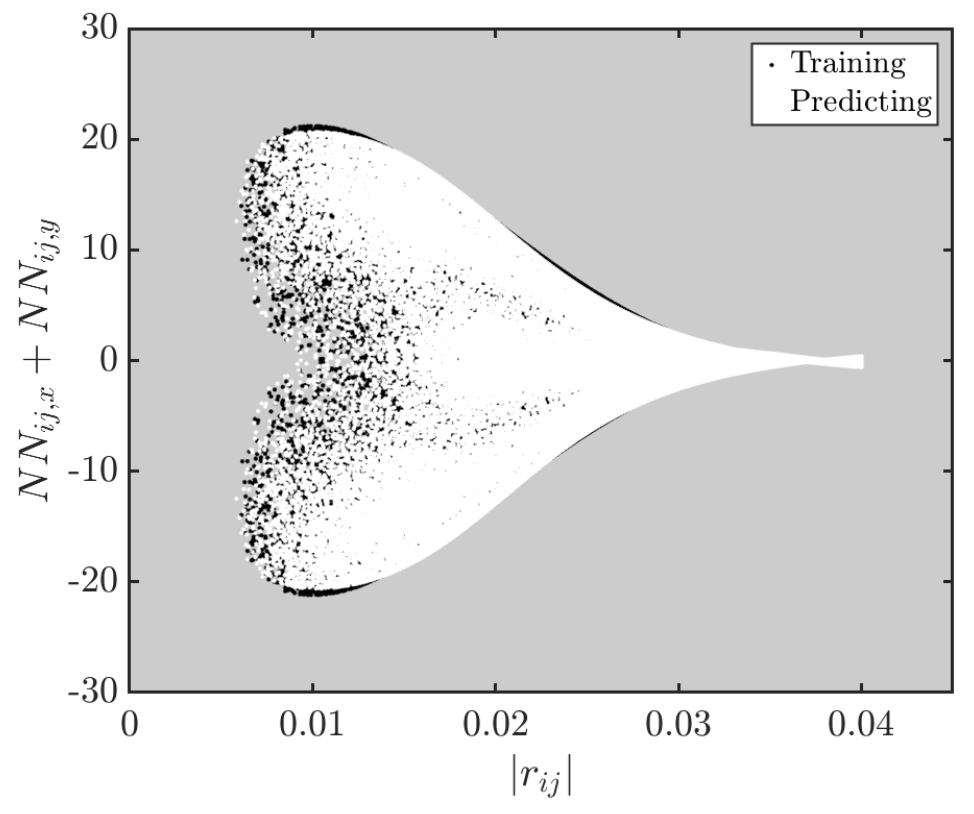}
		\caption{\centering\footnotesize Quantum gradients with norm distance for hybrid crossed-QMLP architecture.}
		\label{fig:6a}
	\end{subfigure}
	\begin{subfigure}[t]{0.4\textwidth}
		\centering
		\includegraphics[scale=0.33]{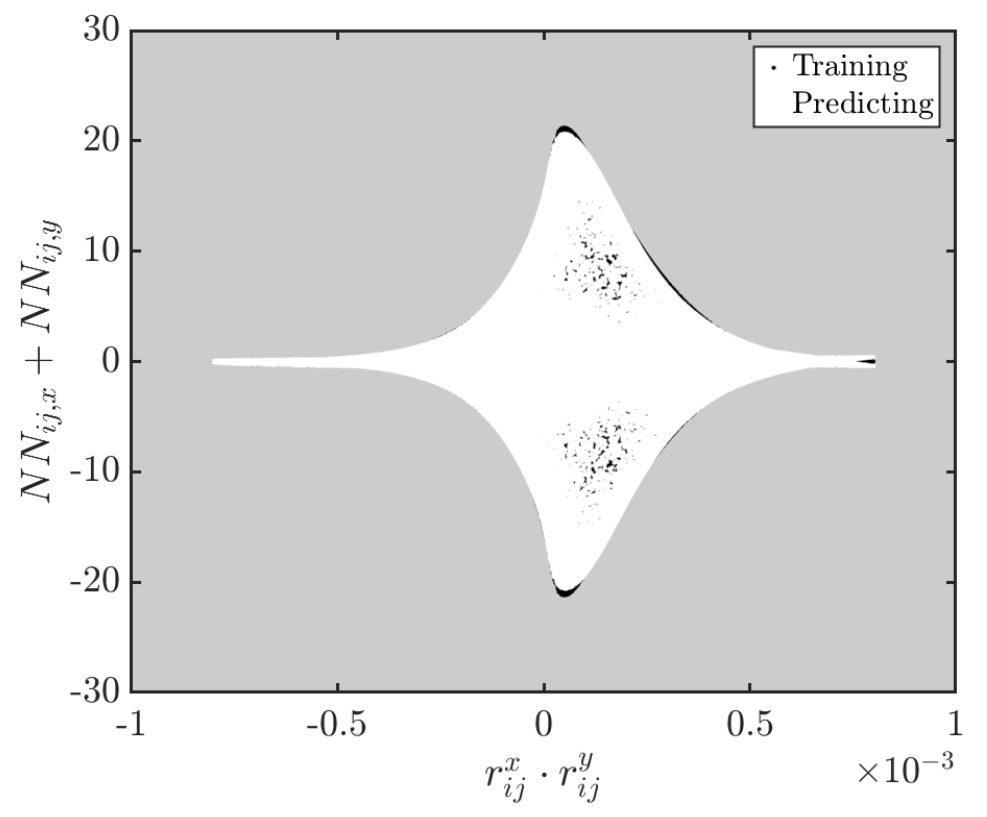}
		\caption{\centering\footnotesize Quantum gradients with inner distance for hybrid crossed-QMLP architecture.}  
		\label{fig:6b}
	\end{subfigure}
	\begin{subfigure}[t]{0.4\textwidth}
		\centering
		\includegraphics[scale=0.33]{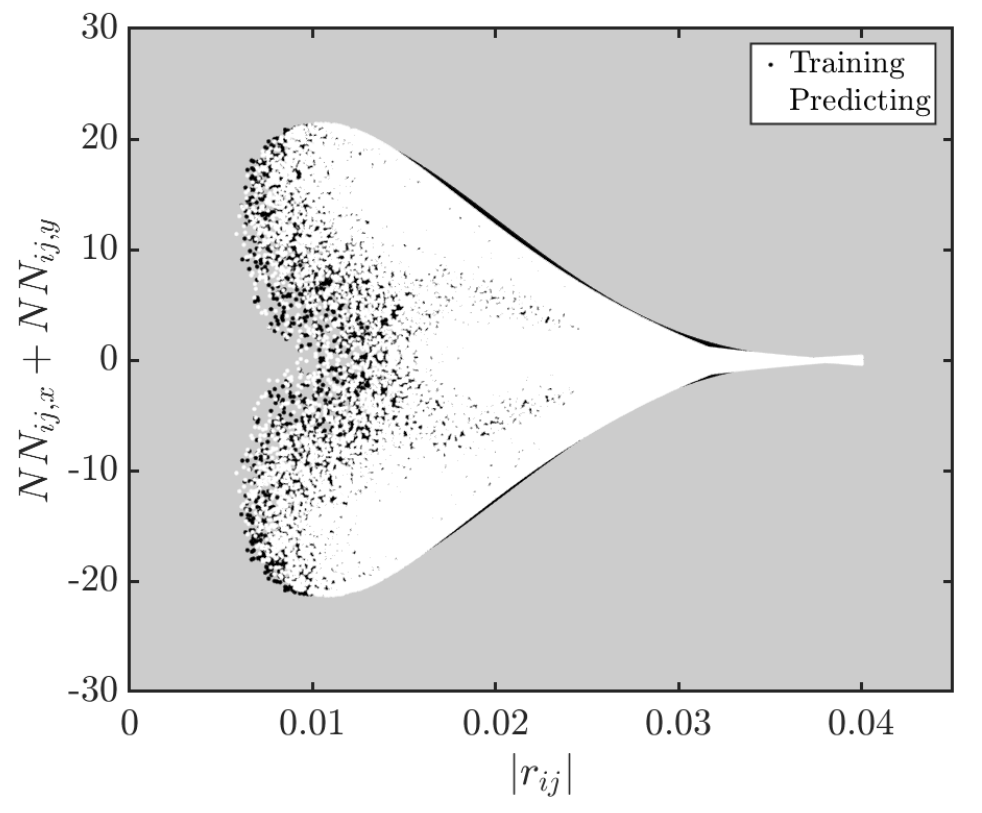}
		\caption{\centering\footnotesize Quantum gradients with norm distance for forward QMLP hierarchy.}
		\label{fig:6a}
	\end{subfigure}
	\begin{subfigure}[t]{0.4\textwidth}
		\centering
		\includegraphics[scale=0.33]{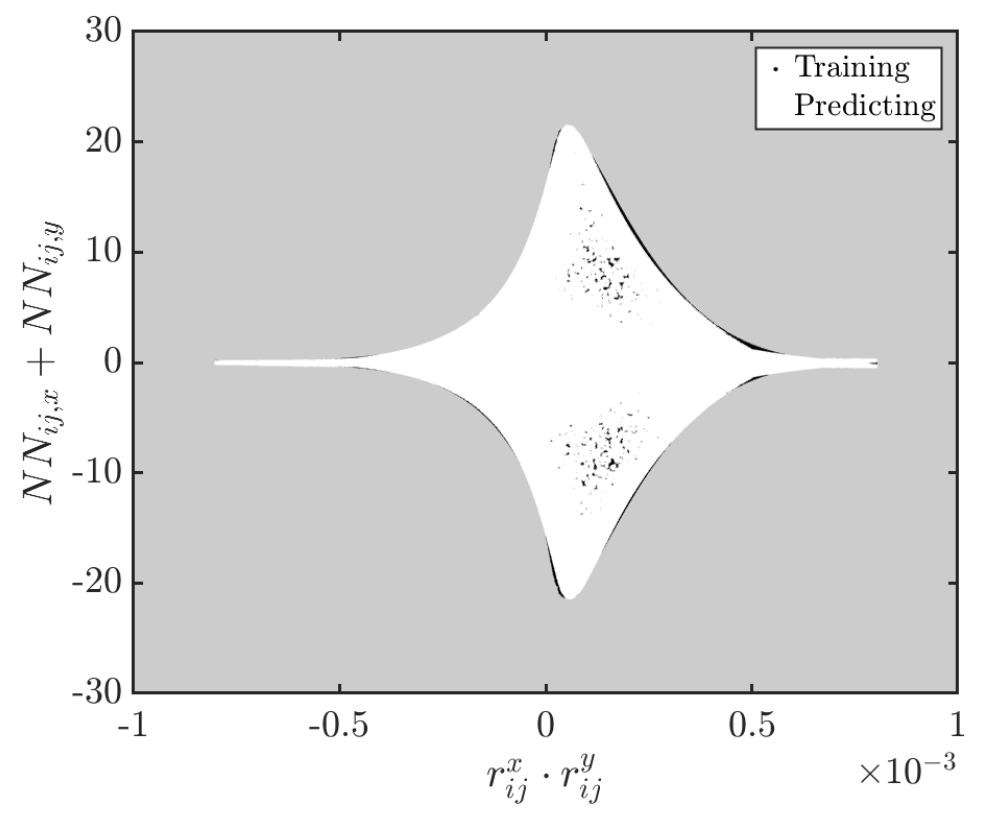}
		\caption{\centering\footnotesize Quantum gradients with inner distance for forward QMLP hierarchy.}  
		\label{fig:6b}
	\end{subfigure}
	\caption{\small Correlations on quantum kernel with SPH ($bs$=640, $lr$=0.01).}
	\label{fig_Case4:smoothingKernel3}
\end{figure}

Furthermore, we extracted the kernel information residing in quantum space, which originates from the present quantum kernel networks on SPH. The resulting data are presented in the following figures. Proceeding to the smoothed kernel conditions, the directional error distributions across varying network configurations are presented in Fig.(\ref{fig_Case4:smoothingKernel2}). It characterizes the correlation between directional distance in the smoothing kernel space and its directional derivatives, as well as the effectiveness and accuracy under quantum kernel circuit learning. Fig.(\ref{fig_Case4:smoothingKernel3}) also depicts the correlations on quantum kernel with SPH, which can be observed that these quantum circuits with forward networks both possess advantageous accuracy in the quantum smoothed kernel space.

Finally, dynamic analysis comparing forward-only smoothing networks against hybrid crossed architectures clearly indicates that pure quantum gates inherently lack strong generalization capabilities for complex continuous variations. By sequentially sandwiching the deep parameterized quantum components between classical dense neural layers, this bidirectional classical compression provides a necessary data-guiding buffer. This setup effectively adapts the parameter-specific generalization of quantum gates, minimizing relative errors comprehensively and ensuring stable, time-variant quantum feature extraction in demanding transient SPH simulations.

	\section{Concluding Remarks}
	\normalsize \hspace{10pt}
	In this work, we have pioneered a systematic integration of quantum intelligence with smoothed particle hydrodynamics (SPH) by developing a hierarchy of Lagrangian quantum network models. The proposed framework bridges the gap between meshfree particle methods and quantum computing through three progressive architectural levels: a single quantum circuit baseline, forward quantum network hierarchies, and hybrid crossed network architectures. The key findings and contributions are summarized as follows.
	
	First, we demonstrated that elementary quantum circuits lack generalization capability for complex, unstructured SPH kernel spaces to an extent. To overcome this limitation, a forward neural network (hidden layers) is introduced to adapt the parameter‑specific generalization of quantum gates, significantly enhancing data fitting performance. 
	
	Second, the hybrid crossed‑QMLP architecture, which sequentially sandwiches deep parameterized quantum components between classical dense layers, exhibits superior accuracy and convergence stability compared to both pure quantum circuits and forward hierarchies. This design leverages bidirectional classical compression to mitigate barren plateaus and quantum noise, enabling robust gradient‑based optimization. 
	
	Third, extensive numerical tests on static multi‑level nebula vortex reconstructions and spatio-temporal scalar field advection demonstrate that the proposed quantum‑intelligent SPH paradigm achieves fitting accuracy comparable to classical SPH while offering a foundation for future exponential acceleration via fault‑tolerant quantum hardware.
	
	Nevertheless, the current study faces limitations in computational efficiency and hardware implementation. All experiments are conducted on quantum simulators due to the decoherence and gate errors of NISQ devices. The overhead of classical‑quantum data encoding and parameter tuning remains nontrivial. Future work will focus on three directions: (i) implementing the proposed models on real quantum processors with error mitigation techniques; (ii) exploring genuine quantum speedup for neighbor search and pairwise force calculations using Grover‑amplitude amplification and quantum linear system algorithms; (iii) extending the Lagrangian quantum network hierarchy to three‑dimensional, multi‑physics SPH simulations.
	
	In summary, this research establishes the first theoretical and experimental foundation for quantum enhanced meshfree/particle methods, mapping unstructured Lagrangian particle topologies into quantum circuits. Certainly, this insight is derived solely from the present study, and further improvements require additional extensions and in-depth considerations. Although still in its infancy, the proposed quantum‑intelligent SPH paradigm opens a new scientific direction at the intersection of computational mechanics, quantum computing and artificial intelligence.
	\\
	\\
	\noindent  \textbf{Authorship contribution statement}\quad\small
	{\textit{Yudong Li:} Writing – review \& editing, Writing – original draft, Visualization, Validation, Methodology, Investigation, Data curation. 
		\textit{Wenkui Shi:} Writing – review \& editing, Investigation, Data curation, Validation. 
		\textit{Chunfa Wang:} Investigation, Formal analysis. 
		\textit{Zhihao Qian:} Resources, Investigation, Funding acquisition. 
		\textit{Zhiqiang Feng:} Supervision, Software, Resources, Project administration, Investigation, Funding acquisition, Formal analysis.
		\textit{Moubin Liu:} Writing – review \& editing, Supervision, Resources, Methodology, Investigation, Formal analysis, Conceptualization. }
	\\
	\\
	\noindent  \textbf{Declaration of competing interest}\quad\small
	{The authors declare that they have no known competing financial
		interests or personal relationships that could have appeared to influence
		the work reported in this paper.}
	\\
	\\
	\noindent  \textbf{Acknowledgements}\quad\small
	{This work has been partially supported by the National Natural Science Foundation of China [Grant Nos. 12602300, 12032002 and U22A20256] and the Sino-German Mobility Programme [No. M-0210]. The experiments were executed on the Origin Quantum Cloud and quantum simulators.}
	
	\bibliographystyle{elsarticle-num} 
	\begin{footnotesize }
		\bibliography{reference}
	\end{footnotesize }
	
\end{document}